\documentclass[12pt,letterpaper]{article}
\usepackage{amsmath,amssymb,array,calc,rotating,epsfig,psfrag,amscd, cite}

\numberwithin{equation}{section}

\makeatletter
\renewcommand\section{\@startsection {section}{1}{\z@}%
                                   {-3.5ex \@plus -1ex \@minus -.2ex}
                                   {2.3ex \@plus.2ex}%
                                   {\normalfont\large\bfseries}}
\renewcommand\subsection{\@startsection{subsection}{2}{\z@}%
                                     {-3.25ex\@plus -1ex \@minus -.2ex}%
                                     {1.5ex \@plus .2ex}%
                                     {\normalfont\bfseries}}
\makeatother

\let\non\nonumber

\let\a=\alpha\let\b=\beta

\let\S=\Sigma

\newcommand{\bea}{\begin{eqnarray}}
\newcommand{\eea}{\end{eqnarray}}
\newcommand{\be}{\begin{equation}}
\newcommand{\ee}{\end{equation}}

\newcommand{\m}{\mu}

\newcommand{\p}{\partial}

\newcommand{\C}[1]{$(\ref{#1})$}

\def\IZ{\relax\ifmmode\mathchoice
{\hbox{\cmss Z\kern-.4em Z}}{\hbox{\cmss Z\kern-.4em Z}}
{\lower.9pt\hbox{\cmsss Z\kern-.4em Z}} {\lower1.2pt\hbox{\cmsss
Z\kern-.4em Z}}\else{\cmss Z\kern-.4em Z}\fi}
\def\IR{\relax{\rm I\kern-.18em R}}

\def\one{{\hbox{ 1\kern-.8mm l}}}

\newlength{\bredde}
\def\slash#1{\settowidth{\bredde}{$#1$}\ifmmode\,\raisebox{.15ex}{/}
\hspace*{-\bredde} #1\else$\,\raisebox{.15ex}{/}\hspace*{-\bredde}
#1$\fi}

\newsavebox{\zzzbar}
\sbox{\zzzbar}
  {\setlength{\unitlength}{0.9em}
  \begin{picture}(0.6,0.7)
  \thinlines
  \put(0,0){\line(1,0){0.6}}
  \put(0,0.75){\line(1,0){0.575}}
  \multiput(0,0)(0.0125,0.025){30}{\rule{0.3pt}{0.3pt}}
  \multiput(0.2,0)(0.0125,0.025){30}{\rule{0.3pt}{0.3pt}}
  \put(0,0.75){\line(0,-1){0.15}}
  \put(0.015,0.75){\line(0,-1){0.1}}
  \put(0.03,0.75){\line(0,-1){0.075}}
  \put(0.045,0.75){\line(0,-1){0.05}}
  \put(0.05,0.75){\line(0,-1){0.025}}
  \put(0.6,0){\line(0,1){0.15}}
  \put(0.585,0){\line(0,1){0.1}}
  \put(0.57,0){\line(0,1){0.075}}
  \put(0.555,0){\line(0,1){0.05}}
  \put(0.55,0){\line(0,1){0.025}}
  \end{picture}}

\newcommand{\ena}{\end{eqnarray}}
\newcommand{\beqa}{\begin{eqnarray}}
\newcommand{\eeqa}{\end{eqnarray}}

\renewcommand{\b}{\beta}

\def\a{\alpha}
\def\b{\beta}

\def\m{\mu}

\def\S{\Sigma}

\begin{document}
\begin{titlepage}

\begin{center}



\vskip 2 cm
{\Large \bf Elliptic Modular Graphs, Eigenvalue Equations and Algebraic Identities}\\
\vskip 1.25 cm { Anirban Basu\footnote{email address:
    anirbanbasu@hri.res.in} } \\
{\vskip 0.5cm  Harish--Chandra Research Institute, A CI of Homi Bhabha National
Institute, \\ Chhatnag Road, Jhusi, Prayagraj 211019, India}

\end{center}

\vskip 2 cm

\begin{abstract}
\baselineskip=18pt

We obtain eigenvalue equations satisfied by various elliptic modular graphs with five links where two of the vertices are unintegrated. Solving them leads to several non--trivial algebraic identities between these graphs.

\end{abstract}

\end{titlepage}


\section{Introduction}

Modular graph functions~\cite{DHoker:2015gmr,DHoker:2015wxz} arise in the analysis of the low momentum expansion of string amplitudes at genus one. The links of these graphs are given by the scalar Green function or its worldsheet derivatives on the toroidal worldsheet $\Sigma$, while the vertices are integrated over $\Sigma$. In fact, using translational invariance on the torus, one of the vertices can be fixed and hence unintegrated. These graphs are invariant under $SL(2,\mathbb{Z})$ transformations of the complex structure $\tau$ of the torus. More generally, one can define modular graph forms that are $SL(2,\mathbb{Z})$ covariant, but we shall not consider them in this paper. 

Let us consider modular graph functions where the links of the graphs are given by the scalar Green function, and not its derivatives. These graphs satisfy various eigenvalue equations, as well as algebraic identities among themselves, illustrating that many of them are not independent. This leads to a rich underlying structure satisfied by them (see \cite{Berkovits:2022ivl,Dorigoni:2022iem,DHoker:2022dxx} for reviews).  While one can analyze various properties of these graphs, our primary focus will be on the algebraic identities between them, which arise on solving the
eigenvalue equations they satisfy\footnote{See \cite{Dorigoni:2022npe} for a recent review and discussion of various methods in the literature to obtain these results, and \cite{Dorigoni:2022bcx} for an analysis of asymptotic expansions.}. The various techniques used to obtain these results, very schematically, all essentially involve taking appropriate derivatives of these graphs with respect to the complex structure modulus of the torus, and simplifying the resulting expressions to get the answer. 
In fact for the various cases that have been considered, in the intermediate steps of the analysis, one generically also has graphs where the links are given by the derivatives of the Green function, and the final answer involves cancellations between such contributions to yield expressions involving graphs where the links are given only by the Green function. 
While this has led to many interesting results involving such graphs, in general obtaining algebraic identities between graphs of distinct topologies with arbitrary number of links remains a challenging problem. Hence understanding various properties of these graphs, as well as those whose links are also given by the derivatives of the Green function, remains an interesting arena to explore.                     

A natural generalization of modular graphs are elliptic modular graphs~\cite{DHoker:2015wxz,DHoker:2017pvk,DHoker:2018mys}. These arise in the asymptotic expansion of genus two graphs around the non--separating node on the moduli space of genus two Riemann surfaces. These objects are interesting in their own right and can be studied without any reference to their genus two origin, which is the viewpoint we shall take. In an elliptic modular graph, at least two of the vertices are unintegrated over the toroidal worldsheet $\S$. In fact, we shall consider the simplest case where only two of the vertices are unintegrated, which we label as 0 and $v$ on $\S$. These graphs are invariant under the $SL(2,\mathbb{Z})$ transformation
\be \tau \rightarrow \frac{a\tau+b}{c\tau+d}, \quad v \rightarrow \frac{v}{c\tau+d} ,\ee
where $a,b,c,d \in \mathbb{Z}$ and $ad-bc=1$.
Note that on identifying the two unintegrated vertices, the elliptic modular graph reduces to a modular graph. Various properties of elliptic modular graphs have been studied in~\cite{DHoker:2020tcq,Basu:2020pey,Basu:2020iok,DHoker:2020hlp,Basu:2021xdt,Hidding:2022vjf,Basu:2022hck}. Much less is known about the elliptic modular graphs compared to modular graphs.          

Given the discussions above, it is very natural to ask how much of the structure of the eigenvalue equations satisfied by the modular graphs as well as the various algebraic identities between themselves, generalize to the case of elliptic modular graphs. The aim of this paper is to analyze this issue in detail for specific cases. While it is not obvious which elliptic graphs to look at, there is a very natural set of graphs worth considering: they are the ones that are obtained by cutting open in all possible ways modular graphs such that two of the vertices are unintegrated, where the parent modular graphs are the ones that satisfy algebraic relations among themselves. In fact, the results of~\cite{DHoker:2020tcq,Basu:2020pey,DHoker:2020hlp,Hidding:2022vjf} provide a simple example of an algebraic relation involving elliptic modular graphs with up to four links where the graphs can be obtained by cutting open modular graphs that arise in the low momentum expansion of the four graviton amplitude in type II string theory. This amplitude yields the simplest set of modular graphs as a consequence of maximal supersymmetry, where the links are given by the Green function. Thus going beyond graphs with four links, the first non--trivial case is to consider modular graphs with five links that arise in the four graviton amplitude, cut them open in all possible ways and try to obtain algebraic identities involving the elliptic modular graphs that arise in the process. This is what we shall accomplish in this paper. While the techniques certainly allow generalizations to graphs with more links, obtaining such identities involving generic graphs remains to be understood with potentially important consequences, for example, constructing a basis of graphs\footnote{We have recently obtained one such non--trivial algebraic identity involving a family of one parameter graphs in~\cite{Basu:2022hck}. While this a far cry from the general case, it is a first step.}.                

\begin{figure}[ht]
\label{a}
\begin{center}
\[
\mbox{\begin{picture}(250,295)(0,0)
\includegraphics[scale=.8]{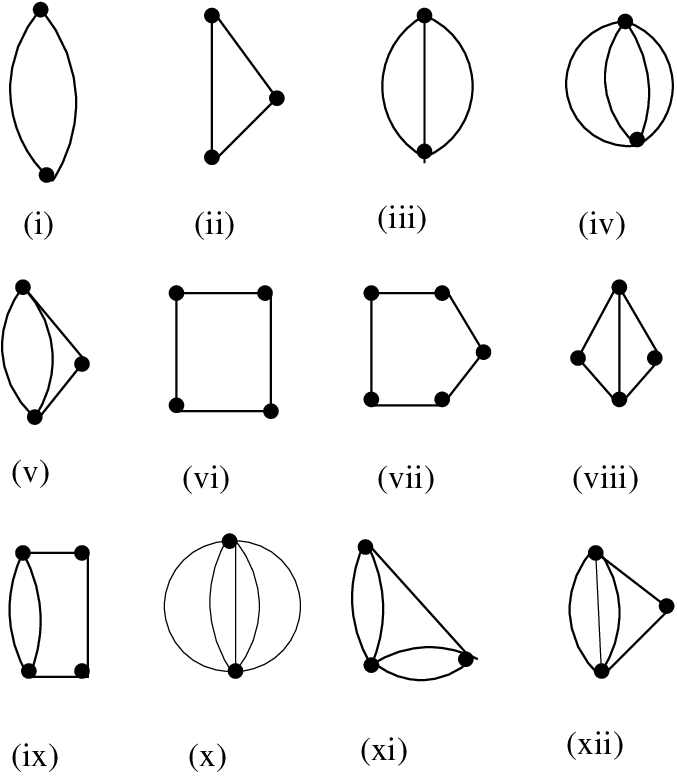}
\end{picture}}
\]
\caption{Modular graphs: (i) $E_2$ (ii) $E_3$ (iii) $D_3$ (iv) $D_4$ (v) $C_{1,1,2}$ (vi) $E_4$ (vii) $E_5$ (viii) $C_{1,2,2}$ (ix) $C_{1,1,3}$ (x) $D_5$ (xi) $D_{1,2,2}$ (xii) $D_{1,1,3}$} 
\end{center}
\end{figure}

Keeping this in mind,  we first consider all the modular graphs that have up to five links given by the scalar Green function that arise in the low momentum expansion of the four graviton amplitude at genus one in type II string theory. These include the ones that arise in this expansion up to the $D^{10}\mathcal{R}^4$ interaction~\cite{Green:1999pv,Green:2008uj}, where schematically $\mathcal{R}$ represents the Riemann tensor, and $D$ is a derivative. While one can analyze various properties of these graphs, for reasons discussed above, we shall focus on the eigenvalue equations they satisfy, and the various algebraic identities between themselves. The various graphs that are relevant for our purposes are given in figure \ref{a}. While some of them arise in the four graviton amplitude, the rest arise in the relations involving them. 

These modular graphs satisfy various eigenvalue equations, as well as several algebraic relations between themselves~\cite{Green:2008uj,DHoker:2015gmr,Basu:2015ayg,DHoker:2016mwo,Basu:2016xrt,Basu:2016kli,Basu:2016mmk,DHoker:2016quv,Kleinschmidt:2017ege,Broedel:2018izr,DHoker:2019blr,Basu:2019idd,Gerken:2019cxz,Gerken:2020yii,Gerken:2020aju} demonstrating that many of them are not independent. Among them, the one loop graph with $s$ links satisfies the eigenvalue equation ($s \geq 2$)
\be \label{Eisen}\Delta E_s = s(s-1)E_s \ee 
where $\Delta$ is the Laplacian defined by
\be \label{Lap}\Delta = 4\tau_2^2\frac{\p^2}{\p\tau \p\overline\tau}\ee
and $E_s$ is the non--holomorphic Eisenstein series. The cases $2 \leq s \leq 5$ will be relevant for our purposes. We now consider the equations involving the other graphs.

While eigenvalue equations have been obtained for certain families of dihedral graphs defined as $C_{a,b,c}$ and $C_{a,b,c,d}$~\cite{DHoker:2015gmr,Basu:2019idd}, we shall consider those that have less than or equal to five links, as well as several others that are not included in these families. We now list the eigenvalue equations satisfied by the various graphs (apart from $E_s$) as well as the algebraic relations between them. 
  
With three links, we have the eigenvalue equation
\be \label{D3}\Delta D_3 = 6 E_3 \ee
and the algebraic identity
\be D_3 = E_3 +\zeta(3)\ee
between the graphs. For this case as well as the others to follow, the algebraic identities can be obtained by solving the eigenvalue equations along with a knowledge of the asymptotic expansion of the graphs around the cusp $\tau_2 \rightarrow \infty$.  

With four links, the eigenvalue equations and algebraic identities are
\bea \label{e4} &&\Big(\Delta -2\Big) C_{1,1,2} = 9 E_4- E_2^2 , \non \\ &&\Big(\Delta -2\Big) \Big(D_4 - 3 E_2^2\Big) = 36 E_4 -24 E_2^2\eea
and
\be \label{a4} D_4 -3 E_2^2 = 24 C_{1,1,2}  - 18 E_4 \ee
respectively. 
Finally with five links, we similarly obtain the eigenvalue equations
\bea \label{e5}&&\Delta C_{1,2,2} = 8 E_5, \non \\  &&\Big(\Delta -6\Big) C_{1,1,3} = \frac{86}{5} E_5 - 4E_2 E_3+\frac{\zeta(5)}{10}, \non \\ &&\Big(\Delta -6\Big) 
D_{1,2,2} = \frac{144}{5}E_5-8E_2E_3 -\frac{8}{5} \zeta(5),\non \\
&&\Big(\Delta -6\Big) \Big(D_{1,1,3} - 3 E_2 E_3\Big) = \frac{162}{5}E_5 - 30 E_2 E_3 -\frac{3}{10}\zeta(5),\non \\ &&\Big(\Delta -6\Big) \Big(D_5 - 10 E_2 D_3\Big) = 360 E_5 -240 E_2 E_3 -90 \zeta(5),\eea
as well as the algebraic identities
\bea \label{a5}&& C_{1,2,2} = \frac{2}{5} E_5 + \frac{\zeta(5)}{30},\non \\ && 10 D_{1,2,2} = 20 C_{1,1,3} - 4 E_5 +3\zeta(5),\non \\  && 40 \Big( D_{1,1,3}-3 E_2E_3\Big) = 300 C_{1,1,3}  -276 E_5 + 7\zeta(5), \non \\ && D_5 - 10 E_2 D_3 = 60 C_{1,1,3}  -48 E_5 + 16 \zeta(5). \eea
Thus analyzing these graphs leads to a rich structure of eigenvalue equations they satisfy, as well as non--trivial algebraic identities between themselves which demonstrate that these graphs are all not independent. Our aim is to generalize these results for elliptic modular graphs by cutting open the various graphs mentioned above, such that only two vertices are unintegrated.  
We shall obtain eigenvalue equations they satisfy, and solve them to get several non--trivial algebraic identities between them.        

Note that various aspects of elliptic modular graphs, including the issues mentioned above, have been discussed in~\cite{DHoker:2020hlp,Hidding:2022vjf}, where the action of complex structure modulus as well as $v$ derivatives on them have been studied in detail. The analysis has been done using the holomorphic subgraph reduction technique, the technique of generating series and also by expressing the graphs in terms of iterated modular integrals, generalizing the analysis of~\cite{DHoker:2016mwo,Gerken:2020yii} to the case of elliptic graphs. The techniques we use are purely graphical, and very different from the algebraic techniques used in these references\footnote{This strategy to obtain the eigenvalue equations has been used in~\cite{Basu:2015ayg,Basu:2016xrt,Basu:2016kli,Kleinschmidt:2017ege,Basu:2019idd,Basu:2020pey,Basu:2021xdt,Basu:2022hck}. It has also been used to obtain relations between graphs with and without links involving derivatives of Green functions in~\cite{Basu:2016mmk,Basu:2020iok}.}. Needless to say, there is sometimes common ground in the intermediate steps, for example, lattice momentum conservation identities in the algebraic approach which change the labels of the various graphs  follow from integration by parts in the graphical approach, which is evident from the details to follow. Also it is worth mentioning that our analysis only involves elementary properties of the Green function and nothing more, and modularity of the various expressions is always manifest. While one can generalize the techniques we use for generic elliptic modular graphs and obtain eigenvalue equations which generically involve graphs with links also given by derivatives of the Green function, this is not our primary aim, as we want to solve these equations to obtain algebraic relations between them, and hence we stick to a tractable set of graphs with a fixed number of links. It turns out this set of graphs only involve those whose links are given by the Green function and not its derivatives.                              

Finally, note that we list a large number of graphs in this paper, making the strategy appear very cumbersome, which is actually not quite the case. In fact, the variations of the graphs produce most of them directly, and we list them graphically rather than write algebraic expressions for brevity, since it is most convenient to write every equation graphically. The non--trivial graphs in the intermediate stages are the auxiliary graphs as detailed later, which are much smaller in number. Also very importantly, many of these graphs can be simplified in terms of those with links given only by the Green function. This can be trivially used to simplify each eigenvalue equation considerably. However we refrain from doing so in order to easily see various cancellations between the different eigenvalue equations. Thus enormous simplifications are achieved right at the end.           

Given the length of the paper, we now briefly summarize its contents for the convenience of the reader:

$\bullet$ In section 2, we review useful properties of the Green function, outline the strategy to obtain the eigenvalue equations and introduce the various elliptic modular graphs.

$\bullet$ Section 3 gives the details of calculations of the eigenvalue equations systematically. The role of auxiliary graphs in obtaining simplifications in the intermediate stages of the calculations is elucidated.

$\bullet$ In section 4, these eigenvalue equations are considered further, and simplified. These lead to the very simple eigenvalue equations \C{6}, \C{2} and \C{Eg} where the eigenvalues are of the form $s(s-1)$ for $s=3,2$ and $0$ respectively. Solving them yields the algebraic identities given in figures 99, 100 and 101 respectively. These eigenvalue equations and the algebraic relations are the key results of this paper.  

$\bullet$ In the various appendices, several graphs are simplified. In appendix G, we provide a derivation of the eigenvalue equation of the graph denoted by $C_{a,b,c} (Z)$ in section 3.5.1 of \cite{DHoker:2020hlp} using the techniques in this paper.  

\section{Green function, Laplacian and elliptic modular graphs}

The coordinate on the torus, $z$, is defined by
\be -\frac{1}{2} \leq {\rm Re} z \leq \frac{1}{2} , \quad 0 \leq {\rm Im} z \leq \tau_2.\ee
The integration measure is given by $d^2 z= d{\rm Re}z d{\rm Im} z$, while the Dirac delta function is normalized to satisfy $\int_{\S}d^2 z \delta^2 (z)=1$. 

The links of the graphs we consider in this paper are given by the scalar Green function between the points $z$ and $w$ on the toroidal worldsheet $\S$, which we denote as either $G(z,w)$ or $G(z-w)$, with the $\tau$ dependence being implicit. 
This is given by~\cite{Lerche:1987qk,Green:1999pv}               
\be \label{Green} G(z) = \frac{1}{\pi} \sum_{(m,n) \neq (0,0)} \frac{\tau_2}{\vert m\tau+n\vert^2}e^{\pi[\bar{z}(m\tau+n)- z(m\bar\tau+n)]/\tau_2} ,\ee
which is modular invariant and doubly periodic on the torus\footnote{The sum in \C{Green} is conditionally convergent, and is performed by first considering the $m=0, n \neq 0$ sector, and then the $m\neq 0$ sector where one Poisson resums $n$ to the label $\widehat{n}$, and then considers the $\widehat{n} =0$ and $\widehat{n} \neq 0$ sectors. In fact, there is a logarithmic divergence from the $m\neq 0, \widehat{n}=0$ sector in \C{Green} when $z=0$ which has to be removed to get the finite contribution. These issues will not bother us as we shall always directly work with the expression \C{Green}. }. It is single valued, and hence we can integrate by parts in the various expressions wherever appropriate, and neglect total derivatives which proves to be very useful in our analysis. In the modular invariant graphs in figure 1 as well as the ones that arise later, the vertices are integrated with the $SL(2,\mathbb{Z})$ invariant measure $d^2z/\tau_2$ over $\S$.  

Now from \C{Green} we have that
\be \int_{\S} d^2 z G(z,w)=0\ee
and hence we cannot have a graph where a link representing the Green function ends on an integrated vertex.   

Also the Green function satisfies the equations
\bea \label{maineqn}\overline\p_w \p_z G(z,w) &=& \pi \delta^2 (z-w)- \frac{\pi}{\tau_2}, \non \\ \overline\p_z \p_z G(z,w) &=& -\pi \delta^2 (z-w)+ \frac{\pi}{\tau_2}.\eea
which we shall often use.

To find the eigenvalue equations satisfied by the various graphs, rather that directly acting on them by $\Delta$ in \C{Lap} we analyze the variation of the complex structure deformation parametrized by the variation of the Beltrami differential\footnote{For relevant details of deformation theory relating the variation of the complex structure to the variation of the Beltrami differential, see section 2.6 of~\cite{DHoker:2015gmr} or the review~\cite{DHoker:1988pdl}. }.  
To do so, we use the variations~\cite{Verlinde:1986kw,DHoker:1988pdl,DHoker:2015gmr}
\be \p_\m G(z_1,z_2) = -\frac{1}{\pi} \int_{\S}d^2 z \p_z G(z,z_1)\p_z G(z,z_2), \ee
as well as
\be \overline\p_\m \p_\m G(z_1,z_2)=0\ee
where $\mu$ is the Beltrami differential\footnote{While the Beltrami differential is constant on the torus, its variations are non--trivial.}.

The Laplacian is expressed in terms of the variations as
\be \label{Belt}\Delta = \overline\p_\m \p_\m \ee
which we shall use in obtaining the eigenvalue equations.

Let us now consider the various elliptic modular graphs we shall use in our analysis. From now onwards, for the sake of brevity, we shall refer to all graphs, elliptic or otherwise, as simply graphs.
 
To start with, consider the iterated Green function $G_s (v)$ defined recursively by ($s \geq 1$) 
\be G_{s+1}(v) = \int_{\S} \frac{d^2 z}{\tau_2} G(v,z) G_s (z)\ee
where $G_1 (z)$ is the Green function \C{Green} (thus $G_s (0) = E_s$ for $s \geq 2$). They satisfy
\be \label{Gs}\Delta G_s (v) = s(s-1) G_s (v)\ee
which generalizes \C{Eisen}. We shall need $G_s (v)$ for $s \leq 5$ in our analysis.

Another family of graphs $D_l^{(k)} (v)$ is defined by ($l \geq k$)
\be D_l^{(k)} (v) = \int_{\S} \frac{d^2 z}{\tau_2} G(v,z)^k G(z)^{l-k}= D_l^{(l-k)} (v).\ee
We shall also consider the family of graphs $D^{(s_1,s_2,s_3)} (v)$ defined by
\be D^{(s_1,s_2,s_3)} (v) = \int_{\S} \frac{d^2 z}{\tau_2} G_{s_1} (v,z) G_{s_2} (v,z) G_{s_3} (z)= D^{(s_2,s_1,s_3)} (v),\ee
as well as the family defined by
\be D^{(s_1,s_2,s_3,s_4)} (v) = \int_{\S} \frac{d^2 z}{\tau_2} G_{s_1} (v,z) G_{s_2} (v,z) G_{s_3} (v,z) G_{s_4} (z)\ee
which is symmetric under the permutations of $s_1,s_2$ and $s_3$. 

Let us now mention the graphs having up to five links we shall need in our analysis, apart from $G_s (v)$. 

\begin{figure}[ht]
\label{b}
\begin{center}
\[
\mbox{\begin{picture}(270,90)(0,0)
\includegraphics[scale=.8]{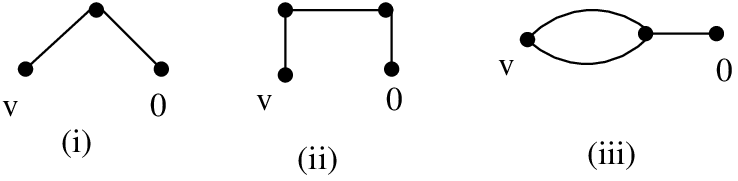}
\end{picture}}
\]
\caption{Elliptic modular graphs with two links: (i) $G_2 (v)$, with three links: (ii) $G_3 (v)$, (iii) $D_3^{(1)} (v)$ } 
\end{center}
\end{figure}

With three links we consider $D_3^{(1)} (v) = D^{(1,1,1)} (v)$ which is obtained from cutting open $D_3$ and hence $D^{(1)}_3 (0) = D_3$, as given in figure \ref{b}.

With four links, we have $D_4^{(1)} (v)$ and $D_4^{(2)} (v)$ which are obtained by cutting open $D_4$, and hence $D_4^{(1)} (0) = D_4^{(2)} (0) = D_4$. We also have $D^{(1,2,1)} (v)$ as well as $D^{(1,1,2)} (v)$ from cutting open $C_{1,1,2}$, and hence $D^{(1,2,1)} (0) = D^{(1,1,2)} (0) =C_{1,1,2}$ as given in figure 3.      

We now consider the graphs with five links\footnote{All the elliptic modular graphs with five links we shall need are given in figure 4.} which forms the central part of the analysis of the paper. While some of them are from the families of graphs mentioned above, the others go beyond. We obtain:

(i) $D^{(2,2,1)} (v)$ and $D^{(1,2,2)} (v)$ from cutting open $C_{1,2,2}$, and hence $D^{(2,2,1)} (0) = D^{(1,2,2)} (0) = C_{1,2,2}$.

(ii) $D^{(1,3,1)} (v)$ and $D^{(1,1,3)} (v)$ from cutting open $C_{1,1,3}$, and hence $D^{(1,3,1)} (0) = D^{(1,1,3)} (0) = C_{1,1,3}$.  

(iii) $D_5^{(1)} (v)$ and $D_5^{(2)} (v)$ from cutting open $D_5$, and hence $D_5^{(1)} (0) = D_5^{(2)} (0) = D_5$,

(iv) $D^{(1,1,1,2)} (v)$, $D^{(1,1,2,1)} (v)$ and $D_5^{(1,2,2)} (v)$ from cutting open $D_{1,1,3}$, and hence $D^{(1,1,1,2)} (0) = D^{(1,1,2,1)} (0) = D_5^{(1,2,2)} (0) = D_{1,1,3}$,

(v) $D_5^{(2;2;1)} (v)$, $D_5^{(1,2,1;1)} (v)$ and $D_5^{(1;2;2)} (v)$ from cutting open $D_{1,2,2}$, and hence $D^{(2;2;1)} (0) = D^{(1,2,1;1)} (0) = D_5^{(1;2;2)} (0) = D_{1,2,2}$.

\begin{figure}[ht]
\label{c}
\begin{center}
\[
\mbox{\begin{picture}(360,150)(0,0)
\includegraphics[scale=.7]{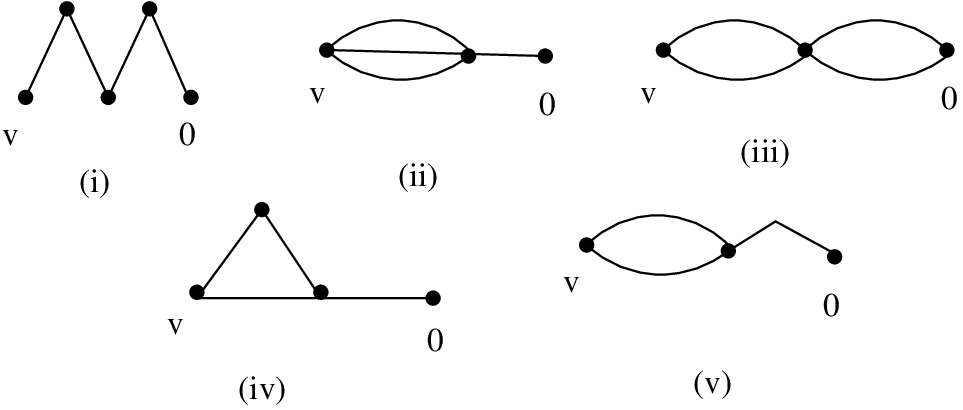}
\end{picture}}
\]
\caption{Elliptic modular graphs with four links: (i) $G_4 (v)$, (ii) $D_4^{(1)} (v)$, (iii) $D_4^{(2)} (v)$, (iv) $D^{(1,2,1)} (v)$, (v) $D^{(1,1,2)} (v)$} 
\end{center}
\end{figure}

Thus this list includes all graphs that can be obtained from cutting open the graphs listed in figure 1, and hence is the complete list of elliptic modular graphs in which the links are given by the Green function and not its derivatives. 

Note that the labels $0$ and $v$ for the unintegrated vertices can be interchanged for each of these graphs.

\section{Eigenvalue equations satisfied by the elliptic modular graphs}

We now consider the eigenvalue equations satisfied by the various elliptic modular graphs. While $G_s (v)$ satisfies an elementary equation given by \C{Gs}, the others satisfy equations which reveal a far richer structure, which is one of the central themes of our analysis.

\begin{figure}[ht]
\begin{center}
\[
\mbox{\begin{picture}(395,375)(0,0)
\includegraphics[scale=.75]{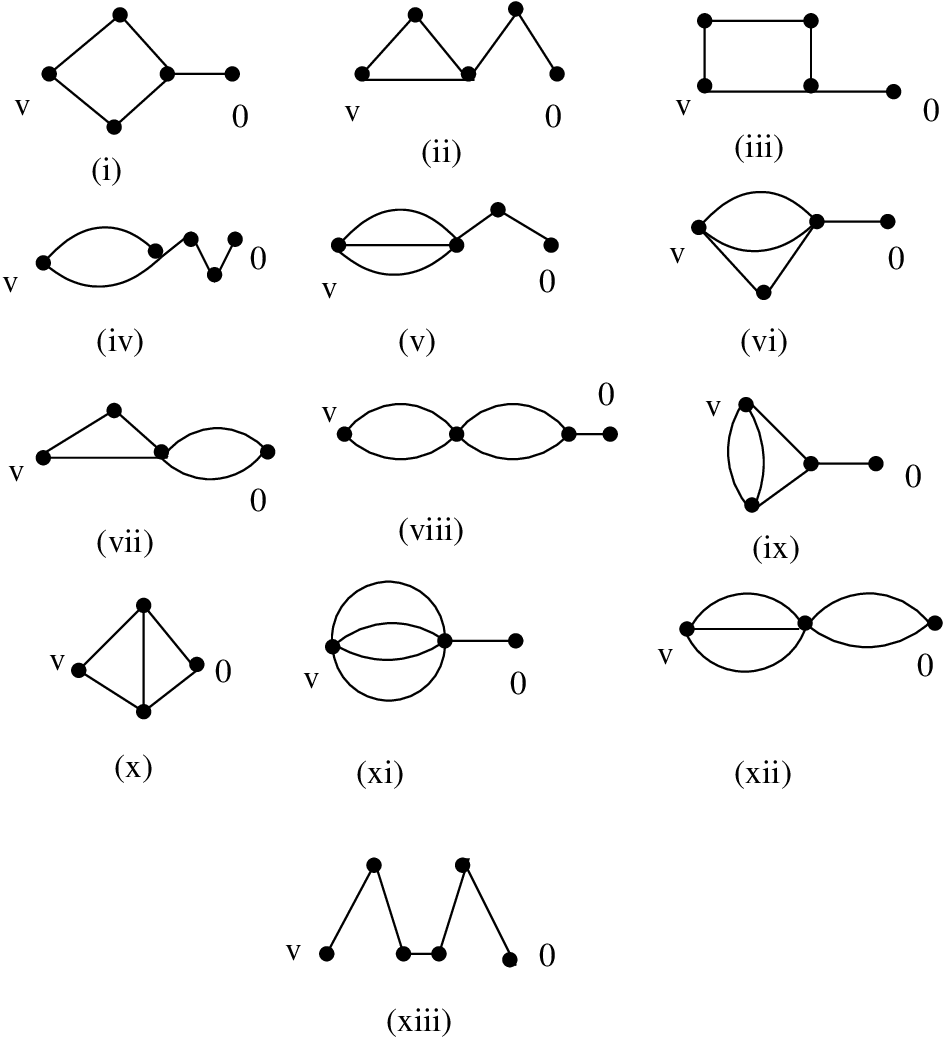}
\end{picture}}
\]
\caption{Elliptic modular graphs with five links: (i) $D^{(2,2,1)} (v)$, (ii) $D^{(1,2,2)} (v)$, (iii) $D^{(1,3,1)} (v)$, (iv) $D^{(1,1,3)} (v)$, (v) $D^{(1,1,1,2)} (v)$, (vi) $D^{(1,1,2,1)} (v)$, (vii) $D_5^{(1,2,2)}(v)$, (viii)$D_5^{(2;2;1)}(v)$, (ix) $D_5^{(1,2,1;1)}(v)$, (x) $D_5^{(1;2;2)}(v)$, (xi) $D_5^{(1)} (v)$, (xii) $D_5^{(2)} (v)$, (xiii) $G_5 (v)$} 
\end{center}
\end{figure}

First we consider the graph $D_3^{(1)} (v)$ with three links. It satisfies the eigenvalue equation~\cite{Basu:2021xdt}
\be \label{D31}\Delta D_3^{(1)} (v)= 2E_3 + 4 G_3 (v) -\frac{2\tau_2}{\pi} \p_v G_2 (v) \overline\p_v G_2 (v).\ee
Thus given the eigenvalue equation satisfied by $G_3 (v)$ we see that there is no algebraic relation between them.

We next consider the graphs with four links. The graph $D^{(1,1,2)} (v)$ satisfies the eigenvalue equation~\cite{Basu:2020pey}
\be \Big(\Delta - 2\Big) D^{(1,1,2)} (v) = 10 G_4 (v) - E_2^2 - E_4 + {\mathcal{F}}_2 (v)^2\ee
where ${\mathcal{F}}_2 (v) = E_2 - G_2 (v)$. Proceeding along the same lines, we get that $D^{(1,2,1)} (v)$ satisfies the eigenvalue equation
\be \Big(\Delta -2\Big) D^{(1,2,1)} (v) = 5 E_4 + 4 G_4 (v) - G_2(v)^2  -\frac{2\tau_2}{\pi} \Big(\p_v G_2 (v) \overline\p_v G_3 (v) + c.c.\Big). \ee

While $D_4^{(1)} (v)$ and $D_4^{(2)} (v)$ each satisfies an eigenvalue equation involving graphs with links given by derivatives of the Green function that is far more involved than the ones above, all the derivatives cancel in an eigenvalue equation involving the combination\footnote{In fact the combination $D_4^{(1)}(v)- 3 D_4^{(2)}(v)/4$ is good enough, however, ${\mathcal{F}}_4(v)$ yields a compact looking equation.}  
\be  {\mathcal{F}}_4 (v) = \frac{D_4}{12} - \frac{D_4^{(1)}(v)}{3} +\frac{D_4^{(2)} (v)}{4},\ee
leading to~\cite{DHoker:2020tcq,Basu:2020pey,DHoker:2020hlp}
\be
\Big( \Delta-2\Big) \Big({\mathcal{F}}_4 (v) - \frac{1}{2} {\mathcal{F}}_2 (v)^2\Big) = - 2 {\mathcal{F}}_2 (v)^2.\ee

In fact analyzing the eigenvalue equations satisfied by $D^{(1,1,2)} (v)$, ${\mathcal{F}}_4 (v)$ and $D_4$, we obtain a non--trivial algebraic identity between the various graphs given by~\cite{Basu:2020pey,Hidding:2022vjf}\footnote{This follows from the fact that $\Phi(v)$ satisfies the equation
\be \Big(\Delta -2 \Big)\Phi (v)=0 \ee
where $\Phi(v)$ is the expression on the left hand side of \C{Phi}. The identity \C{Phi} has been obtained in~\cite{Hidding:2022vjf} using different techniques.} 
\be \label{Phi}D_4^{(1,1,2)} (v) - \frac{D_4^{(1)}(v)}{6} +\frac{D_4^{(2)}(v)}{8} - G_4 (v) - \frac{G_2(v)^2}{4} +\frac{1}{2} E_2 G_2 (v) +\frac{E_4}{4} -\frac{E_2^2}{8}=0.\ee

Let us highlight some of the features in the analysis involving graphs with four links. Generic graphs satisfy eigenvalue equations which involve graphs with links given by derivatives of Green functions, and not just Green functions. However, manipulating a subset of these equations judiciously, we obtain equations where terms involving derivatives of Green functions cancel. Solving these equations eventually leads to the algebraic identity between the graphs. 

It is natural to expect that this phenomenon persists for elliptic modular graphs with arbitrary number of links, given that many modular graphs have been analyzed which satisfy eigenvalue equations involving graphs with links given by Green functions, solving which lead to algebraic identities among themselves. We shall see this expectation is indeed true for the case of elliptic modular graphs with five links, which generalize the structure of \C{e5} and \C{a5} to graphs with two unintegrated vertices.

Thus, we now systematically obtain the eigenvalue equations satisfied by the various graphs with five links. Very briefly, we shall consider the action of \C{Belt} on any graph, and manipulate it to get the final expression, on making heavy use of \C{maineqn}. Naturally, it will be very useful to denote various expressions graphically. Derivatives of Green functions will be denoted as given in figure 5.  

\begin{figure}[ht]
\begin{center}
\[
\mbox{\begin{picture}(260,85)(0,0)
\includegraphics[scale=.7]{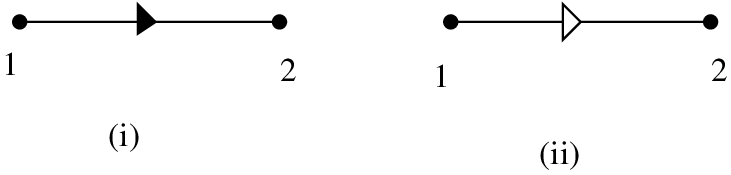}
\end{picture}}
\]
\caption{(i) $\p_{z_2} G(z_1,z_2) = -\p_{z_1} G(z_1,z_2)$, (ii)  $\overline\p_{z_2} G(z_1,z_2) = -\overline\p_{z_1} G(z_1,z_2)$} 
\end{center}
\end{figure}

Also for the sake of brevity, we shall denote the integral over $z_i$ as $\int_{i}$. The measure ($d^2 z_i$ or $d^2 z_i/\tau_2$) will be left implicit. At intermediate stages of the analysis, we use normal ordering to set $G(z,z)=0$ to regularize whenever this expression arises, and hence such divergences arising from coincident points do not lie on the moduli space of these graphs.

To start with, we obtain the eigenvalue equations for $D^{(2,2,1)} (v)$ and $D^{(1,2,2)} (v)$ which we obtain from cutting open $C_{1,2,2}$. Note that we have not included the graph on the left hand side of figure 6 in the list because of the equality is satisfies, which follows from figure 7 using elementary integration by parts and \C{maineqn}.  

\begin{figure}[ht]
\begin{center}
\[
\mbox{\begin{picture}(260,70)(0,0)
\includegraphics[scale=.7]{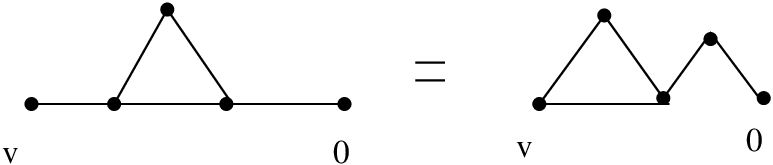}
\end{picture}}
\]
\caption{An identity among graphs} 
\end{center}
\end{figure}

\begin{figure}[ht]
\begin{center}
\[
\mbox{\begin{picture}(260,70)(0,0)
\includegraphics[scale=.7]{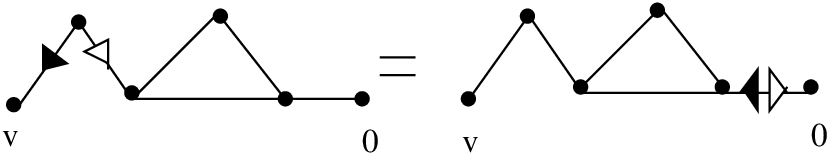}
\end{picture}}
\]
\caption{Proof of figure 6} 
\end{center}
\end{figure}

\subsection{Eigenvalue equation for $D^{(2,2,1)} (v)$}

To start with, we consider the graph $D^{(2,2,1)} (v)$. We get that

\bea &&\Big(\Delta -4\Big) D^{(2,2,1)} (v) = 8 \int_{zwu} G(v,z) G(v,w) \p_\m G(z,u) \overline\p_\m G(w,u) G(u) \non \\
&&+4\int_{zwu}\Big(G(v,z)G(v,w)\p_\m G(z,u) G(w,u) \overline\p_\m G(u)+c.c.\Big) \non \\ &&= 8E_5 -4D^{(2,2,1)}(v)+\frac{4}{\pi}\Big(F_1 (v) -F_2 (v)+ c.c.\Big) -\frac{8\tau_2}{\pi} \p_v G_3 (v) \overline\p_v G_3 (v), \eea
where the graphs $F_1 (v)$ and $F_2 (v)$ are given in figure 8\footnote{For these graphs as well as the ones to be considered later on, the vertices are integrated with the measure $d^2z/\tau_2$, while there is an overall factor of $\tau_2$ for every factor of $\p  \overline\p $ in the graph. Hence they are modular invariant.}. 

\begin{figure}[ht]
\begin{center}
\[
\mbox{\begin{picture}(240,110)(0,0)
\includegraphics[scale=.75]{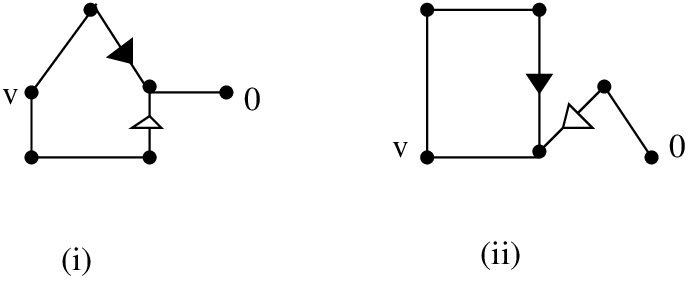}
\end{picture}}
\]
\caption{(i) $F_1 (v)$, (ii) $F_2 (v)$} 
\end{center}
\end{figure}

Using the results in appendix A, this gives us that
\bea \label{D221}\Big(\Delta -4\Big) D^{(2,2,1)} (v) &=& 12 E_5 - 4 G_5 (v) + 4 E_3 G_2 (v) - 4 D^{(1,2,2)} (v) \non \\ &&- 4 G_2 (v) G_3 (v) - \frac{8\tau_2}{\pi} \p_v G_3 (v) \overline\p_v G_3 (v).\eea

\subsection{Eigenvalue equation for $D^{(1,2,2)} (v)$}

We next consider the graph $D^{(1,2,2)} (v)$, which yields
\bea \Big(\Delta -4\Big) D^{(1,2,2)} (v) &=& 2\int_{zw}\Big(G(v,w)\p_\m G(w,z) \overline\p_\m G(v,z) G_2 (z)+c.c.\Big) \non \\ &&+ 4\int_{zwu}\Big(G(v,w) G(v,z) \p_\m G(w,z) G(z,u) \overline\p_\m G(u)+c.c.\Big) \non \\ &&+ 2\int_{zw}\Big( G_2(v,z) \p_\m G(v,z) G(z,w) \overline\p_\m G(w)+c.c.\Big), \eea 
leading to
\bea \Delta D^{(1,2,2)} (v) = 8 E_5 -\frac{2}{\pi}\Big(F_1 (v) - F_2 (v)+c.c.\Big)  .\eea
This gives us
\bea \label{D122}\Big(\Delta -2\Big) D^{(1,2,2)} (v) = 10 G_5 (v) + 2 G_2 (v) G_3 (v) -2 E_5 - 2 D^{(2,2,1)} (v)- 2 E_3 G_2(v)\eea
on using appendix A. This eigenvalue equation does not contain any graph with links given by derivatives of Green functions. 

We next consider the graphs $D^{(1,3,1)} (v)$ and $D^{(1,1,3)} (v)$ that arise from cutting open $C_{1,1,3}$. We do not include the graph on the left hand side of figure 9 in the list because of the equality given in the figure.   

\begin{figure}[ht]
\begin{center}
\[
\mbox{\begin{picture}(300,50)(0,0)
\includegraphics[scale=.7]{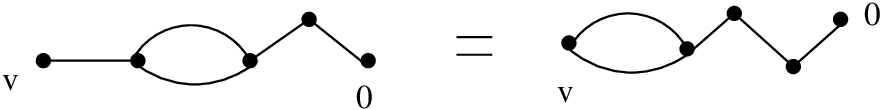}
\end{picture}}
\]
\caption{An identity among graphs} 
\end{center}
\end{figure}

\subsection{Eigenvalue equation for $D^{(1,3,1)} (v)$}

Starting with $D^{(1,3,1)}$, we have that
\bea \Big(\Delta -6\Big) D^{(1,3,1)} (v) &=& 3\int_{zw}\Big(G_2(v,z) \p_\m G(z,w) \overline\p_\m G(v,w) G(w)+c.c.\Big) \non \\ &&+3\int_{zw}\Big(G_2(v,z) \p_\m G(z,w) G(v,w) \overline\p_\m G(w)+c.c.\Big) \non \\ &&+\int_z\Big(G_3(v,z) \p_\m G(v,z) \overline\p_\m G(z)+c.c.\Big),\eea
leading to
\bea &&\Big(\Delta -2\Big) D^{(1,3,1)} (v) = 6 E_5 - D^{(2,2,1)} (v) + 6 G_5 (v) \non \\ &&- \frac{3\tau_2}{\pi}\Big(\p_v G_2 (v) \overline\p_v G_4 (v)+c.c.\Big) +\frac{2}{\pi} \Big(F_1 (v) +F_2 (v)+c.c.\Big).\eea
Again using results in appendix A, we get that
\bea \label{D131}\Big(\Delta -6\Big) D^{(1,3,1)}(v) &=& 8 E_5 +D^{(2,2,1)} (v) + 8 G_5 + 2 D^{(1,2,2)} (v)-2 E_3 G_2 (v)\non \\ &&-2 G_2 (v)G_3 (v) - \frac{3\tau_2}{\pi}\Big(\p_v G_2 (v) \overline\p_v G_4 (v)+c.c.\Big).  \non \\  \eea

\subsection{Eigenvalue equation for $D^{(1,1,3)} (v)$}

For $D^{(1,1,3)}(v)$, we get that
\bea \Big(\Delta -6\Big) D^{(1,1,3)}(v) &=& 2 \int_z \p_\m G(v,z) \overline\p_\m G(v,z) G_3 (z) \non \\ &&+ 6\int_{zw}\Big(G(v,z)\p_\m G(v,z) \overline\p_\m G(z,w) G_2(w)+c.c.\Big),\eea
yielding
\bea \Big(\Delta -10\Big) D^{(1,1,3)} (v) = 12 G_5 (v) -\frac{2}{\pi} F_3 (v) -\frac{4}{\pi}\Big(F_4 (v) +c.c.\Big)\eea
where the graphs $F_3(v)$ and $F_4 (v)$ are given in figure 10.

\begin{figure}[ht]
\begin{center}
\[
\mbox{\begin{picture}(300,100)(0,0)
\includegraphics[scale=.75]{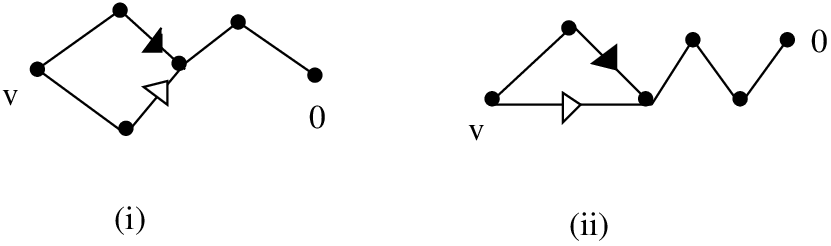}
\end{picture}}
\]
\caption{(i) $F_3 (v)$, (ii) $F_4 (v)$} 
\end{center}
\end{figure}

Using the results in appendix A, this gives us
\bea \label{D113}\Big(\Delta-6\Big) D^{(1,1,3)} (v) = 2 D^{(1,2,2)} (v) + D^{(2,2,1)} (v)-4 E_2 G_3 (v) + 16 G_5 (v).\eea
This is another instance of a graph whose eigenvalue equation does not involve graphs with links given by derivatives of Green functions.  

We next consider the graphs $D^{(1,1,1,2)} (v)$, $D^{(1,1,2,1)} (v)$ and $D_5^{(1,2,2)} (v)$ that are obtained by cutting open $D_{1,1,3}$. We do not consider the graph on the left hand side of figure 11 because of the equality it satisfies. 

\begin{figure}[ht]
\begin{center}
\[
\mbox{\begin{picture}(300,50)(0,0)
\includegraphics[scale=.7]{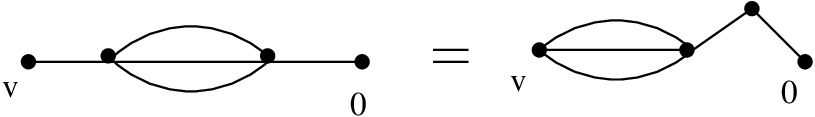}
\end{picture}}
\]
\caption{An identity among graphs} 
\end{center}
\end{figure}

\subsection{Eigenvalue equation for $D^{(1,1,1,2)} (v)$}

For $D^{(1,1,1,2)} (v)$, we have that
\bea \label{e1}\frac{1}{6}\Big(\Delta -2\Big) D^{(1,1,1,2)} (v) &=& \int_z G(v,z) \p_\m G(v,z) \overline\p_\m G(v,z) G_2 (z) \non \\ &&+\int_{zw} \Big(G^2 (v,z) \p_\m G(v,z) \overline\p_\m G(z,w) G(w)+c.c.\Big).\eea

\begin{figure}[ht]
\begin{center}
\[
\mbox{\begin{picture}(440,90)(0,0)
\includegraphics[scale=.65]{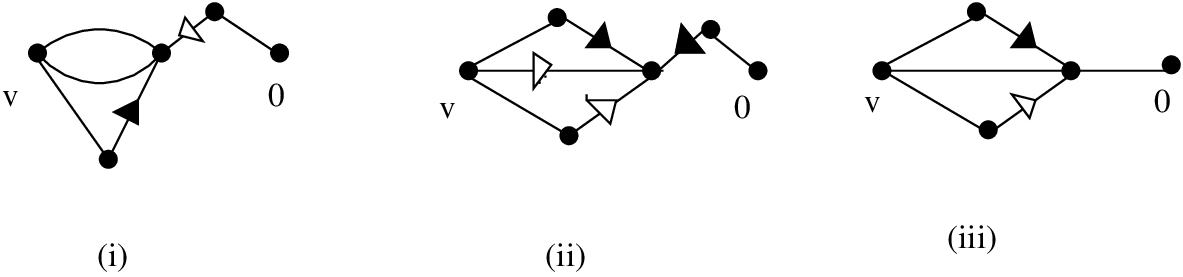}
\end{picture}}
\]
\caption{(i) $F_5 (v)$, (ii) $F_6 (v)$, (iii) $F_7 (v)$} 
\end{center}
\end{figure}

Now the first term on the right hand side of \C{e1} gives
\bea &&\int_z G(v,z) \p_\m G(v,z) \overline\p_\m G(v,z) G_2 (z) = D^{(1,2,2)} (v) -\frac{1}{2} D^{(2,2,1)} (v) \non \\ &&-\frac{1}{2\pi} \Big(F_5 (v)+c.c.\Big) +\frac{1}{\pi^2}\Big(F_6 (v)+c.c.\Big) -\frac{1}{\pi} F_7 (v),\eea
where the graphs $F_5 (v)$, $F_6 (v)$ and $F_7 (v)$ are given in figure 12. 

While $F_5 (v)$ and $F_7 (v)$ are graphs with two derivatives, $F_6 (v)$ has four derivatives. We now express $F_6 (v)$ differently which will be very useful for our purposes\footnote{From now onwards, we shall often say that we ``simplify'' various graphs. This will either mean that we express a graph in terms of graphs with lesser number of derivatives along the links, or express it in terms of graphs with the same number of derivatives along the links. However, in the second case, the resulting graphs will be either from among those that have already appeared in the analysis before, or have a special structure. As we shall see, this ``simplification'' is crucial to obtain the algebraic identities from solving the eigenvalue equations.}.  
To do so, for this case as well as for the several others to follow, we introduce auxiliary graphs. These graphs are related to the original graph using \C{maineqn} very easily. On the other hand, they can be manipulated in a different way by moving the derivatives along the links by integrating by parts, to obtain the desired relation between the original graph and several others. This technique has proved very useful in obtaining the eigenvalue equations in \cite{Basu:2016xrt,Basu:2016kli,Basu:2019idd,Basu:2020pey}.  

\begin{figure}[ht]
\begin{center}
\[
\mbox{\begin{picture}(160,60)(0,0)
\includegraphics[scale=.5]{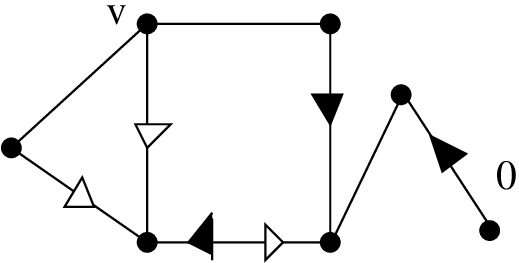}
\end{picture}}
\]
\caption{Auxiliary graph $F_8 (v)$} 
\end{center}
\end{figure}

For $F_6 (v)$, we introduce the auxiliary graph $F_8 (v)$ given in figure 13. Using \C{maineqn} for the link having both the $\p$ and $\overline\p$ derivatives, we see that 
\be \label{F8}\frac{1}{\pi}F_8 (v) = F_6 (v) - P_1 (v) P_2^*,\ee
where the graphs $P_1 (v)$ and $P_2$ are given in appendix B.

\begin{figure}[ht]
\begin{center}
\[
\mbox{\begin{picture}(200,100)(0,0)
\includegraphics[scale=.55]{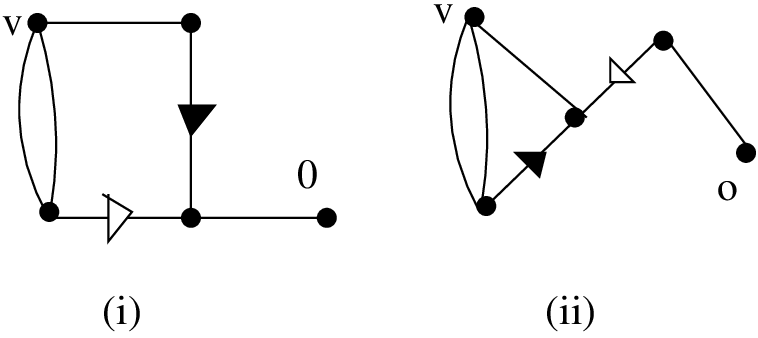}
\end{picture}}
\]
\caption{(i) $F_9 (v)$, (ii) $F_{10} (v)$} 
\end{center}
\end{figure}

Thus the second term on the right hand side of \C{F8} has a special structure and factorizes into two contributions, one of which has two $\p$s while the other has two $\overline\p$s. This will be generically true of such manipulations involving auxiliary graphs, and we shall see that these terms play a very important role in obtaining the eigenvalue equations. Alternatively, we can calculate $F_8 (v)$ in another way by moving the derivatives differently around the circuit, leading to the relation
\bea \Big(F_6 (v) +c.c.\Big) - \Big(P_1(v) P_2^*+c.c.\Big) = -\pi\Big(F_1 (v) + F_2 (v) -\frac{F_9(v)}{2} -\frac{F_{10}(v)}{2}+ c.c.\Big), \eea
where the graphs $F_9 (v)$ and $F_{10} (v)$ are given in figure 14, which can be simplified using the results in appendix A. 

\begin{figure}[ht]
\begin{center}
\[
\mbox{\begin{picture}(160,60)(0,0)
\includegraphics[scale=.8]{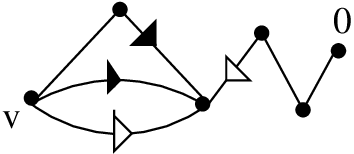}
\end{picture}}
\]
\caption{$F_{11}(v)$} 
\end{center}
\end{figure}

We now consider the second term on the right hand side of \C{e1}, which gives
\bea \int_{zw} \Big(G^2 (v,z) \p_\m G(v,z) \overline\p_\m G(z,w) G(w)+c.c.\Big) = -\frac{4}{3} D^{(1,1,1,2)}(v) \non \\ +4D^{(1,1,3)}(v) +\frac{1}{\pi} \Big(F_5(v)+c.c.\Big) -\frac{2}{\pi}\Big(F_4(v)-\frac{F_{11}(v)}{\pi}+c.c.\Big),  \eea 
where the graph $F_{11}(v)$ is given in figure 15. 
\begin{figure}[ht]
\begin{center}
\[
\mbox{\begin{picture}(120,80)(0,0)
\includegraphics[scale=.7]{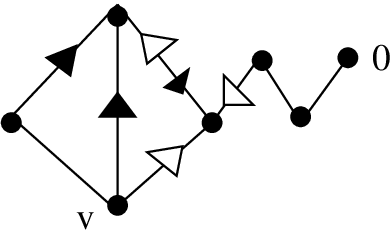}
\end{picture}}
\]
\caption{Auxiliary graph $F_{12}(v)$} 
\end{center}
\end{figure}

To simplify $F_{11} (v)$ we start with the auxiliary graph $F_{12} (v)$ given in figure 16, to get
\bea &&\Big(F_{11}(v) + c.c.\Big)- \Big(P_1 (v)P_2^* + c.c.\Big) = D^{(1,1,1,2)} (v) -2 D^{(1,2,2)} (v)\non \\ &&-D^{(1,1,3)} (v)  -E_2 G_3 (v)+ 2 G_5 (v) +\frac{1}{\pi}\Big(F_2(v) -\frac{F_{10}(v)}{2}+c.c.\Big).   \eea 

Adding the various contributions, we get the eigenvalue equation
\bea \label{D1112}&&\frac{1}{6}\Big(\Delta -6\Big) D^{(1,1,1,2)} (v) = \frac{3}{\pi^2} \Big(P_1(v)P_2^*+c.c.\Big)+2 D^{(1,1,3)} (v) -3 D^{(1,2,2)}(v) \non \\ &&- \frac{1}{2} D^{(2,2,1)} (v) - 2 E_2 G_3 (v) + 4 G_5 (v)+\frac{1}{2\pi} \Big(F_5 (v) + c.c.\Big) -\frac{1}{\pi}F_7 (v) \non \\&& -\frac{1}{\pi}\Big(F_1(v) - F_2(v) +2 F_4(v)-\frac{F_{9}(v)}{2} +\frac{F_{10}(v)}{2} +c.c.\Big).\eea
Now we can simplify the final expression in \C{D1112} using the identities in appendix A\footnote{In fact, $F_5(v)+c.c.$ can be simplified also, as we show in appendix C.}, but we refrain from doing so as it is not particularly illuminating. 

\begin{figure}[ht]
\begin{center}
\[
\mbox{\begin{picture}(260,130)(0,0)
\includegraphics[scale=.75]{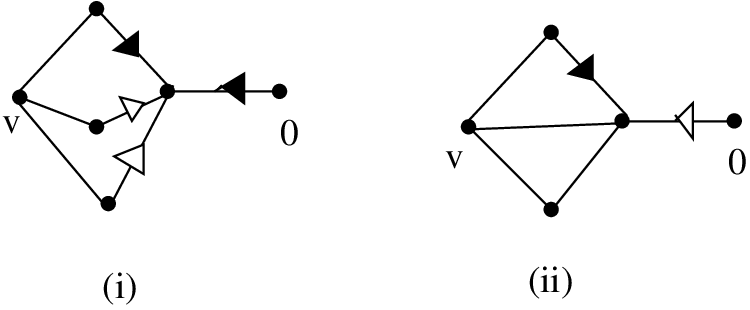}
\end{picture}}
\]
\caption{(i) $F_{13}(v)$, (ii) $F_{14}(v)$} 
\end{center}
\end{figure}

This will be the strategy we shall follow for the remaining eigenvalue equations as well. We shall simply collect the various contributions and write down the equation. However, we shall postpone simplifying various contributions we can simplify using the results in the appendices. One of the obvious reasons is that the resulting cumbersome expression is not readily useful. However, the principal reason is that we shall combine the various eigenvalue equations later on to obtain much more tractable equations we shall solve, and further striking simplifications happen during the process of combining the contributions from the various eigenvalue equations. This makes simplifying at this stage unnecessary.           

Thus we see that the eigenvalue equation we have obtained for $D^{(1,1,1,2)} (v)$ is significantly more involved than the ones we have obtained before. This will also be true of the eigenvalue equations we shall obtain for the remaining graphs.

\subsection{Eigenvalue equation for $D^{(1,1,2,1)} (v)$}

We next consider the eigenvalue equation for $D^{(1,1,2,1)}(v)$. We get that 
\begin{figure}[ht]
\begin{center}
\[
\mbox{\begin{picture}(120,90)(0,0)
\includegraphics[scale=.7]{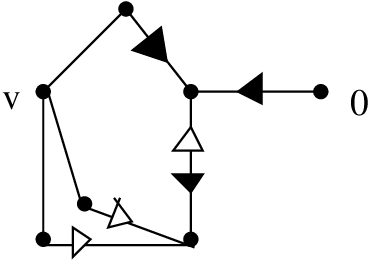}
\end{picture}}
\]
\caption{Auxiliary graph $F_{15}(v)$} 
\end{center}
\end{figure}
\bea \label{e2}\frac{1}{2} \Big(\Delta -2\Big) D^{(1,1,2,1)} (v) &=& \int_z G_2 (v,z)\p_\m G(v,z) \overline\p_\m G(v,z)G(z) \non \\ &&+2\int_{zw} \Big(G(v,z) \p_\m G(v,z) G(v,w) \overline\p_\m G(w,z)G(z)+c.c.\Big)\non \\ &&+\int_z\Big(G_2(v,z) G(v,z) \p_\m G(v,z) \overline\p_\m G(z)+c.c.\Big) \non \\ &&+\int_{zw}\Big(G^2(v,z) G(v,w) \p_\m G(w,z)\overline\p_\m G(z)+c.c.\Big).\eea

\begin{figure}[ht]
\begin{center}
\[
\mbox{\begin{picture}(220,130)(0,0)
\includegraphics[scale=.75]{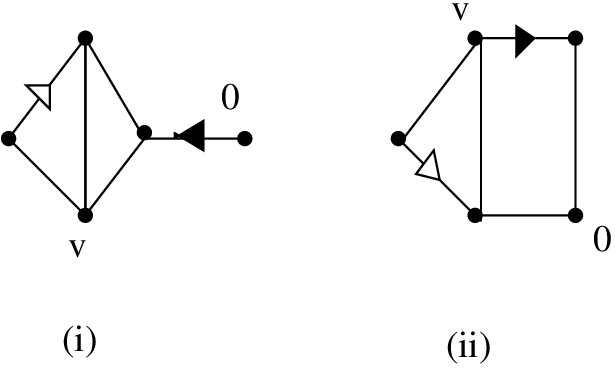}
\end{picture}}
\]
\caption{(i) $F_{16}(v)$, (ii) $F_{17}(v)$} 
\end{center}
\end{figure}

The first term on the right hand side of \C{e2} gives
\bea &&\int_z G_2 (v,z)\p_\m G(v,z) \overline\p_\m G(v,z)G(z) = D^{(1,1,2,1)} (v) +\frac{1}{2} C_{1,2,2} \non \\ &&-\frac{\tau_2}{\pi} G_2(v) \p_v G_2(v) \overline\p_v G_2 (v)  -\frac{3}{\pi}F_7 (v)+\frac{1}{\pi}\Big(\frac{F_{13}(v)}{\pi} - F_{14}(v)+c.c.\Big),\eea
where the graphs $F_{13}(v)$ and $F_{14} (v)$ are given in figure 17. 

\begin{figure}[ht]
\begin{center}
\[
\mbox{\begin{picture}(320,115)(0,0)
\includegraphics[scale=.75]{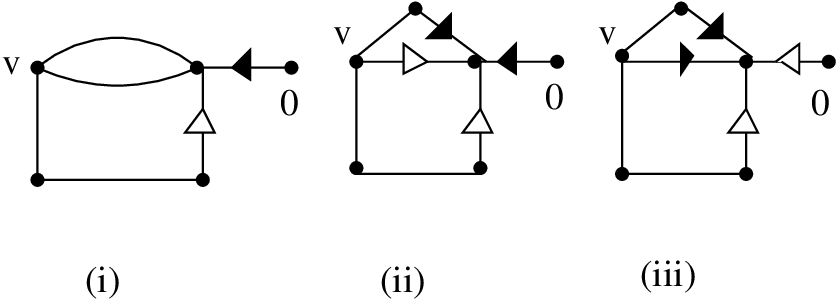}
\end{picture}}
\]
\caption{(i) $F_{18}(v)$, (ii) $F_{19}(v)$, (iii) $F_{20}(v)$} 
\end{center}
\end{figure}

To simplify $F_{13}(v)$, we start with the auxiliary graph $F_{15}(v)$ given in figure 18, leading to
\bea \Big( F_{13}(v)+c.c. \Big) - \Big(P_3(v)P_4^* +c.c.\Big)= 2\pi \Big( F_{16} (v) + F_{17}(v)- Q_1 +c.c.\Big),\eea
where the graphs $P_3 (v),$ $P_4$ and $Q_1$ are given in appendix B, and the graphs $F_{16} (v)$ and $F_{17} (v)$ are given in figure 19\footnote{Note that $F_{17}(v)$ factorizes into a  product of two graphs. This will be true of several graphs that arise in our analysis, which factorize into a product of several graphs.}.

The second term on the right hand side of \C{e2} gives
\bea &&\int_{zw} \Big(G(v,z)\p_\m G(v,z) G(v,w) \overline\p_\m G(w,z)G(z) +c.c.\Big)=-D^{(1,1,2,1)}(v) -\frac{1}{2\pi}\Big(F_{18}(v)+c.c.\Big)  \non \\&&
+\frac{1}{\pi}\Big(Q_1+c.c.\Big)-\frac{1}{\pi}G(v) \Big(\p_v G_2 (v) \overline\p_v G_3 (v)+c.c.\Big)  +\frac{1}{\pi} \Big(F_1(v)+c.c.\Big) +\frac{2F_7(v)}{\pi}\non \\ &&+\frac{1}{\pi^2}\Big(F_{19}(v)+F_{20}(v)+c.c.\Big),\eea
where the graphs $F_{18}(v)$, $F_{19}(v)$ and $F_{20}(v)$ are given in figure 20.

\begin{figure}[ht]
\begin{center}
\[
\mbox{\begin{picture}(260,120)(0,0)
\includegraphics[scale=.6]{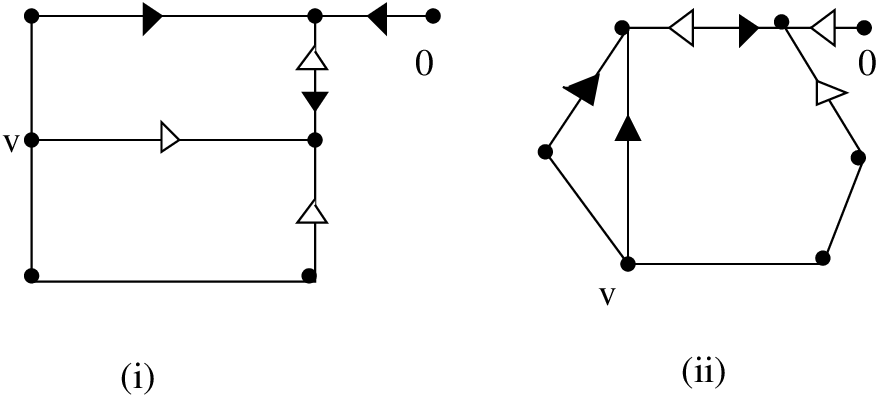}
\end{picture}}
\]
\caption{Auxiliary graphs (i) $F_{21}(v)$, (ii) $F_{22}(v)$} 
\end{center}
\end{figure}

To simplify $F_{19} (v)$ and $F_{20} (v)$, we introduce the auxiliary graphs $F_{21} (v)$ and $F_{22}(v)$ respectively as given in figure 21.

\begin{figure}[ht]
\begin{center}
\[
\mbox{\begin{picture}(320,150)(0,0)
\includegraphics[scale=.8]{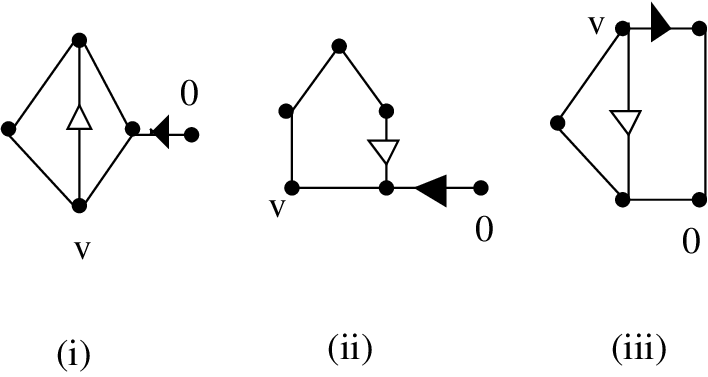}
\end{picture}}
\]
\caption{(i) $F_{23}(v)$, (ii) $F_{24}(v)$, (iii) $F_{25}(v)$} 
\end{center}
\end{figure}

For $F_{19}(v)$ this gives us
\bea &&\Big(F_{19} (v) + c.c.\Big)-\Big(P_3(v)P_4^* +c.c.\Big)= \pi  \Big(F_{23}(v)-F_{24}(v)+F_{25}(v)+Q_1+c.c.\Big) \non \\ && -\pi\tau_2\Big(\p_v G_2(v) \overline\p_v G_4(v)+c.c.\Big)  +2\pi^2 E_5 -2\pi^2 C_{1,2,2},\eea
where the graphs $F_{23}(v)$, $F_{24}(v)$ and $F_{25}(v)$ are given in figure 22.

\begin{figure}[ht]
\begin{center}
\[
\mbox{\begin{picture}(80,75)(0,0)
\includegraphics[scale=.65]{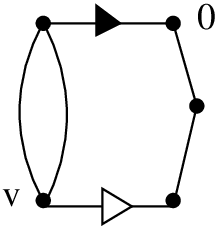}
\end{picture}}
\]
\caption{$F_{26}(v)$} 
\end{center}
\end{figure}

For $F_{20}(v)$, we get
\bea &&\Big(F_{20}(v)+c.c.\Big)-\Big(P_1(v)P_2^*+c.c.\Big) = 2\pi^2 E_5 -\pi^2 C_{1,1,3} +\pi^2 D^{(1,1,2,1)}(v) \non \\ &&-2\pi^2 D^{(2,2,1)}(v)  -\pi^2 E_2 G_3(v)-2\pi\tau_2\p_v G_3(v)\overline\p_vG_3(v) \non \\ &&+\pi \Big(F_1(v) -\frac{F_9(v)}{2} +\frac{F_{26}(v)}{2}+c.c.\Big),\eea
where the graph $F_{26}(v)$ is given in figure 23.

\begin{figure}[ht]
\begin{center}
\[
\mbox{\begin{picture}(140,70)(0,0)
\includegraphics[scale=.85]{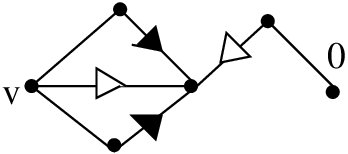}
\end{picture}}
\]
\caption{$F_{27}(v)$} 
\end{center}
\end{figure}

Now the third term on the right hand side of \C{e2} gives
\bea &&\int_z \Big(G_2(v,z) G(v,z) \p_\m G(v,z) \overline\p_\m G(z)+c.c.\Big)= D^{(2,2,1)} (v) - D^{(1,1,2,1)} (v)\non \\ &&+\frac{1}{\pi}\Big(F_{14}(v) -\frac{3F_5(v)}{2}+\frac{F_6(v)}{\pi}+\frac{F_{27}(v)}{\pi}+c.c.\Big),\eea
where the graph $F_{27}(v)$ is given in figure 24. 

\begin{figure}[ht]
\begin{center}
\[
\mbox{\begin{picture}(100,110)(0,0)
\includegraphics[scale=.85]{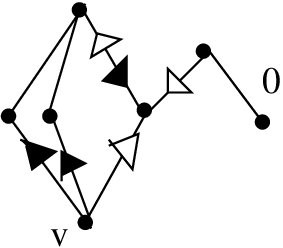}
\end{picture}}
\]
\caption{Auxiliary graph $F_{28}(v)$} 
\end{center}
\end{figure}

To simplify $F_{27}(v)$, we consider the auxiliary graph $F_{28}(v)$ given in figure 25, which leads to
\bea \Big(F_{27}(v) +c.c.\Big) - \Big(P_3(v)P_4^*+c.c.\Big) = -4\pi^2 D^{(1,1,3)} (v) +2\pi\Big(F_4(v)+ F_{29}(v)+c.c.\Big),\eea
where the graph $F_{29}(v)$ is given in figure 26.

\begin{figure}[ht]
\begin{center}
\[
\mbox{\begin{picture}(100,70)(0,0)
\includegraphics[scale=.55]{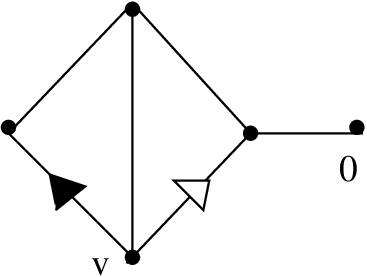}
\end{picture}}
\]
\caption{$F_{29}(v)$} 
\end{center}
\end{figure}

Finally, the fourth term on the right hand side of \C{e2} is given by
\bea \int_{zw}\Big(G^2(v,z) G(v,w) \p_\m G(w,z)\overline\p_\m G(z)+c.c.\Big) = -2D^{(1,1,2,1)} (v) \non \\ +\frac{1}{\pi}\Big(2F_2(v) + F_5 (v) + F_{18}(v)+\frac{2F_{30}(v)}{\pi}+c.c.\Big),\eea
where the graph $F_{30}(v)$ is given in figure 27.

\begin{figure}[ht]
\begin{center}
\[
\mbox{\begin{picture}(120,60)(0,0)
\includegraphics[scale=.6]{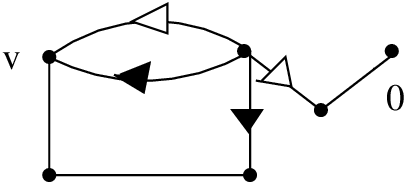}
\end{picture}}
\]
\caption{$F_{30}(v)$} 
\end{center}
\end{figure}
To simplify $F_{30} (v)$, we start with the auxiliary graph $F_{31}(v)$ given in figure 28, to obtain
\begin{figure}[ht]
\begin{center}
\[
\mbox{\begin{picture}(120,80)(0,0)
\includegraphics[scale=.6]{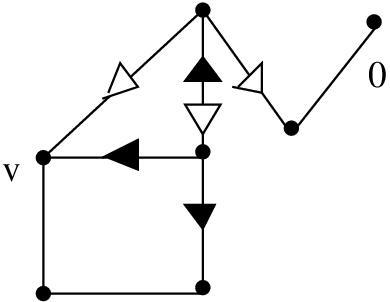}
\end{picture}}
\]
\caption{Auxiliary graph $F_{31}(v)$} 
\end{center}
\end{figure}

\bea \Big(F_{30}(v) + c.c.\Big) - \Big(P_3(v)P_4^*+c.c.\Big) = 4\pi^2 G_5 (v) -2\pi^2 D^{(1,3,1)} (v)- 2\pi^2 E_2 G_3(v)\non \\ +\pi \Big(F_4(v)+F_{24}(v)+F_{32}(v)+c.c.\Big),\eea
where the graph $F_{32}(v)$ is given in figure 29.

\begin{figure}[ht]
\begin{center}
\[
\mbox{\begin{picture}(120,80)(0,0)
\includegraphics[scale=.6]{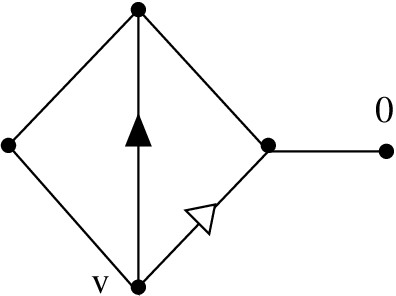}
\end{picture}}
\]
\caption{$F_{32}(v)$} 
\end{center}
\end{figure}

Thus to obtain the final expression for the eigenvalue equation, we add the various contributions, to get that
\bea \label{L1}&&\frac{1}{2}\Big(\Delta +2\Big) D^{(1,1,2,1)} (v) = \frac{3}{\pi^2}\Big(P_1(v)P_2^* +c.c.\Big)+ \frac{6}{\pi^2}\Big(P_3(v) P_4^* +c.c.\Big) +\frac{1}{\pi} F_7(v) \non \\ &&+\frac{1}{\pi}\Big(3F_1(v)+F_2(v)-\frac{F_5(v)}{2} -\frac{F_9(v)}{2}+\frac{F_{10}(v)}{2}+F_{26}(v) +2F_{33}(v)+2F_{34}(v)\non \\ &&+2F_{35}(v)+c.c.\Big)  + 10 E_5 -2 E_2 E_3  + 4 G_5 (v) -2E_2 G_3(v) -4 D^{(1,2,2)}(v) \non \\&&-4 D^{(1,3,1)}(v)  - 3 D^{(2,2,1)} (v) -\frac{3}{2} C_{1,2,2}-\frac{\tau_2}{\pi}G_2(v) \p_v G_2(v) \overline\p_v G_2(v)  \non \\ &&-\frac{2\tau_2}{\pi}G(v)\Big(\p_v G_2(v)\overline\p_vG_3(v)+c.c.\Big) -\frac{2\tau_2}{\pi} \Big(\p_v G_2 (v) \overline\p_v G_4 (v)+c.c.\Big)\non \\ &&-\frac{4\tau_2}{\pi} \p_v G_3 (v) \overline\p_v G_3 (v) ,    \eea
where the graphs $F_{33} (v)$, $F_{34} (v)$ and $F_{35} (v)$ are given in figure 30.

\begin{figure}[ht]
\begin{center}
\[
\mbox{\begin{picture}(320,120)(0,0)
\includegraphics[scale=.75]{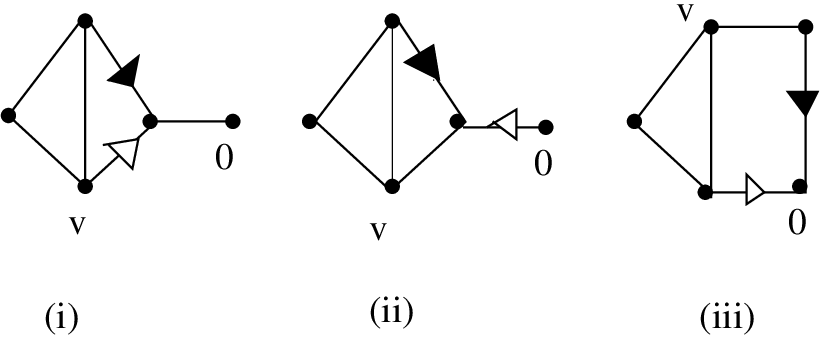}
\end{picture}}
\]
\caption{(i) $F_{33}(v)$, (ii) $F_{34}(v)$, (iii) $F_{35}(v)$} 
\end{center}
\end{figure}

In obtaining \C{L1}, we have performed some simplifications by adding some graphs where the derivatives of the Green functions on the links are arranged appropriately, and we have also used the expression involving $Q_1$ in appendix B\footnote{Also note that $F_{33} (v)+F_{34}(v) = \pi D^{(1,1,2,1)}(v) -\pi E_3 G_2(v)$.}.

\subsection{Eigenvalue equation for $D_5^{(1,2,2)} (v)$}

\begin{figure}[ht]
\begin{center}
\[
\mbox{\begin{picture}(360,160)(0,0)
\includegraphics[scale=.8]{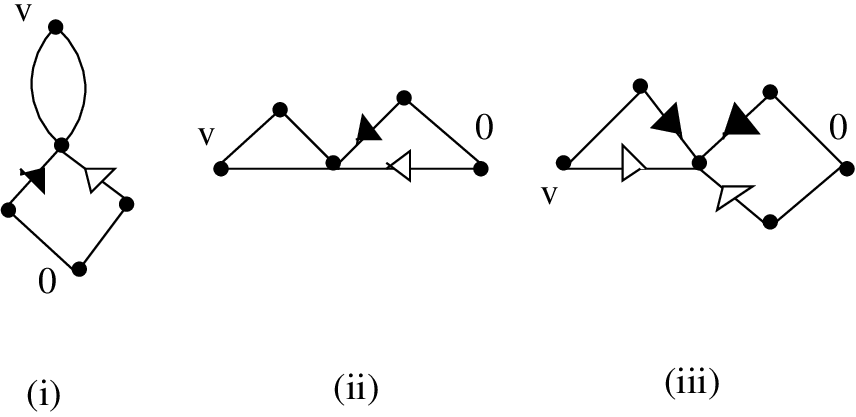}
\end{picture}}
\]
\caption{(i) $F_{36}(v)$, (ii) $F_{37}(v)$, (iii) $F_{38}(v)$} 
\end{center}
\end{figure}

For $D_5^{(1,2,2)}(v)$, we have that

\bea \label{e3}\frac{1}{2} \Big(\Delta -2\Big)D_5^{(1,2,2)} (v) &=& \int_z G_2(v,z) G(v,z) \p_\m G(z) \overline\p_\m G(z) \non \\ &&+ \int_{zw} \Big(G(v,z) \p_\m G(z,w) \overline\p_\m G(v,w) G^2(w)+c.c.\Big) \non \\&& +2\int_{zw} \Big(G(v,z) G(v,w) \p_\m G(z,w) G(w) \overline\p_\m G(w)+c.c.\Big) \non \\ &&+\int_z\Big(G_2(v,z) \p_\m G(v,z) G(z) \overline\p_\m G(z)+c.c.\Big).\eea

\begin{figure}[ht]
\begin{center}
\[
\mbox{\begin{picture}(140,70)(0,0)
\includegraphics[scale=.65]{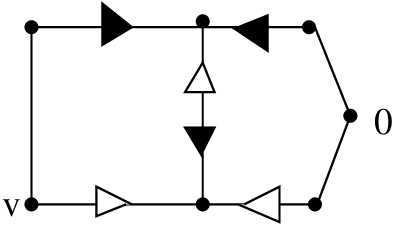}
\end{picture}}
\]
\caption{Auxiliary graph $F_{39}(v)$} 
\end{center}
\end{figure}

The first term on the right hand side of \C{e3} gives us

\bea &&\int_z G_2(v,z) G(v,z) \p_\m G(z) \overline\p_\m G(z)= -D_5^{(1,2,2)} (v) -\frac{\tau_2}{\pi} E_2 \p_v G_2(v) \overline\p_v G_2(v) \non \\ &&+\frac{1}{\pi}\Big(F_3(v)-F_{36}(v)\Big) +\frac{1}{\pi}\Big(F_{37}(v)+\frac{F_{38}(v)}{\pi}+c.c.\Big), \eea
where the graphs $F_{36}(v)$, $F_{37}(v)$ and $F_{38}(v)$ are given in figure 31\footnote{$F_{36}(v)$ has been simplified in appendix D which will be very useful for us later.}. 
\begin{figure}[ht]
\begin{center}
\[
\mbox{\begin{picture}(180,120)(0,0)
\includegraphics[scale=.75]{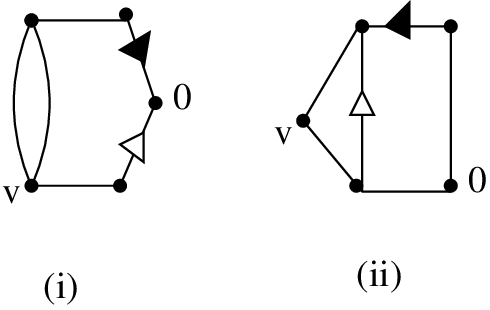}
\end{picture}}
\]
\caption{(i) $F_{40}(v)$, (ii) $F_{41}(v)$} 
\end{center}
\end{figure}

To simplify $F_{38}(v)$, we start with the auxiliary graph $F_{39}(v)$ given by figure 32, to get that 
\bea \Big(F_{38}(v) +c.c.\Big) - \Big(P_1(v) P_3(v)^*+c.c.\Big)&=& -\pi\Big(F_1(v) +F_2(v) +F_{17}(v)\non \\ &&-F_{40}(v) -F_{41}(v)+c.c.\Big),
\eea
where the graphs $F_{40}(v)$ and $F_{41}(v)$ are given in figure 33.

\begin{figure}[ht]
\begin{center}
\[
\mbox{\begin{picture}(200,120)(0,0)
\includegraphics[scale=.8]{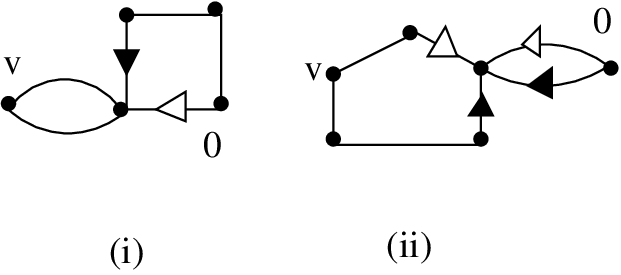}
\end{picture}}
\]
\caption{(i) $F_{42}(v)$, (ii) $F_{43}(v)$} 
\end{center}
\end{figure}

The second term on the right hand side of \C{e3} gives us

\bea &&\int_{zw} \Big(G(v,z) \p_\m G(z,w) \overline\p_\m G(v,w) G^2(w)+c.c.\Big) = -2 D_5^{(1,2,2)} (v)+\frac{2}{\pi}F_{36}(v)  \non \\
&&+\frac{1}{\pi}\Big(2F_1(v) +F_{42}(v)+\frac{2}{\pi}F_{43}(v)+c.c.\Big),  \eea
where the graphs $F_{42}(v)$ and $F_{43}(v)$ are given in figure 34. 

\begin{figure}[ht]
\begin{center}
\[
\mbox{\begin{picture}(120,90)(0,0)
\includegraphics[scale=.6]{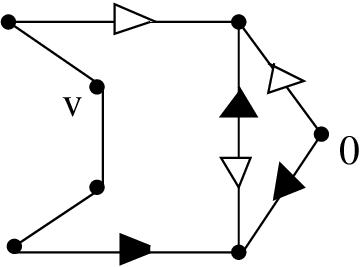}
\end{picture}}
\]
\caption{Auxiliary graph $F_{44}(v)$} 
\end{center}
\end{figure}

To simplify $F_{43}(v)$, we start with the auxiliary graph $F_{44}(v)$ given in figure 35, to get 
\bea &&\Big(F_{43}(v)+c.c.\Big) -\Big(P_1(v)P_3(v)^*+c.c.\Big) = 2\pi^2 E_5 -2\pi^2 D^{(2,2,1)} (v) \non \\ &&+\pi\Big(F_1(v) -F_{24}(v) -\frac{F_{26}(v)}{2}-\frac{F_{40}(v)}{2} +F_{45}(v)+c.c.\Big) \non \\ &&-2\pi \tau_2\p_v G_3(v)\overline\p_v G_3(v) -\pi\tau_2\Big(\p_vG_2(v)\overline\p_v G_4(v)+c.c.\Big)    ,\eea
where the graph $F_{45}(v)$ is given in figure 36.

\begin{figure}[ht]
\begin{center}
\[
\mbox{\begin{picture}(120,70)(0,0)
\includegraphics[scale=.55]{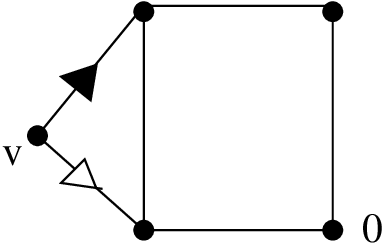}
\end{picture}}
\]
\caption{$F_{45}(v)$} 
\end{center}
\end{figure}

The third term on the right hand side of \C{e3} gives us
\bea &&\int_{zw} \Big(G(v,z) G(v,w) \p_\m G(z,w) G(w) \overline\p_\m G(w)+c.c.\Big) =-D_5^{(1,2,2)} (v) +2 D^{(1,1,3)} (v)  \non \\ &&+\frac{1}{\pi}\Big(F_2(v)-F_4(v)-\frac{F_{42}(v)}{2}+\frac{F_{47}(v)}{\pi}+\frac{F_{48}(v)}{\pi} +c.c.\Big)+\frac{2F_{46}(v)}{\pi},\eea
where the graphs $F_{46}(v)$, $F_{47}(v)$ and $F_{48}(v)$ are given in figure 37.

\begin{figure}[ht]
\begin{center}
\[
\mbox{\begin{picture}(320,120)(0,0)
\includegraphics[scale=.8]{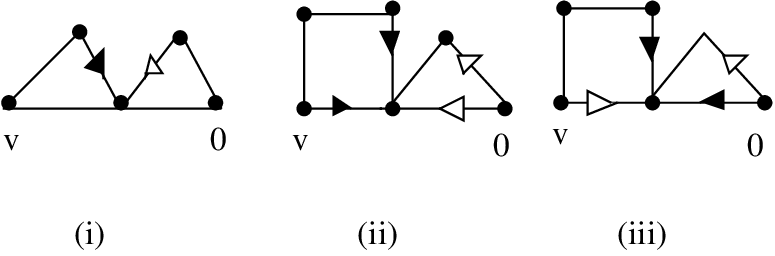}
\end{picture}}
\]
\caption{(i) $F_{46}(v)$, (ii) $F_{47}(v)$, (iii) $F_{48}(v)$} 
\end{center}
\end{figure}

In order to simplify $F_{47} (v)$ and $F_{48} (v)$, we start with the auxiliary graphs $F_{49}(v)$ and $F_{50}(v)$ respectively, as given in figure 38.

\begin{figure}[ht]
\begin{center}
\[
\mbox{\begin{picture}(270,100)(0,0)
\includegraphics[scale=.7]{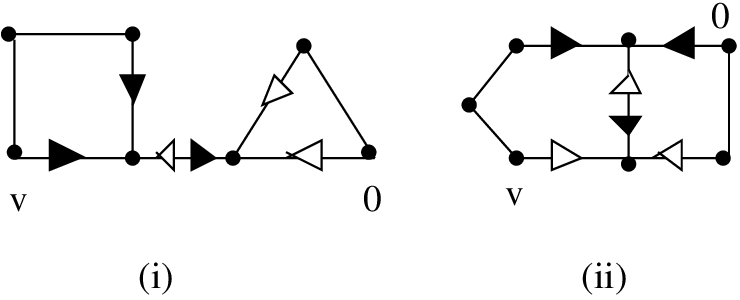}
\end{picture}}
\]
\caption{Auxiliary graphs (i) $F_{49}(v)$, (ii) $F_{50}(v)$} 
\end{center}
\end{figure}

For $F_{47}(v)$, this leads to
\bea &&\Big(F_{47}(v)+c.c.\Big) - \Big(P_2 P_4^*+c.c.\Big)= 4\pi^2 E_5 -2\pi^2 E_2 G_3 (v)\non \\ &&-\pi^2 D^{(1,1,3)} (v)  +\pi\Big(F_4(v)+\frac{F_{51}(v)}{2}+c.c.\Big),\eea
where the graph $F_{51}(v)$ is given in figure 39. 

\begin{figure}[ht]
\begin{center}
\[
\mbox{\begin{picture}(140,55)(0,0)
\includegraphics[scale=.7]{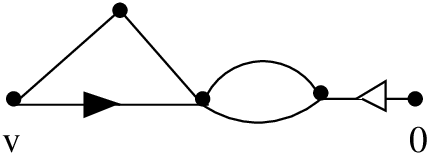}
\end{picture}}
\]
\caption{$F_{51}(v)$} 
\end{center}
\end{figure}

For $F_{48}(v)$, we get that
\bea &&\Big(F_{48}(v)+c.c.\Big) - \Big(P_1(v) P_3(v)^*+c.c.\Big)= 2\pi^2 G_5(v) -2\pi^2 D^{(1,2,2)} (v) \non \\ &&-2\pi^2 D^{(1,3,1)} (v) +\pi\Big(F_2(v)+F_{24}(v) +F_{26}(v) +F_{35}(v) +F_{52}(v)   +c.c.\Big) \non \\ &&-\tau_2 G(v)\Big(\p_vG_2(v)\overline\p_v G_3(v) + c.c.\Big),\eea
where the graph $F_{52}(v)$ is given in figure 40.

\begin{figure}[ht]
\begin{center}
\[
\mbox{\begin{picture}(110,65)(0,0)
\includegraphics[scale=.65]{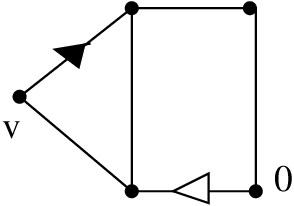}
\end{picture}}
\]
\caption{$F_{52}(v)$} 
\end{center}
\end{figure}

Lastly, the fourth term on the right hand side of \C{e3} gives us
\bea &&\int_z\Big(G_2(v,z) \p_\m G(v,z) G(z) \overline\p_\m G(z)+c.c.\Big)=\pi^2 D^{(2,2,1)}(v) +\pi^2 D_5^{(1,2,2)}(v) \non \eea
\bea &&-\pi\Big(\frac{F_{36}(v)}{2}+F_{37}(v) -\frac{F_{38}(v)}{\pi} - \frac{F_{53}(v)}{\pi}+c.c.\Big)-4\pi F_{46}(v),\eea
where the graph $F_{53}(v)$ is given in figure 41.  

To simplify $F_{53}(v)$, we introduce the auxiliary graph $F_{54}(v)$ in figure 42, which leads to  
\begin{figure}[ht]
\begin{center}
\[
\mbox{\begin{picture}(120,80)(0,0)
\includegraphics[scale=.7]{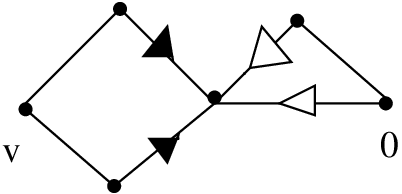}
\end{picture}}
\]
\caption{$F_{53}(v)$} 
\end{center}
\end{figure}
\begin{figure}[ht]
\begin{center}
\[
\mbox{\begin{picture}(150,60)(0,0)
\includegraphics[scale=.8]{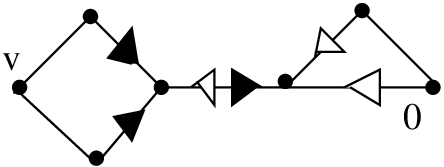}
\end{picture}}
\]
\caption{Auxiliary graph $F_{54}(v)$} 
\end{center}
\end{figure}
\bea \Big(F_{53}(v) +c.c.\Big)-\Big(P_2 P_4^*+c.c.\Big)& =& 2\pi^2 D_5^{(1,2,2)} (v)-4\pi^2 D^{(1,1,3)} (v)\non \\ &&-2\pi^2 E_2 E_3 +\pi\Big(2F_4(v)-F_{51}(v)+c.c.\Big).\eea

Thus adding the various contributions, we get that 
\bea &&\frac{1}{2} \Big(\Delta -6\Big) D_5^{(1,2,2)} (v) = \frac{6}{\pi^2}\Big(P_1(v)P_3(v)^* +c.c.\Big) +\frac{3}{\pi^2}\Big(P_2 P_4^* +c.c.\Big)\non \\&& -\frac{2\tau_2}{\pi}\Big(\p_v G_2(v)\overline\p_v G_4(v)+c.c.\Big)-\frac{2\tau_2}{\pi}G(v)\Big(\p_v G_2(v)\overline\p_v G_3(v) +c.c.\Big)\non \\ &&-\frac{4\tau_2}{\pi} \p_v G_3(v)\overline\p_v G_3(v)
-\frac{\tau_2}{\pi}E_2 \p_v G_2(v) \overline\p_v G_2 (v) +\frac{1}{\pi}\Big(2F_1(v)+2F_2(v) \non \\ &&+2F_4(v)+2F_{25}(v)+F_{26}(v)+F_{40}(v)+c.c.\Big)  -4 G_2(v) G_3(v) \non \\ &&-4 E_2 G_3(v)+12 G_5(v)+4 E_5 -2 E_2E_3  -\frac{7}{2} D^{(2,2,1)} (v) -4 D^{(1,3,1)} (v) \non \\ &&-3 D^{(1,2,2)} (v) - 2 D^{(1,1,3)}(v).\eea

We next consider the graphs $D_5^{(2;2;1)} (v)$, $D_5^{(1,2,1;1)} (v)$ and $D_5^{(1;2,2)} (v)$ that we obtain from cutting open $D_{1,2,2}$. We do not consider the graph on the left hand side of figure 43 because of the equality it satisfies. 

\begin{figure}[ht]
\begin{center}
\[
\mbox{\begin{picture}(200,60)(0,0)
\includegraphics[scale=.75]{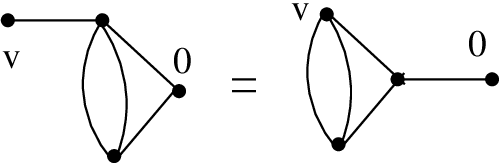}
\end{picture}}
\]
\caption{An identity among graphs} 
\end{center}
\end{figure}

\subsection{Eigenvalue equation for $D_5^{(2;2;1)} (v)$}

For $D_5^{(2;2;1)}(v)$, we get that\footnote{The second term on the right hand side of \C{f1} is real.}

\bea \label{f1}&&\frac{1}{4} \Delta D_5^{(2;2;1)} (v)= \int_{zw} \p_\m G(v,z) \overline\p_\m G(v,z) G(z,w)^2 G(w)\non \\ &&+2\int_{zw} G(v,z) \p_\m G(v,z) G(z,w) \overline\p_\m G(z,w) G(w)\non \\ &&+\int_{zw}\Big(G(v,z)^2 G(z,w) \p_\m G(z,w) \overline\p_\m G(w)+c.c.\Big).\eea
In obtaining \C{f1}, we have used the equalities given in figure 44. 

\begin{figure}[ht]
\begin{center}
\[
\mbox{\begin{picture}(340,110)(0,0)
\includegraphics[scale=.7]{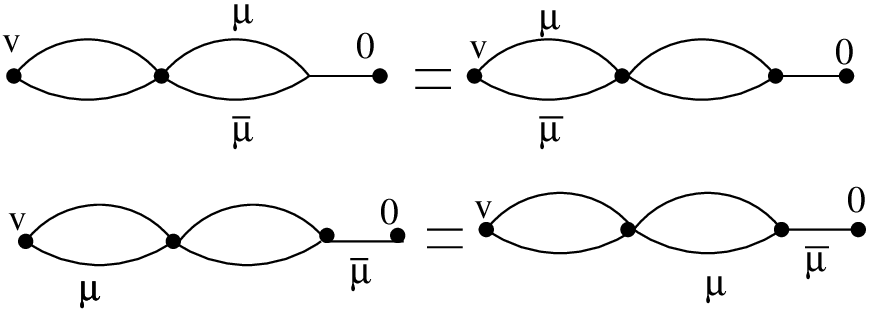}
\end{picture}}
\]
\caption{Identities among graphs} 
\end{center}
\end{figure}
In these  figures, $\m$ ($\overline\m$) along a link of the graph actually stands for $\p_\m G$ ($\overline\p_\m G$) along that link.

\begin{figure}[ht]
\begin{center}
\[
\mbox{\begin{picture}(150,65)(0,0)
\includegraphics[scale=.75]{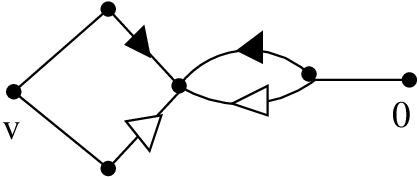}
\end{picture}}
\]
\caption{$F_{55}(v)$} 
\end{center}
\end{figure}

We now consider the first term on the right hand side of \C{f1}, which gives us
\bea \int_{zw} \p_\m G(v,z) \overline\p_\m G(v,z) G(z,w)^2 G(w) &=& D_5^{(2;2;1)}(v)-2D_5^{(1,2,2)} (v)+2 E_2 E_3  +\frac{2}{\pi}F_3(v) \non \\ &&+\frac{1}{\pi}\Big(F_{51}(v) +\frac{2F_{55}(v)}{\pi}+c.c.\Big),\eea 
where the graph $F_{55}(v)$ is given in figure 45.

To simplify $F_{55}(v)$, we start with the auxiliary graph $F_{56}(v)$ given in figure 46, to get 
\begin{figure}[ht]
\begin{center}
\[
\mbox{\begin{picture}(100,70)(0,0)
\includegraphics[scale=.6]{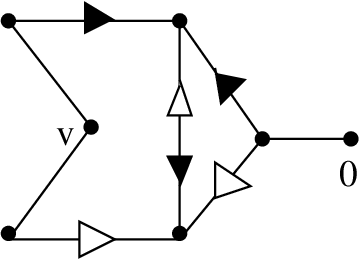}
\end{picture}}
\]
\caption{Auxiliary graph $F_{56}(v)$} 
\end{center}
\end{figure}
\begin{figure}[ht]
\begin{center}
\[
\mbox{\begin{picture}(240,55)(0,0)
\includegraphics[scale=.7]{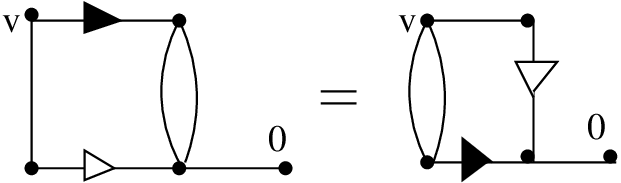}
\end{picture}}
\]
\caption{An identity among graphs} 
\end{center}
\end{figure}
\bea \label{F55}F_{55}(v) &=& -\pi^2 D^{(2,2,1)} (v)+\pi^2 E_5 -\pi \tau_2 \p_v G_3 (v) \overline\p_v G_3 (v) +\pi F_{57}(v) \non \\ &&-\pi\Big(F_2(v) + \frac{F_9(v)}{2} +c.c.\Big),\eea
where we have used the identity given in figure 47\footnote{Similar elementary identities have been used elsewhere in the calculation as well.}, and the graph $F_{57}(v)$ is given in figure 48.

\begin{figure}[ht]
\begin{center}
\[
\mbox{\begin{picture}(120,70)(0,0)
\includegraphics[scale=.7]{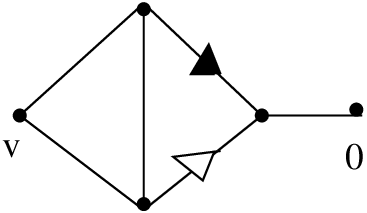}
\end{picture}}
\]
\caption{$F_{57}(v)$} 
\end{center}
\end{figure}
\begin{figure}[ht]
\begin{center}
\[
\mbox{\begin{picture}(320,80)(0,0)
\includegraphics[scale=.75]{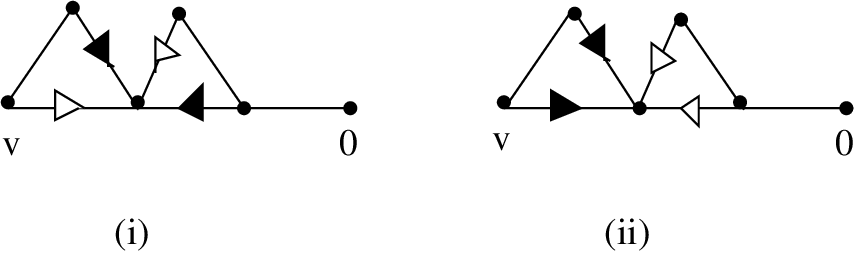}
\end{picture}}
\]
\caption{(i) $F_{58}(v)$, (ii) $F_{59}(v)$} 
\end{center}
\end{figure}

The second term on the right hand side of \C{f1} gives us
\bea &&\int_{zw}G(v,z) \p_\m G(v,z) G(z,w) \overline\p_\m G(z,w) G(w)=2D^{(1,1,3)} (v)-D_5^{(2;2;1)} (v) \non \\ &&+D_5^{(1,2,2)} (v)- E_2 E_3 -\frac{1}{\pi}\Big(F_4(v)+\frac{F_{51}(v)}{2}+c.c.\Big) +\frac{1}{\pi^2}\Big(F_{58}(v) + F_{59}(v)\Big),\eea
where the graphs $F_{58}(v)$ and $F_{59}(v)$ are given in figure 49. Note that both of them are real.

To simplify $F_{58}(v)$, we consider the auxiliary graph $F_{60}(v)$ in figure 50, which leads to
\begin{figure}[ht]
\begin{center}
\[
\mbox{\begin{picture}(170,80)(0,0)
\includegraphics[scale=.65]{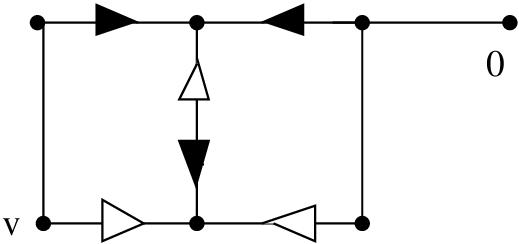}
\end{picture}}
\]
\caption{Auxiliary graph $F_{60}(v)$} 
\end{center}
\end{figure}
\begin{figure}[ht]
\begin{center}
\[
\mbox{\begin{picture}(140,80)(0,0)
\includegraphics[scale=.7]{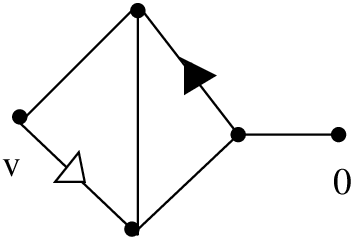}
\end{picture}}
\]
\caption{$F_{61}(v)$} 
\end{center}
\end{figure}
\bea &&F_{58}(v) = \pi^2 G_5 (v) + \pi^2 E_5 -2\pi^2 D^{(1,2,2)} (v) -\pi^2 D^{(2,2,1)} (v)-\pi F_7(v) \non \\ &&+\pi\Big(F_1(v)+F_2(v)+F_9(v)+c.c.\Big) +\pi F_{61}(v)-\pi \tau_2 \p_v G_3(v) \overline\p_v G_3(v),\eea
where the graph $F_{61}(v)$ is given in figure 51.

Also to simplify $F_{59}(v)$, we consider the auxiliary graph $F_{62}(v)$ in figure 52, which leads to
\begin{figure}[ht]
\begin{center}
\[
\mbox{\begin{picture}(140,65)(0,0)
\includegraphics[scale=.7]{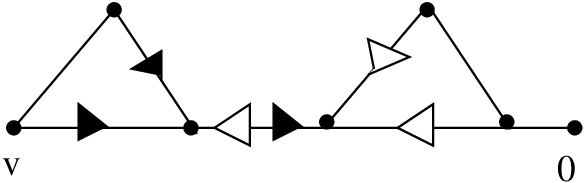}
\end{picture}}
\]
\caption{Auxiliary graph $F_{62}(v)$} 
\end{center}
\end{figure}
\bea F_{59}(v) = \pi^2 G_5(v) +\frac{\pi^2}{4} D_5^{(2;2;1)} (v) -\pi^2 D^{(1,1,3)} (v).\eea

Finally, the third term on the right hand side of \C{f1} gives 
\bea &&\int_{zw}\Big(G(v,z)^2 G(z,w) \p_\m G(z,w) \overline\p_\m G(w)+c.c.\Big) = -D_5^{(2;2;1)} (v) +2D^{(1,1,3)} (v) \non \\ && + 2 D_5^{(1,2,2)} (v) -2 E_2 E_3 -\frac{1}{\pi}\Big(F_{51}(v)+c.c.\Big).\eea

Thus adding the various contributions, we get that
\bea \label{D2;2;1}&&\frac{1}{4}\Big(\Delta -2\Big) D_5^{(2;2;1)} (v) = 4 G_5(v) +4D^{(1,1,3)} (v)-4 D^{(1,2,2)} (v) -4D^{(2,2,1)} (v) \non \\ &&+4 E_5  +\frac{2F_3(v)}{\pi}-\frac{2F_7(v)}{\pi} -\frac{4\tau_2}{\pi} \p_v G_3 (v) \overline\p_v G_3 (v) +\frac{1}{\pi}\Big(2F_1(v)-2F_4(v)\non \\ && +F_9(v)+F_{63}(v)-F_{64}(v)+c.c.\Big),\eea
where the graphs $F_{63}(v)$ and $F_{64}(v)$ are given in figure 53\footnote{Note that $F_{64}(v)= - F_{51}(v)^* +\pi E_2 D_3^{(1)} (v) -\pi D^{(1,1,3)}(v)$.}. 

\begin{figure}[ht]
\begin{center}
\[
\mbox{\begin{picture}(260,95)(0,0)
\includegraphics[scale=.7]{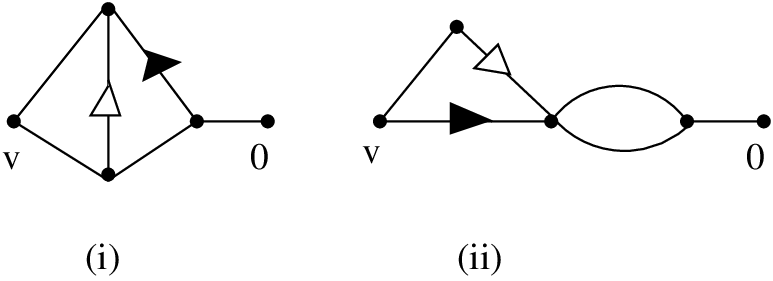}
\end{picture}}
\]
\caption{(i) $F_{63}(v)$, (ii) $F_{64}(v)$} 
\end{center}
\end{figure}

\subsection{Eigenvalue equation for $D_5^{(1,2,1;1)} (v)$}

For $D_5^{(1,2,1;1)}(v)$, we get
\bea \label{g1} \Delta D_5^{(1,2,1;1)}(v) &=& 2\int_{zw} \p_\m G(v,z) \overline\p_\m G(v,z) G(v,w) G(z,w)G(w)\non \\ &&+2\int_{zw} \Big(G(v,z) \p_\m G(v,z) G(z,w)\overline\p_\m G(v,w)G(w)+c.c.\Big)  \non \\ &&+\int_{zw}\Big( G(v,z)^2 \p_\m G(v,w) \overline\p_\m G(z,w)G(w)+c.c.\Big) \non \\ &&+ 2\int_{zw}\Big( G(v,z)\p_\m G(v,z) G(v,w) \overline\p_\m G(z,w)G(w)+c.c.\Big) \non \\&&+ 2\int_{zw}\Big(G(v,z) \p_\m G(v,z) G(z,w) G(v,w)\overline\p_\m G(w) +c.c.\Big)\non \\&&+ \int_{zw}\Big(G(v,z)^2 G(z,w)\p_\m G(v,w) \overline\p_\m G(w)+c.c.\Big)\non \\ &&+\int_{zw}\Big(G(v,z)^2 G(v,w) \p_\m G(z,w) \overline\p_\m G(w)+c.c.\Big).\eea

We now write down the contributions from the seven terms on the right hand side of \C{g1}. 

\begin{figure}[ht]
\begin{center}
\[
\mbox{\begin{picture}(120,90)(0,0)
\includegraphics[scale=.7]{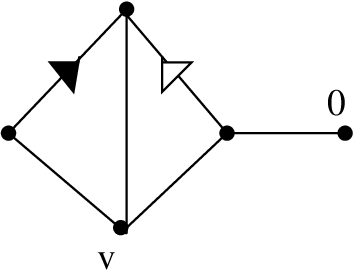}
\end{picture}}
\]
\caption{$F_{65}(v)$} 
\end{center}
\end{figure}

The first term gives
\bea \int_{zw} \p_\m G(v,z) \overline\p_\m G(v,z) G(v,w) G(z,w)G(w) &=& D_5^{(1,2,1;1)}(v) + E_3 G_2 (v) \non \\ &&-\frac{F_7(v)}{\pi} -\frac{1}{\pi}\Big(F_{65}(v)+c.c.\Big),\eea
where the graph $F_{65}(v)$ is given in figure 54\footnote {Note that $F_{65}(v) = F_{16}(v)^*+F_{29}(v)$.}.

In calculating the second term, we encounter the graph $F_{66}(v)$ which we simplify using the auxiliary graph $F_{67}(v)$, both of which are given in figure 55. 
\begin{figure}[ht]
\begin{center}
\[
\mbox{\begin{picture}(270,115)(0,0)
\includegraphics[scale=.7]{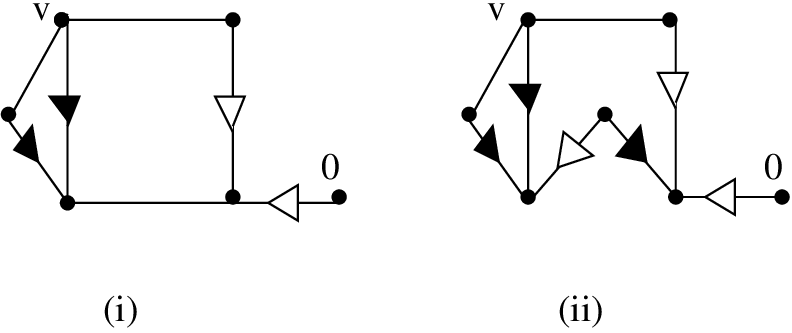}
\end{picture}}
\]
\caption{(i) $F_{66}(v)$ and (ii) auxiliary graph $F_{67}(v)$} 
\end{center}
\end{figure}
This gives
\bea &&\int_{zw} \Big(G(v,z) \p_\m G(v,z) G(z,w)\overline\p_\m G(v,w)G(w)+c.c.\Big) =2E_5 -C_{1,1,3} +\frac{2F_7(v)}{\pi}\non \\ &&+\frac{1}{\pi}\Big(Q_1 +F_1(v) -\frac{F_9(v)}{2} -F_{16}(v) -F_{17}(v) -F_{24}(v)+\frac{F_{40}(v)}{2} +\frac{F_{68}(v)}{2} +c.c.\Big)\non \\ &&- \frac{\tau_2}{\pi}\Big(\p_v G_2(v) \overline\p_v G_4(v)+c.c.\Big), \eea
where the graph $F_{68}(v)$ is given in figure 56.
\begin{figure}[ht]
\begin{center}
\[
\mbox{\begin{picture}(120,95)(0,0)
\includegraphics[scale=.8]{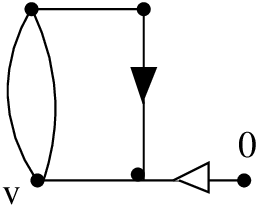}
\end{picture}}
\]
\caption{$F_{68}(v)$} 
\end{center}
\end{figure}

The third term gives
\bea &&\int_{zw}\Big( G(v,z)^2 \p_\m G(v,w) \overline\p_\m G(z,w)G(w)+c.c.\Big) \non \\ &&=2C_{1,1,3} +\frac{1}{\pi}\Big(F_9(v)-F_{40}(v)-F_{68}(v)+c.c.\Big),\eea
while the fourth term gives us
\bea &&\int_{zw}\Big( G(v,z)\p_\m G(v,z) G(v,w) \overline\p_\m G(z,w)G(w)+c.c.\Big) \non \\ &&=2D^{(1,3,1)} (v) - D_5^{(1,2,1;1)} (v) +\frac{1}{\pi}\Big(F_{65}(v)+c.c.\Big). \eea

To calculate the fifth term, we encounter the graph $F_{69}(v)$ which we simplify using the auxiliary graph $F_{70}(v)$, both of which are given in figure 57. 

\begin{figure}[ht]
\begin{center}
\[
\mbox{\begin{picture}(300,145)(0,0)
\includegraphics[scale=.8]{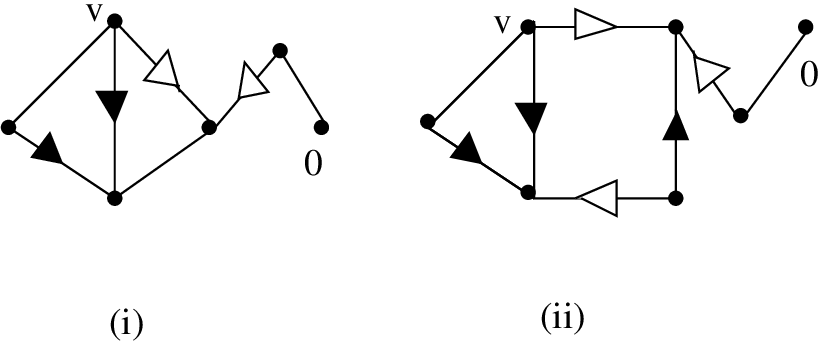}
\end{picture}}
\]
\caption{(i) $F_{69}(v)$ and (ii) auxiliary graph $F_{70}(v)$} 
\end{center}
\end{figure}
This leads to
\bea &&\int_{zw}\Big(G(v,z) \p_\m G(v,z) G(z,w) G(v,w)\overline\p_\m G(w) +c.c.\Big) = D^{(1,1,3)} (v) + 2 G_5 (v) \non \\ &&-2 D^{(1,3,1)} (v)+ D_5^{(1,2,1;1)} (v) +\frac{1}{\pi}\Big(F_2 (v) -F_4(v) + F_5(v)  -\frac{F_{10}(v)}{2} +F_{24}(v)\non \\ &&-F_{29}(v)-\frac{F_{68}(v)}{2}+c.c.\Big).\eea

Next, the sixth term yields
\bea &&\int_{zw}\Big(G(v,z)^2 G(z,w)\p_\m G(v,w) \overline\p_\m G(w)+c.c.\Big)= 2D_5^{(1,2,1;1)} (v) + 2 E_2 G_3(v)\non \\ &&-\frac{1}{\pi}\Big(F_5(v) + F_9(v)+F_{10}(v)+c.c.\Big).\eea

Finally, the seventh term gives us 
\bea &&\int_{zw}\Big(G(v,z)^2 G(v,w) \p_\m G(z,w) \overline\p_\m G(w)+c.c.\Big)= 2D^{(1,1,3)} (v) -2 D_5^{(1,2,1;1)}(v) \non \\ &&+\frac{1}{\pi}\Big(F_{10}(v)+F_{68}(v)+c.c.\Big).\eea

Thus adding the several contributions, we get that
\bea&&\Big(\Delta -2\Big)D_5^{(1,2,1;1)} (v) = 4D^{(1,1,3)}(v)+4 E_5 + 4 G_5 (v) +2 E_3 G_2(v)+ 2 E_2 G_3(v)\non \\ &&+\frac{1}{\pi}\Big(2 Q_1+2F_1(v)+2F_2(v)-2F_4(v)+F_5(v)-F_9(v)-F_{10}(v)-2F_{17}(v)\non \\&& -2F_{65}(v)+c.c.\Big)+\frac{2F_7(v)}{\pi} -\frac{2\tau_2}{\pi}\Big(\p_vG_2(v)\overline\p_v G_4(v)+c.c.\Big).\eea

\subsection{Eigenvalue equation for $D_5^{(1;2;2)} (v)$}

\begin{figure}[ht]
\begin{center}
\[
\mbox{\begin{picture}(220,100)(0,0)
\includegraphics[scale=.8]{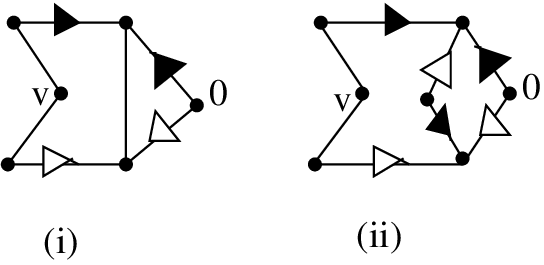}
\end{picture}}
\]
\caption{(i) $F_{71}(v)$ and (ii) auxiliary graph $F_{72}(v)$} 
\end{center}
\end{figure}

For $D_5^{(1;2;2)} (v)$, we have that
\bea \label{h1}\frac{1}{4}\Delta D_5^{(1;2;2)} (v) &=& \int_{zw} \p_\m G(v,z) \overline\p_\m G(v,w) G(z,w) G(z)G(w)\non \\ &&+ \int_{zw} G(v,w) \p_\m G(v,z) G(z,w)G(w) \overline\p_\m G(z) \non \\ &&+\int_{zw} G(v,w) \p_\m G(v,z) G(z,w) G(z)\overline\p_\m G(w)\non \\&& +\int_{zw}\Big(G(v,w) \p_\m G(v,z) \overline\p_\m G(z,w)G(w)G(z)+c.c.\Big).\eea

We now consider the contributions from the four terms on the right hand side of \C{h1}.

\begin{figure}[ht]
\begin{center}
\[
\mbox{\begin{picture}(220,110)(0,0)
\includegraphics[scale=.7]{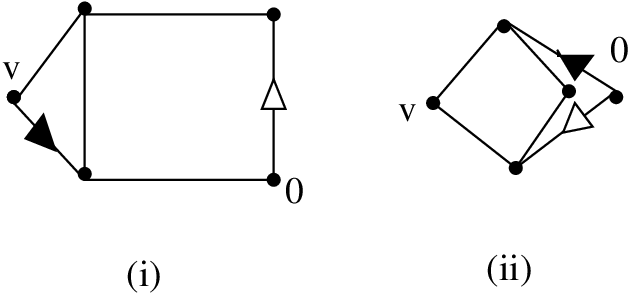}
\end{picture}}
\]
\caption{(i) $F_{73}(v)$, (ii) $F_{74}(v)$} 
\end{center}
\end{figure}

To calculate the first term, we encounter the graph $F_{71}(v)$ which we simplify using the auxiliary graph $F_{72}(v)$, both of which are given in figure 58, to get  
\bea &&\int_{zw} \p_\m G(v,z) \overline\p_\m G(v,w) G(z,w) G(z)G(w)= E_5 +\frac{F_{36}(v)}{\pi}+\frac{F_{74}(v)}{\pi} \non \\ &&+\frac{1}{\pi}\Big(F_1 (v)-F_{24}(v) -F_{25}(v) -F_{40}(v)-F_{73}(v)+c.c.\Big)\non \\ &&+\frac{\tau_2}{\pi} E_2 \p_v G_2(v) \overline\p_v G_2(v) -\frac{\tau_2}{\pi}\p_v G_3(v) \overline\p_v G_3(v)-\frac{\tau_2}{\pi}\Big(\p_vG_2(v) \overline\p_v G_4(v)+c.c.\Big),\eea
where the graphs $F_{73}(v)$ and $F_{74}(v)$ are given in figure 59. 

\begin{figure}[ht]
\begin{center}
\[
\mbox{\begin{picture}(140,75)(0,0)
\includegraphics[scale=.6]{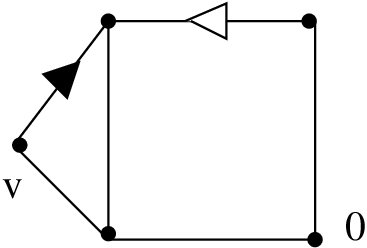}
\end{picture}}
\]
\caption{$F_{75}(v)$} 
\end{center}
\end{figure}

The second term gives
\bea \int_{zw} G(v,w) \p_\m G(v,z) G(z,w)G(w) \overline\p_\m G(z) &=& G_2(v)G_3(v) - D_5^{(1;2;2)} (v)\non \\ &&-\frac{F_{46}(v)}{\pi}+\frac{1}{\pi}\Big(F_{75}(v)+c.c.\Big),\eea
where the graph $F_{75}(v)$ is given in figure 60.

\begin{figure}[ht]
\begin{center}
\[
\mbox{\begin{picture}(200,130)(0,0)
\includegraphics[scale=.6]{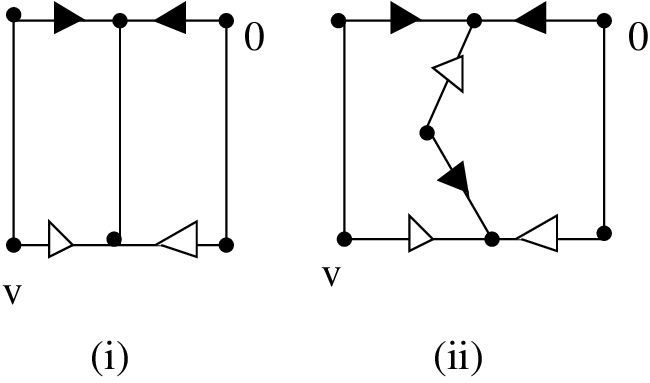}
\end{picture}}
\]
\caption{(i) $F_{76}(v)$ and (ii) auxiliary graph $F_{77}(v)$} 
\end{center}
\end{figure}

To evaluate the third term, we encounter the graph $F_{76}(v)$ which we simplify using the auxiliary graph $F_{77}(v)$, both of which are given in figure 61, to get  
\bea &&\int_{zw} G(v,w) \p_\m G(v,z) G(z,w) G(z)\overline\p_\m G(w) = G_5(v) -2D^{(1,3,1)}(v)+\frac{F_{46}(v)}{\pi}\non \\ &&+\frac{F_{78}(v)}{\pi}+\frac{1}{\pi}\Big(F_2(v) +F_{17}(v)+F_{24}(v)+F_{35}(v)-F_{41}(v)+F_{73}(v)+c.c.\Big) \non \\ &&+\frac{\tau_2}{\pi} \p_v G_3(v) \overline\p_v G_3(v) - \frac{\tau_2}{\pi} G_2(v) \p_v G_2(v) \overline\p_v G_2(v),\eea
where the graph $F_{78}(v)$ is given in figure 62.

\begin{figure}[ht]
\begin{center}
\[
\mbox{\begin{picture}(140,75)(0,0)
\includegraphics[scale=.75]{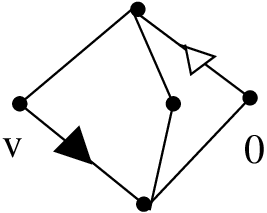}
\end{picture}}
\]
\caption{$F_{78}(v)$} 
\end{center}
\end{figure}

Finally, the fourth term yields
\bea &&\int_{zw}\Big(G(v,w) \p_\m G(v,z) \overline\p_\m G(z,w)G(w)G(z)+c.c.\Big) = 2D^{(1,3,1)} (v)\non \\&&+\frac{1}{\pi}\Big(F_{25}(v)-F_{35}(v)+F_{41}(v)+c.c.\Big)-\frac{2F_{79}(v)}{\pi},\eea
where the graph $F_{79}(v)$ is given in figure 63.

\begin{figure}[ht]
\begin{center}
\[
\mbox{\begin{picture}(100,85)(0,0)
\includegraphics[scale=.8]{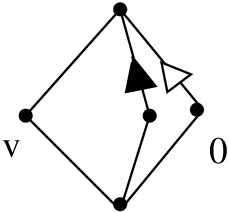}
\end{picture}}
\]
\caption{$F_{79}(v)$} 
\end{center}
\end{figure}

Thus adding the various contributions, we get that
\bea &&\frac{1}{4}\Big(\Delta -2\Big)D_5^{(1;2;2)} (v) = E_5 + G_5 (v)+ G_2(v)G_3(v) +\frac{\tau_2}{\pi}E_2 \p_v G_2(v) \overline\p_v G_2(v)\non \\ &&- \frac{\tau_2}{\pi}G_2(v) \p_v G_2(v) \overline\p_v G_2(v)+\frac{F_{36}(v)}{\pi}+\frac{1}{\pi}\Big(F_1(v)+F_2(v)+F_{17}(v)-F_{40}(v)\non \\ &&-F_{41}(v)+c.c.\Big)-\frac{\tau_2}{\pi}\Big(\p_v G_2(v)\overline\p_v G_4(v)+c.c.\Big).\eea

We now consider the eigenvalue equations for $D_5^{(1)} (v)$ and $D_5^{(2)}(v)$ which are obtained by cutting open $D_5$.

\subsection{Eigenvalue equation for $D_5^{(1)} (v)$}

For $D_5^{(1)}(v)$, we have that
\bea \label{l1}\frac{1}{4}\Delta D_5^{(1)} (v) &=& 3\int_z G(v,z)^2 \p_\m G(v,z) \overline\p_\m G(v,z) G(z)\non \\ &&+ \int_z\Big(G(v,z)^3 \p_\m G(v,z) \overline\p_\m G(z)+c.c.\Big).\eea

\begin{figure}[ht]
\begin{center}
\[
\mbox{\begin{picture}(350,105)(0,0)
\includegraphics[scale=.8]{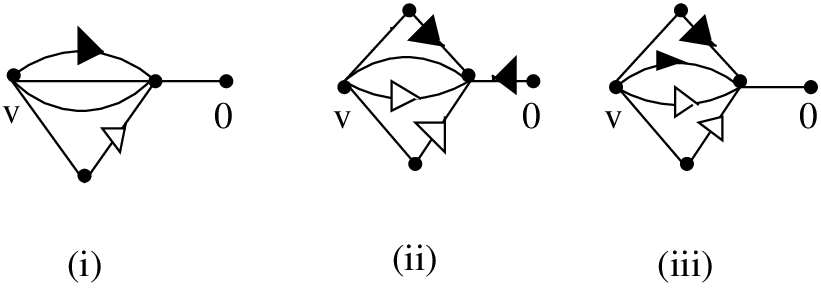}
\end{picture}}
\]
\caption{(i) $F_{80}(v)$, (ii) $F_{81}(v)$, (iii) $F_{82}(v)$} 
\end{center}
\end{figure}

The first term on the right hand side gives
\bea \int_z G(v,z)^2 \p_\m G(v,z) \overline\p_\m G(v,z) G(z) = - D_5^{(1)} (v) +\frac{Q_3}{\pi} -\frac{\tau_2}{\pi} G(v)^2 \p_v G_2(v) \overline\p_v G_2(v) \non \\ +\frac{2F_7(v)}{\pi}+\frac{1}{\pi}\Big(F_{80}(v) +\frac{2}{\pi} F_{81}(v)+c.c.\Big)+\frac{2}{\pi^2}F_{82}(v),\eea
where $Q_3$ is given in appendix B, and the graphs $F_{80}(v)$, $F_{81}(v)$ and $F_{82}(v)$ are given in figure 64. 

\begin{figure}[ht]
\begin{center}
\[
\mbox{\begin{picture}(180,60)(0,0)
\includegraphics[scale=.8]{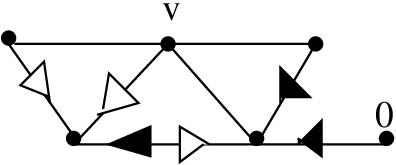}
\end{picture}}
\]
\caption{Auxiliary graph $F_{83}(v)$} 
\end{center}
\end{figure}

Now to simplify $F_{81}(v)$, we start with the auxiliary graph $F_{83}(v)$ in figure 65, leading to
\bea \Big(F_{81}(v)+c.c.\Big)-\Big(P_2 P_5(v)^*+c.c.\Big)= -\pi\tau_2 G(v)\Big(\p_v G_2(v) \overline\p_v G_3(v)+c.c.\Big)\non \\
+\pi\Big(Q_1 -\frac{Q_2}{2}-F_{18}(v)+\frac{F_{84}(v)}{2}+\frac{F_{85}(v)}{2}+c.c.\Big)  +\Big(F_{19}(v)-\frac{F_{86}(v)}{2}+c.c.\Big), \eea
where $P_5(v)$ is given in appendix B, and the graphs $F_{84}(v)$, $F_{85}(v)$ and $F_{86}(v)$ are given in figure 66.  

\begin{figure}[ht]
\begin{center}
\[
\mbox{\begin{picture}(320,110)(0,0)
\includegraphics[scale=.7]{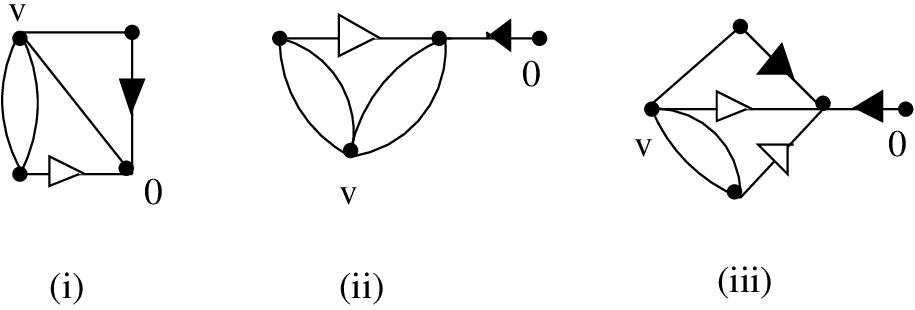}
\end{picture}}
\]
\caption{(i) $F_{84}(v)$, (ii) $F_{85}(v)$, (iii) $F_{86}(v)$} 
\end{center}
\end{figure}

To simplify $F_{86}(v)$, we start with the auxiliary graph $F_{87}(v)$ in figure 67, leading to
\begin{figure}[ht]
\begin{center}
\[
\mbox{\begin{picture}(120,85)(0,0)
\includegraphics[scale=.6]{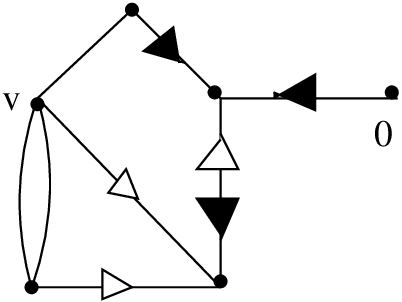}
\end{picture}}
\]
\caption{Auxiliary graph $F_{87}(v)$} 
\end{center}
\end{figure}
\bea \Big(F_{86}(v)+c.c.\Big)- \Big(P_3(v)P_6^*+c.c.\Big) = -\pi\Big(F_{40}(v)+F_{68}(v) -\frac{F_{88}(v)}{3} -\frac{F_{89}(v)}{3}\non \\+E_2 F_{90}(v)+c.c.\Big)+2\pi^2 C_{1,1,3} -\frac{2\pi^2}{3}D_{1,1,3} +2\pi^2 E_2E_3 -2\pi\tau_2 E_2 \p_v G_2(v) \overline\p_v G_2(v),\eea
where $P_6$ is given in appendix B, and the graphs $F_{88}(v)$, $F_{89}(v)$ and $F_{90}(v)$ are given in figure 68. 

\begin{figure}[ht]
\begin{center}
\[
\mbox{\begin{picture}(260,80)(0,0)
\includegraphics[scale=.7]{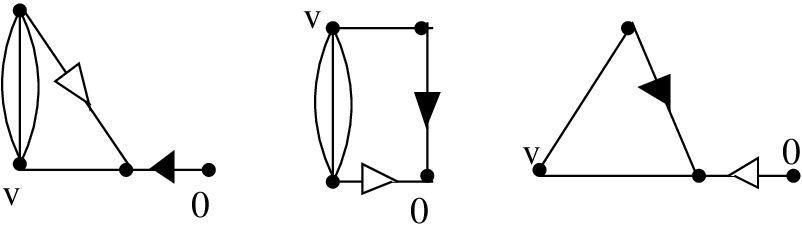}
\end{picture}}
\]
\caption{(i) $F_{88}(v)$, (ii) $F_{89}(v)$, (iii) $F_{90}(v)$} 
\end{center}
\end{figure}

\begin{figure}[ht]
\begin{center}
\[
\mbox{\begin{picture}(100,75)(0,0)
\includegraphics[scale=.55]{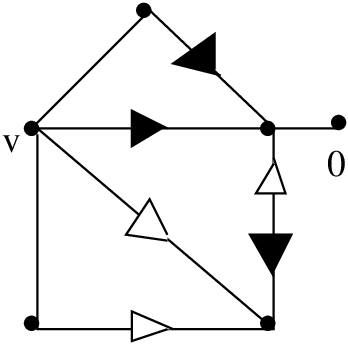}
\end{picture}}
\]
\caption{$F_{91}(v)$} 
\end{center}
\end{figure}

Next to simplify $F_{82}(v)$, we start with the auxiliary graph $F_{91}(v)$ given in figure 69, which leads to
\bea &&F_{82}(v)- \frac{1}{2}\Big(P_2 P_7(v)^*+c.c.\Big)=\frac{\pi^2}{4}D_5^{(1)}(v) - \frac{\pi^2}{4} E_2 D_3^{(1)} (v) -\frac{\pi^2}{2} D^{(1,1,2,1)}(v) \non \\ &&+\frac{\pi}{2}\Big(F_1(v)-\frac{F_9(v)}{2}+\frac{F_{18}(v)}{2}-\frac{F_{85}(v)}{4}  +c.c.\Big)+\frac{1}{2}\Big(F_{20}(v) -\frac{F_{92}(v)}{2}+c.c.\Big),\eea
where $P_7 (v)$ is given in appendix B, and the graph $F_{92}(v)$ is given in figure 70.
\begin{figure}[ht]
\begin{center}
\[
\mbox{\begin{picture}(100,85)(0,0)
\includegraphics[scale=.7]{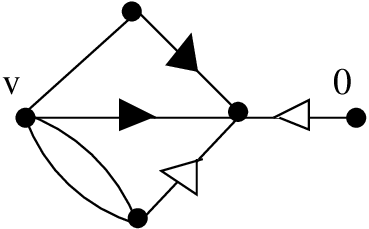}
\end{picture}}
\]
\caption{$F_{92}(v)$} 
\end{center}
\end{figure}

To simplify $F_{92}(v)$, we start with the auxiliary graph $F_{93}(v)$ in figure 71, leading to
\begin{figure}[ht]
\begin{center}
\[
\mbox{\begin{picture}(100,90)(0,0)
\includegraphics[scale=.7]{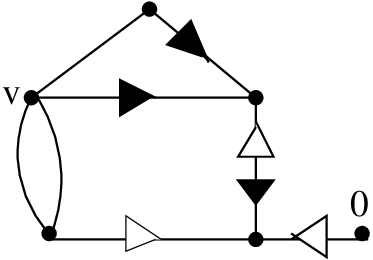}
\end{picture}}
\]
\caption{Auxiliary graph $F_{93}(v)$} 
\end{center}
\end{figure}
\bea &&\Big(F_{92}(v) +c.c.\Big)-\Big(P_2 P_8(v)^*+c.c.\Big)=2\pi^2 C_{1,1,3} -\pi^2 D_{1,2,2}+2\pi^2 E_2 G_3(v) \non\\ &&-\pi^2 E_2 D_3^{(1)}(v)-\pi\Big(F_{18}(v)+F_{26}(v)-\frac{F_{85}(v)}{2}+c.c.\Big) +\pi F_{94}(v),\eea
where $P_8(v)$ is given in appendix B, and the graph $F_{94}(v)$ is given in figure 72.
\begin{figure}[ht]
\begin{center}
\[
\mbox{\begin{picture}(100,70)(0,0)
\includegraphics[scale=.7]{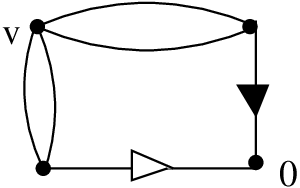}
\end{picture}}
\]
\caption{$F_{94}(v)$} 
\end{center}
\end{figure}

\begin{figure}[ht]
\begin{center}
\[
\mbox{\begin{picture}(120,70)(0,0)
\includegraphics[scale=.7]{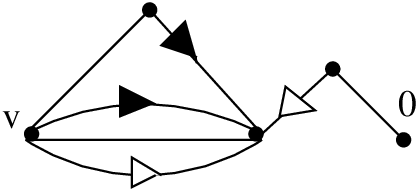}
\end{picture}}
\]
\caption{$F_{95}(v)$} 
\end{center}
\end{figure}

Next the second term on the right hand side of \C{l1} gives us
\bea &&\int_z\Big(G(v,z)^3 \p_\m G(v,z) \overline\p_\m G(z)+c.c.\Big) = \frac{1}{2}D_5^{(1)} (v) \non \\ &&+\frac{3}{\pi}\Big(F_5(v) - F_{80}(v)+\frac{2}{\pi}F_{95}(v)+c.c.\Big),\eea
where the graph $F_{95}(v)$ is given in figure 73. 

To simplify $F_{95}(v)$, we use the auxiliary graph $F_{96}(v)$ in figure 74 to get 
\begin{figure}[ht]
\begin{center}
\[
\mbox{\begin{picture}(120,90)(0,0)
\includegraphics[scale=.7]{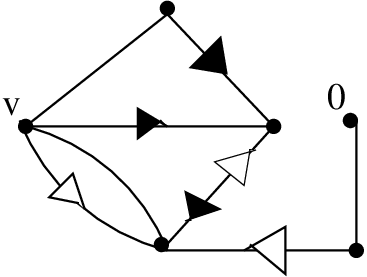}
\end{picture}}
\]
\caption{Auxiliary graph $F_{96}(v)$} 
\end{center}
\end{figure}
\bea &&\Big(F_{95}(v)+c.c.\Big)-\frac{1}{2}\Big(P_2 P_8(v)^*+c.c.\Big) =\frac{\pi^2}{2} D_5^{(1)}(v) -\frac{\pi^2}{2} E_2 D_3^{(1)} (v) -\pi^2 D^{(1,1,2,1)}(v) \non \\ &&+\pi\Big(F_2(v) -\frac{F_{10}(v)}{2} +\frac{F_{18}(v)}{2}+\frac{F_{30}(v)}{\pi}-\frac{F_{85}(v)}{4}-\frac{F_{97}(v)}{2\pi}+c.c.\Big) ,\eea
where the graph $F_{97}(v)$ is given in figure 75. 
\begin{figure}[ht]
\begin{center}
\[
\mbox{\begin{picture}(120,70)(0,0)
\includegraphics[scale=.6]{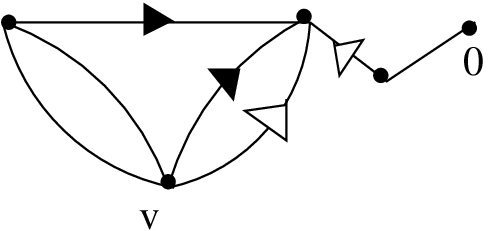}
\end{picture}}
\]
\caption{$F_{97}(v)$} 
\end{center}
\end{figure}
To simplify it, we start with the auxiliary graph $F_{98}(v)$ in figure 76 to get
\begin{figure}[ht]
\begin{center}
\[
\mbox{\begin{picture}(120,70)(0,0)
\includegraphics[scale=.65]{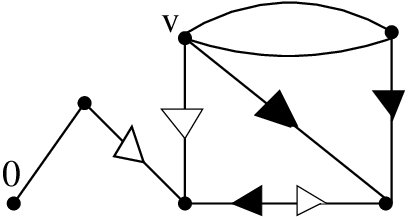}
\end{picture}}
\]
\caption{Auxiliary graph $F_{98}(v)$} 
\end{center}
\end{figure}
\bea &&\Big(F_{97}(v)+c.c.\Big)-\Big(P_3(v)P_6^*+c.c.\Big)= \frac{2\pi^2}{3} D_5^{(1)}(v) -\frac{2\pi^2}{3} D^{(1,1,1,2)} +2\pi^2 D^{(1,1,3)} (v)\non \\  &&-2\pi^2 D_5^{(1,2,1;1)} (v)-\frac{2\pi^2}{3} D_3 G_2(v) -2\pi^2 E_2 D_3^{(1)} (v) +2\pi^2 E_2 G_3(v)  \non \\ &&+\pi\Big(F_{68}(v)-\frac{F_{88}(v)}{3}+E_2 F_{90}(v)+c.c.\Big).\eea

Thus adding the various contributions, we get that
\bea \label{D51}&&\frac{1}{4}\Delta D_5^{(1)} (v) = \frac{1}{\pi^2}\Big(3 P_1 (v) P_2^*+ 6 P_2 P_5(v)^*+3 P_2 P_7(v)^*+ \frac{3}{2} P_2 P_8(v)^*\non \\&&+12 P_3(v)P_4^* 
-6 P_3(v)P_6^*+c.c.\Big) -6 D^{(1,1,2,1)} (v)+ 2D^{(1,1,1,2)} (v)\non \\ &&-12 D^{(1,3,1)} (v)+ 6D_5^{(1,2,1;1)}(v)-6 D^{(2,2,1)} (v)  -6 D^{(1,1,3)}(v)+3 E_2 D_3^{(1)} (v)\non \\&& +2 D_3 G_2(v)-24 E_2 G_3(v)+24 G_5(v)-12 C_{122} -12 C_{1,1,3}+2 D_{1,1,3} \non \\ &&+\frac{3}{2}D_{1,2,2}+18 E_5-6 E_2E_3 +\frac{1}{\pi}\Big(12 Q_1 -3 Q_2 +6F_1(v)  
+6F_2(v)\non \\&&+6F_4(v) +3F_5(v)- 3F_9(v)-3F_{10}(v)+6F_{23}(v)+6F_{25}(v) +3F_{26}(v)\non \\&&+6F_{32}(v)+3F_{40}(v)+3F_{84}(v)-F_{89}(v)+c.c.\Big)+\frac{3}{\pi}Q_3+\frac{6}{\pi}F_7(v)\non \\&&-\frac{3}{2\pi}F_{94}(v)-\frac{3\tau_2}{\pi} G(v)^2 \p_v G_2(v) \overline\p_v G_2(v)+\frac{6\tau_2}{\pi} E_2 \p_v G_2(v) \overline\p_v G_2(v)\non \\&&-\frac{6\tau_2}{\pi} \p_v G_3(v) \overline\p_v G_3(v)-\frac{6\tau_2}{\pi}\Big( \p_v  G_2(v) \overline\p_v G_4(v)+c.c.\Big)\non \\&&-\frac{6\tau_2}{\pi} G(v)\Big(\p_v G_2(v) \overline\p_v G_3(v)+c.c.\Big).\eea

\subsection{Eigenvalue equation for $D_5^{(2)} (v)$}

\begin{figure}[ht]
\begin{center}
\[
\mbox{\begin{picture}(405,100)(0,0)
\includegraphics[scale=.7]{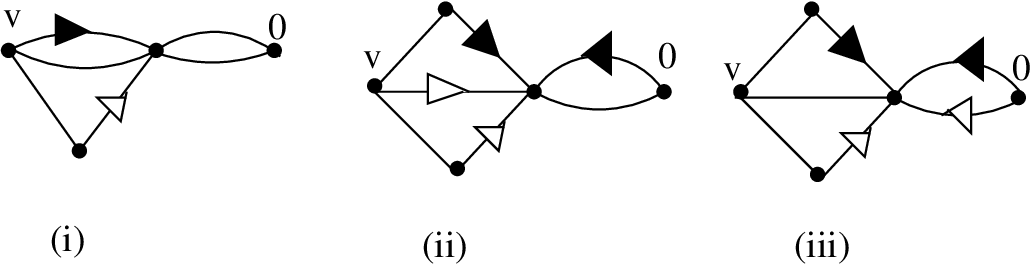}
\end{picture}}
\]
\caption{(i) $F_{99}(v)$, (ii) $F_{100} (v)$, (iii) $F_{101}(v)$} 
\end{center}
\end{figure}

We have that
\bea \label{m1}\frac{1}{2}\Delta D_5^{(2)} (v) &=& 3 \int_z G(v,z) \p_\m G(v,z) \overline\p_\m G(v,z)G(z)^2 + \int_z G(v,z)^3 \p_\m G(z) \overline\p_\m G(z) \non\\ &&+3\int_z\Big(G(v,z)^2 \p_\m G(v,z) G(z) \overline\p_\m G(z)+c.c.\Big).\eea
Since the analysis of this graph is somewhat more involved, we describe the intermediate steps in some detail.

\begin{figure}[ht]
\begin{center}
\[
\mbox{\begin{picture}(240,120)(0,0)
\includegraphics[scale=.65]{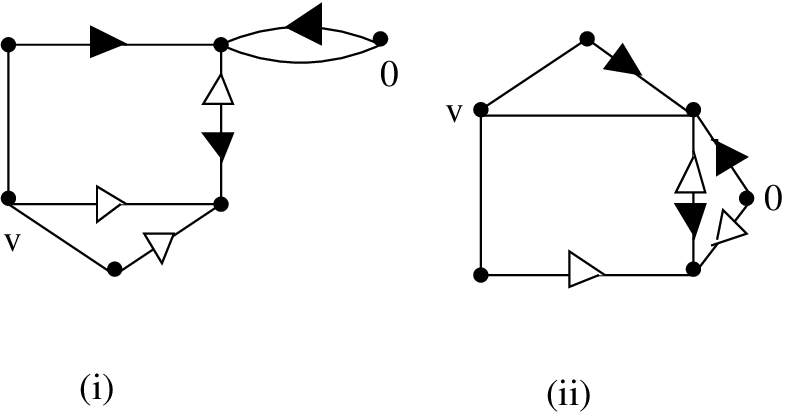}
\end{picture}}
\]
\caption{Auxiliary graphs (i) $F_{102}(v)$, (ii) $F_{103} (v)$} 
\end{center}
\end{figure}

Consider the first term on the right hand side of \C{m1}. We have that
\bea \int_z G(v,z) \p_\m G(v,z) \overline\p_\m G(v,z)G(z)^2 = - D_5^{(2)} (v)+\frac{1}{\pi}\Big(2F_7(v) + F_{36}(v)\Big) \non \\ +\frac{1}{\pi}\Big(F_{99}(v) + \frac{2}{\pi}F_{100}(v)+c.c.\Big)+\frac{2}{\pi^2}F_{101}(v),\eea
where the graphs $F_{99}(v)$, $F_{100}(v)$ and $F_{101}(v)$ are given in figure 77. 

To simplify $F_{100}(v)$ and $F_{101}(v)$, we start with the auxiliary graphs $F_{102}(v)$ and $F_{103}(v)$ respectively, given in figure 78. For $F_{100}(v)$, we get that
\bea &&\Big(F_{100}(v)+c.c.\Big)-\frac{1}{2} \Big(P_2 P_8(v)^*+c.c.\Big)= -\pi^2 D_5^{(1,2,2)}(v) \non \\ &&+\pi \Big(F_1(v) -\frac{F_9(v)}{2} +\frac{F_{42}(v)}{2} +\frac{F_{104}(v)}{2} +\frac{F_{43}(v)}{\pi} -\frac{F_{105}(v)}{2\pi}+c.c.\Big),    \eea 
where the graphs $F_{104}(v)$ and $F_{105}(v)$ are given in figure 79.

\begin{figure}[ht]
\begin{center}
\[
\mbox{\begin{picture}(240,75)(0,0)
\includegraphics[scale=.7]{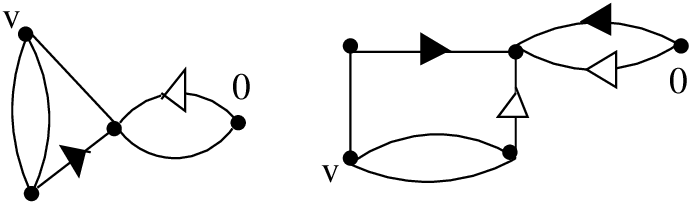}
\end{picture}}
\]
\caption{(i) $F_{104}(v)$, (ii) $F_{105} (v)$} 
\end{center}
\end{figure}
To simplify $F_{105}(v)$, we start with the auxiliary graph $F_{106}(v)$ in figure 80 to obtain
\begin{figure}[ht]
\begin{center}
\[
\mbox{\begin{picture}(110,80)(0,0)
\includegraphics[scale=.6]{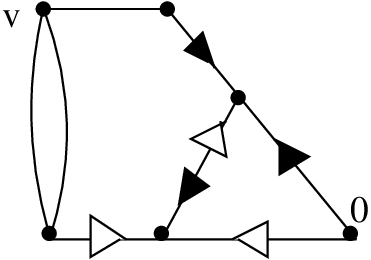}
\end{picture}}
\]
\caption{$F_{106} (v)$} 
\end{center}
\end{figure}
\bea &&\Big(F_{105}(v)+c.c.\Big) -\Big(P_3(v)P_8(v)^*+c.c.\Big) = 2\pi^2 C_{1,1,3} +2\pi^2 E_2 G_3(v)\non \\ &&-2\pi^2  E_2 D_3^{(1)} (v)  -\pi\Big(F_{18}(v) + F_{26}(v)+F_{40}(v) +F_{68}(v)\non \\ &&-E_2 F_{90}(v)+F_{108}(v)- F_{109}(v)+c.c.\Big)-\pi F_{107}(v),\eea
where the graphs $F_{107}(v)$, $F_{108}(v)$ and $F_{109}(v)$ are given in figure 81. 
\begin{figure}[ht]
\begin{center}
\[
\mbox{\begin{picture}(250,100)(0,0)
\includegraphics[scale=.7]{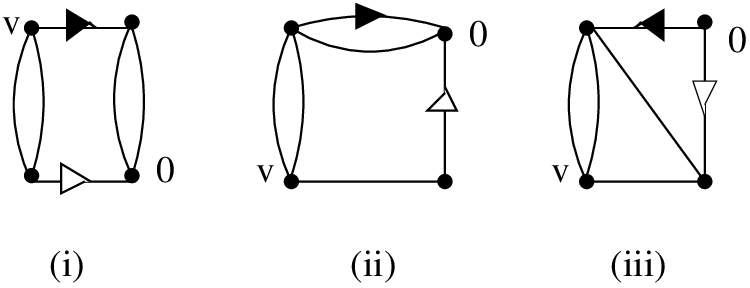}
\end{picture}}
\]
\caption{(i) $F_{107} (v)$, (ii) $F_{108}(v)$, (iii) $F_{109}(v)$} 
\end{center}
\end{figure}

For $F_{101}(v)$, we get
\bea F_{101}(v)-\frac{1}{2}\Big(P_3(v) P_5(v)^*+c.c.\Big)= -\frac{\pi\tau_2}{2}G(v)\Big(\p_v G_2(v)\overline\p_v G_3(v)+c.c.\Big) \non \\ +\frac{1}{2}\Big(\pi Q_1 -\pi F_{17}(v)-\pi F_{18}(v) +F_{19}(v) -\pi F_{108} (v)+ F_{110}(v)+F_{111}(v)+c.c.\Big),\eea
where the graphs $F_{110}(v)$ and $F_{111}(v)$ are given in figure 82. The simplification of the graph $F_{111}(v)$ has been analyzed in appendix E, while we consider $F_{110}(v)$ later.
\begin{figure}[ht]
\begin{center}
\[
\mbox{\begin{picture}(270,110)(0,0)
\includegraphics[scale=.7]{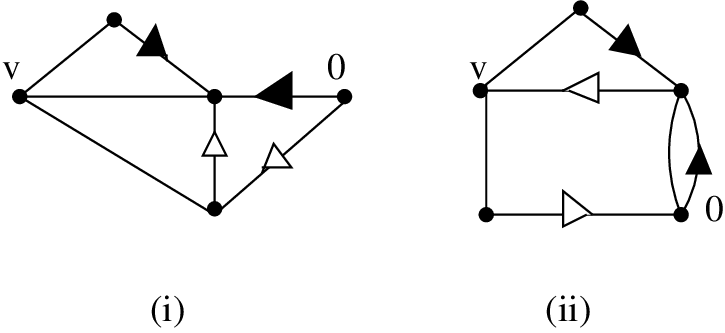}
\end{picture}}
\]
\caption{(i) $F_{110} (v)$, (ii) $F_{111}(v)$} 
\end{center}
\end{figure}

The second term on the right hand side of \C{m1} gives
\bea  \int_z G(v,z)^3 \p_\m G(z) \overline\p_\m G(z)= -D_5^{(2)}(v) +\frac{3}{\pi} F_{36}(v)+\frac{1}{\pi}\Big(F_{112}(v)+c.c.\Big)+\frac{6}{\pi^2}F_{113}(v),\eea
where the graphs $F_{112}(v)$ and $F_{113}(v)$ are given in figure 83. To simplify $F_{113}(v)$, we start with the auxiliary graph $F_{114}(v)$ given in figure 84 giving us
\begin{figure}[ht]
\begin{center}
\[
\mbox{\begin{picture}(290,100)(0,0)
\includegraphics[scale=.75]{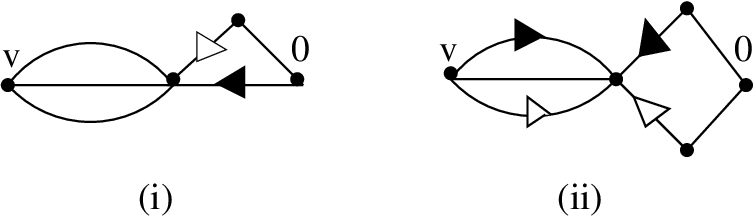}
\end{picture}}
\]
\caption{(i) $F_{112} (v)$, (ii) $F_{113}(v)$} 
\end{center}
\end{figure}
\bea&& F_{113}(v) -\frac{1}{4}\Big(P_3(v) P_8(v)^*+c.c.\Big)= -\frac{\pi^2}{2} D_5^{(1,2,2)}(v) \non \\ &&+\frac{\pi}{2}\Big(F_1(v) -\frac{F_{40}(v)}{2} +\frac{F_{42}(v)}{2} +\frac{F_{43}(v)}{\pi}-\frac{F_{89}(v)}{6} +\frac{F_{115}(v)}{\pi}+c.c.\Big),\eea
where $F_{115}(v)$ is given in figure 85.

\begin{figure}[ht]
\begin{center}
\[
\mbox{\begin{picture}(180,70)(0,0)
\includegraphics[scale=.7]{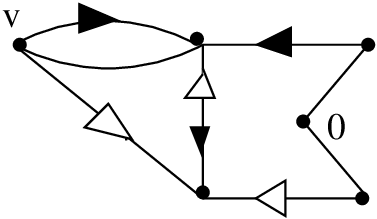}
\end{picture}}
\]
\caption{Auxiliary graph $F_{114} (v)$} 
\end{center}
\end{figure}

Now the graphs $F_{110}(v)$ and $F_{115}(v)$ are not easy to analyze and simplify directly. This follows from trying to move the derivatives along the circuit on integrating by parts. Hence we shall not analyze them directly. However, we shall see that there are contributions from the remaining terms that will prove very useful in enabling us to evaluate the total sum.   
\begin{figure}[ht]
\begin{center}
\[
\mbox{\begin{picture}(160,45)(0,0)
\includegraphics[scale=.65]{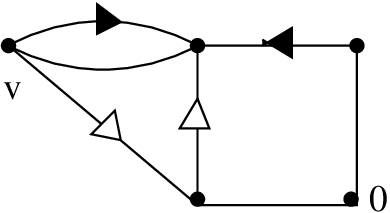}
\end{picture}}
\]
\caption{$F_{115}(v)$} 
\end{center}
\end{figure}

\begin{figure}[ht]
\begin{center}
\[
\mbox{\begin{picture}(440,80)(0,0)
\includegraphics[scale=.75]{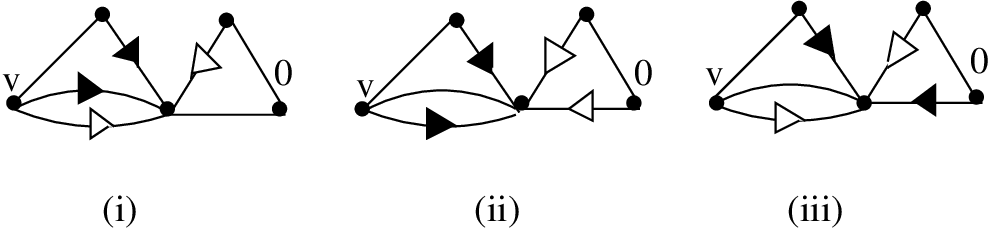}
\end{picture}}
\]
\caption{(i) $F_{116}(v)$, (ii) $F_{117}(v)$, (iii) $F_{118}(v)$} 
\end{center}
\end{figure}

Thus proceeding, the third term on the right hand side of \C{m1} yields
\bea &&\int_z\Big(G(v,z)^2 \p_\m G(v,z) G(z) \overline\p_\m G(z)+c.c.\Big)=-\frac{1}{3} D_5^{(2)}(v) \non \\ &&+\frac{1}{\pi}\Big(F_5(v)-F_{99}(v)-\frac{F_{112}(v)}{3}+c.c.\Big)+\frac{4}{\pi} F_{46}(v)\non \\ &&+\frac{2}{\pi^2}\Big(F_{116}(v)+F_{117}(v)+F_{118}(v)+c.c.\Big),\eea
where the graphs $F_{116}(v)$, $F_{117}(v)$ and $F_{118}(v)$ are given in figure 86.

To simplify $F_{116}(v)$ and $F_{117}(v)$, we use the auxiliary graphs $F_{119}(v)$ and $F_{120}(v)$ respectively, given in figure 87.
\begin{figure}[ht]
\begin{center}
\[
\mbox{\begin{picture}(335,110)(0,0)
\includegraphics[scale=.75]{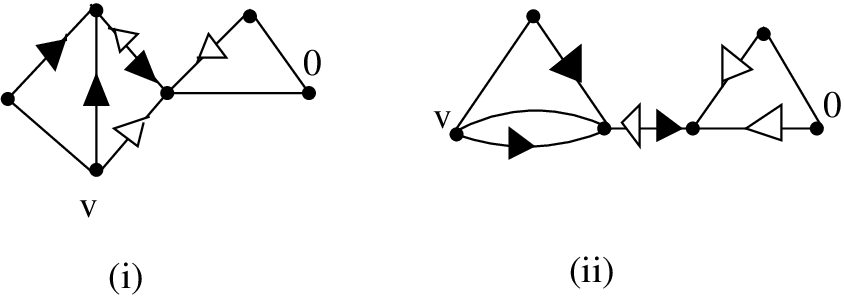}
\end{picture}}
\]
\caption{Auxiliary graphs (i) $F_{119}(v)$, (ii) $F_{120}(v)$} 
\end{center}
\end{figure}

For $F_{116}(v)$, we get that
\bea &&\Big(F_{116}(v)+c.c.\Big) -\Big(P_2 P_5(v)^*+c.c.\Big) = \pi^2 D_5^{(2)} (v) -\pi^2 E_2 D_3^{(1)} (v) \non \\ &&+\pi^2 E_2 E_3  -\pi^2 D_5^{(1,2,2)} (v) +2\pi^2 D^{(1,1,3)} (v) -\pi\Big(F_4(v) + F_{42}(v) \non \\ &&-\frac{F_{48}(v)}{\pi}-\frac{F_{51}(v)}{2}+F_{104}(v)+\frac{F_{121}(v)}{2\pi}+c.c.\Big),\eea
where the graph $F_{121}(v)$ is given in figure 88. To simplify it, we start with the auxiliary graph $F_{122}(v)$ given in figure 89 to get
\begin{figure}[ht]
\begin{center}
\[
\mbox{\begin{picture}(140,80)(0,0)
\includegraphics[scale=.7]{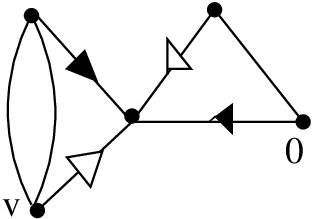}
\end{picture}}
\]
\caption{$F_{121}(v)$} 
\end{center}
\end{figure}
\begin{figure}[ht]
\begin{center}
\[
\mbox{\begin{picture}(140,70)(0,0)
\includegraphics[scale=.7]{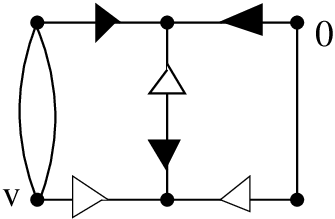}
\end{picture}}
\]
\caption{Auxiliary graph $F_{122}(v)$} 
\end{center}
\end{figure}
\bea &&\Big(F_{121}(v)+c.c.\Big)- \Big(P_3(v)P_8(v)^*+c.c.\Big) = 2\pi^2 D^{(1,1,3)}(v) -2\pi^2 D_5^{(1,2,1;1)}(v)\non \\&& + 2\pi^2 E_2 E_3 +2\pi F_{107}(v) -\pi\Big(F_{42}(v)-F_{68}(v)+F_{84}(v)-F_{89}(v)\non \\ &&+E_2 F_{90}(v)+F_{123}(v)+c.c.\Big)-2\tau_2\pi E_2 \p_v G_2(v) \overline\p_v G_2(v),\eea
where the graph $F_{123}(v)$ is given in figure 90. 
\begin{figure}[ht]
\begin{center}
\[
\mbox{\begin{picture}(100,75)(0,0)
\includegraphics[scale=.75]{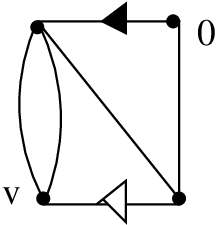}
\end{picture}}
\]
\caption{$F_{123}(v)$} 
\end{center}
\end{figure}

Next for $F_{117}(v)$, we get that
\bea &&\Big(F_{117}(v)+c.c.\Big)- \frac{1}{2}\Big(P_2 P_6^*+c.c.\Big) = \frac{\pi^2}{3} D_5^{(2)} (v)-\frac{\pi^2}{3} E_2 D_3 \non \\ &&-\frac{2\pi^2}{3} D^{(1,1,1,2)} (v)  -\pi^2 D_5^{(1,2,2)}(v) +\pi^2 E_2 E_3 +2\pi^2 D^{(1,1,3)} (v)\non \\&&-\pi\Big(F_4(v) -\frac{F_{51}(v)}{2} -\frac{F_{11}(v)}{\pi}+\frac{F_{124}(v)}{2\pi}+c.c.\Big), \eea
where the graph $F_{124}(v)$ is given in figure 91. 
\begin{figure}[ht]
\begin{center}
\[
\mbox{\begin{picture}(100,70)(0,0)
\includegraphics[scale=.75]{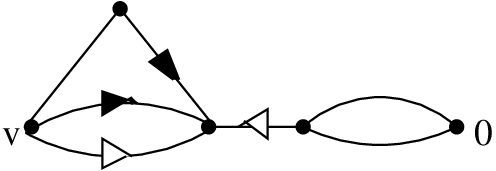}
\end{picture}}
\]
\caption{$F_{124}(v)$} 
\end{center}
\end{figure}
To simplify it, we start with the auxiliary graph $F_{125}(v)$ given in figure 92 to get
\begin{figure}[ht]
\begin{center}
\[
\mbox{\begin{picture}(100,70)(0,0)
\includegraphics[scale=.7]{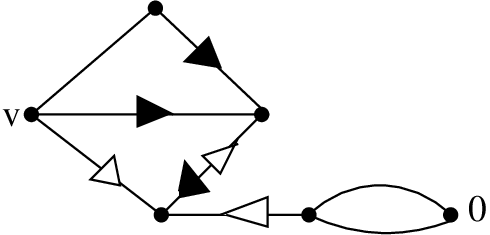}
\end{picture}}
\]
\caption{Auxiliary graph $F_{125}(v)$} 
\end{center}
\end{figure}
\bea \Big(F_{124}(v)+c.c.\Big)-\Big(P_2 P_8(v)^*+c.c.\Big)= \pi^2 D_5^{(2)}(v)-\pi^2 E_2 D_3^{(1)}(v) -\pi^2 E_2 D_3 \non \\ -\pi^2 D_5^{(2;2;1)}(v) +2\pi^2 E_2E_3+2\pi^2 D^{(1,1,3)}(v)-\pi\Big(F_{42}(v)+F_{104}(v)+c.c.\Big).\eea

We next simplify $F_{118}(v)$. It will prove very useful for our purposes to simply write
\be F_{118}(v) = \frac{F_{118}(v)}{2}+\frac{F_{118}(v)}{2},\ee
and evaluate the two equal contributions separately, by starting with the distinct auxiliary graphs $F_{126}(v)$ and $F_{127}(v)$ in figure 93.
\begin{figure}[ht]
\begin{center}
\[
\mbox{\begin{picture}(220,110)(0,0)
\includegraphics[scale=.7]{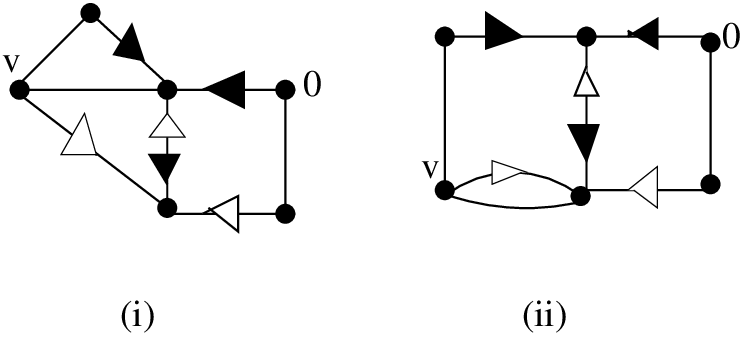}
\end{picture}}
\]
\caption{Auxiliary graphs (i) $F_{126}(v)$, (ii) $F_{127}(v)$} 
\end{center}
\end{figure}

Now $F_{126}(v)$ gives us
\bea &&\Big(F_{118}(v)+c.c.\Big) -\Big(P_3(v)P_5(v)^*+c.c.\Big) = 2\pi^2 D^{(1,1,3)}(v)-\pi\Big(F_4(v)+F_{42}(v)\non \\&&-\frac{F_{89}(v)}{2} -\frac{F_{48}(v)}{\pi}-\frac{F_{128}(v)}{\pi}+c.c.\Big)-\pi\tau_2 G(v)^2 \p_v G_2(v) \overline\p_v G_2(v),\eea
where the graph $F_{128}(v)$ is given in figure 94. 

\begin{figure}[ht]
\begin{center}
\[
\mbox{\begin{picture}(140,70)(0,0)
\includegraphics[scale=.6]{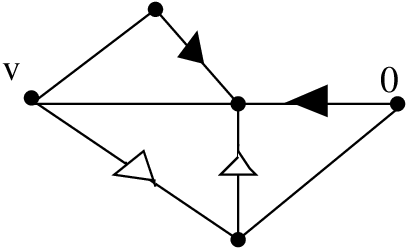}
\end{picture}}
\]
\caption{$F_{128}(v)$} 
\end{center}
\end{figure}

Also $F_{127}(v)$ yields
\bea &&\Big(F_{118}(v)+c.c.\Big)-\frac{1}{2} \Big(P_3(v)P_8(v)^*+c.c.\Big)=-\pi^2 D^{(1,1,2,1)}(v) \non \\ &&+\pi\Big(F_2(v) +\frac{F_{18}(v)}{2}+\frac{F_{30}(v)}{\pi}+\frac{F_{111}(v)}{\pi}+\frac{F_{129}(v)}{\pi}+c.c.\Big),\eea
where the graph $F_{129}(v)$ is given in figure 95. 
\begin{figure}[ht]
\begin{center}
\[
\mbox{\begin{picture}(150,70)(0,0)
\includegraphics[scale=.65]{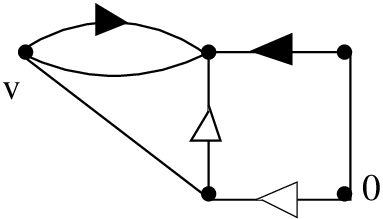}
\end{picture}}
\]
\caption{$F_{129}(v)$} 
\end{center}
\end{figure}

Importantly, the contributions of $F_{110} (v)$, $F_{115}(v)$, $F_{128}(v)$ and $F_{129}(v)$ to the right hand side of \C{m1} is equal to
\be \frac{3}{\pi^2}\Big(F_{110}(v)+F_{128}(v)+c.c.\Big) +\frac{3}{\pi^2}\Big(F_{115}(v)+F_{129}(v)+c.c.\Big).\ee
The combination $F_{110}(v)+F_{128}(v)+c.c.$ gives us
\bea F_{110}(v)+F_{128}(v)+c.c. &=& -\pi\Big(F_{65}(v)-\frac{F_{84}(v)}{2}\non \\ &&-F_{109}(v)+F_{123}(v)+\frac{F_{130}(v)}{\pi}+c.c.\Big),\eea
while the combination $F_{115}(v)+F_{129}(v)+c.c.$ gives us
\bea &&F_{115}(v)+F_{129}(v)+c.c.= \pi^2D_5^{(2)}(v) - \pi^2 G_2(v) D_3^{(1)}(v) \non \\&& -\pi\Big(F_{41}(v)+\frac{F_{109}(v)}{2}-\frac{F_{123}(v)}{2}+\frac{F_{131}(v)}{\pi}+c.c.\Big),\eea
where the graphs $F_{130}(v)$ and $F_{131}(v)$ are given in figure 96. 
\begin{figure}[ht]
\begin{center}
\[
\mbox{\begin{picture}(250,110)(0,0)
\includegraphics[scale=.75]{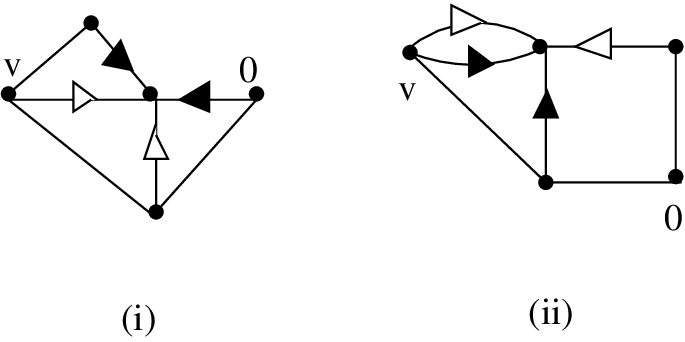}
\end{picture}}
\]
\caption{(i) $F_{130}(v)$ (ii) $F_{131}(v)$} 
\end{center}
\end{figure}
To simplify them, we introduce the auxiliary graphs $F_{132}(v)$ and $F_{133}(v)$ respectively, given in figure 97.
\begin{figure}[ht]
\begin{center}
\[
\mbox{\begin{picture}(250,120)(0,0)
\includegraphics[scale=.7]{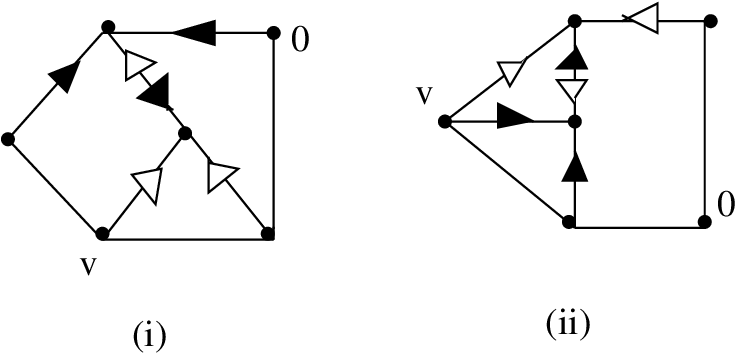}
\end{picture}}
\]
\caption{Auxiliary graphs (i) $F_{132}(v)$ (ii) $F_{133}(v)$} 
\end{center}
\end{figure}

For $F_{130}(v)+c.c.$, we get that
\bea &&\Big(F_{130}(v)+c.c.\Big)- \Big(P_3(v)\Big[P_5(v)+P_7(v)\Big]^*+c.c.\Big)=-\pi^2 D_5^{(1;2;2)}(v) +2\pi^2 D^{(1,3,1)} (v) \non \\ &&-\pi^2 D^{(1,1,2,1)} (v) +2\pi^2 E_3 G_2(v) +\pi F_{94}(v) -2\pi\tau_2 G_2(v) \p_v G_2(v)\overline\p_v G_2(v) -\pi\Big(F_{17}(v) \non \\ &&-\frac{F_{18}(v)}{2} +\frac{F_{84}(v)}{2} +G_2(v)F_{90}(v)-F_{108}(v)+\frac{F_{109}(v)}{2}-\frac{F_{134}(v)}{2}+c.c.\Big),\eea
where the graph $F_{134}(v)$ is given in figure 98.
\begin{figure}[ht]
\begin{center}
\[
\mbox{\begin{picture}(100,80)(0,0)
\includegraphics[scale=.7]{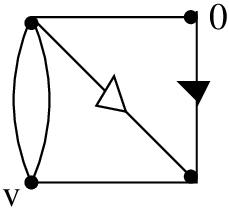}
\end{picture}}
\]
\caption{$F_{134}(v)$} 
\end{center}
\end{figure}

For $F_{131}(v) +c.c.$, we get that
\bea &&\Big(F_{131}(v)+c.c.\Big)- \Big(P_3(v)\Big[P_5(v)+P_7(v)\Big]^*+c.c.\Big) = 2\pi^2 D^{(1,3,1)} (v) \non \\ &&+2\pi^2 G_2(v)G_3(v)  -\pi^2 D_5^{(1;2;2)}(v) -\pi^2 D_5^{(1,2,2)} (v) +\pi^2 D_5^{(2)} (v) -3\pi^2 G_2(v)D_3^{(1)}(v)\non \\&& -\frac{\pi}{2}F_{94}(v) -\pi\Big(F_{17}(v) +\frac{F_{26}(v)}{2} -\frac{F_{42}(v)}{2} - G_2(v) F_{90}(v)+\frac{F_{89}(v)}{3}\non \\ &&-\frac{F_{123}(v)}{2}+\frac{F_{134}(v)}{2}+c.c.\Big).\eea

We could now simply add all the contributions and write down the eigenvalue equation. However this would not be quite useful for our purposes. This is because for this graph (as well as for most of the others) the eigenvalue equation involves several contributions whose links are given by derivatives of Green functions (this is even after simplifying using the several relations in appendices A and B). Our aim is to find suitable linear combinations of these graphs such that the eigenvalue equations contain only graphs whose links are given by Green functions, and not their derivatives. Eventually we would like to solve these equations and find algebraic relations between the graphs\footnote{This is what happened for the analysis involving the graphs with four links}. 

However for $D_5^{(2)}$ the eigenvalue equation involves some graphs which do not arise elsewhere. Hence to reach our final goal, we should be able to express them in terms of already existing graphs. This indeed is the case, and the analysis is given in appendix F.       

Thus, using the relation \C{imp} as well the results in appendices E and F, we get the eigenvalue equation
\bea \label{D52}&&\frac{1}{6}\Delta D_5^{(2)}(v)=\frac{1}{\pi^2}\Big[P_2^*\Big(2P_1(v) +P_6 +2P_5(v)\Big)+2P_3(v)^*\Big(2P_5(v)+2P_7(v)+P_8(v)\Big)\non \\ &&+c.c.\Big] +4 D_5^{(1,2,2)} (v) +2 D^{(1,1,2,1)}(v) +4 D_5^{(1,2,1;1)} (v) +10 D^{(1,1,3)} (v) \non \\ &&-20 D^{(1,2,2)} (v) -6 D^{(2,2,1)}(v) -12 D^{(1,3,1)} (v) +\frac{2}{3} D^{(1,1,1,2)} (v) + D_5^{(2;2;1)} (v) \non \\&&+2 D_5^{(1;2;2)} (v) +2 C_{1,1,3} +10E_5 +16 G_5(v) -6 E_2 G_3(v) -4 E_2E_3-3 E_2D_3^{(1)}(v) \non \\ &&+\frac{1}{3} E_2 D_3 -2 E_3 G_2(v) -8 G_2(v)G_3(v) -\frac{2\tau_2}{\pi}G_2(v) \p_v G_2(v) \overline\p_v G_2(v) \non \\&& +\frac{8\tau_2}{\pi}E_2 \p_v G_2(v)\overline\p_v G_2(v)-\frac{6\tau_2}{\pi} \p_v G_3(v)\overline\p_v G_3(v) -\frac{10\tau_2}{\pi}\Big(\p_v G_2(v)\overline\p_v G_4(v)+c.c.\Big)\non \\ &&-\frac{6\tau_2}{\pi}G(v)\Big(\p_v G_2(v)\overline\p_v G_3(v)+c.c.\Big) -\frac{3\tau_2}{\pi} G(v)^2\p_v G_2(v)\overline\p_v G_2(v) +\frac{1}{\pi}\Big(10 F_1(v)\non \\&&+14F_2(v)-6 F_4(v)+F_5(v)+3F_9(v) -F_{10}(v)+2F_{17}(v)+F_{23}(v) +3 F_{26}(v)\non \\
&&+F_{32}(v)+6 F_{35}(v)+3 F_{40}(v)-4 F_{41}(v)+2 F_{51}(v) - F_{65}(v)+3F_{84}(v) \non \eea
\bea &&- F_{89}(v)+c.c.\Big) -\frac{4}{\pi}F_3(v)-\frac{2}{\pi}F_7(v) -\frac{4}{\pi}F_{57}(v)+\frac{4}{\pi}F_{61}(v) -\frac{3}{2\pi}F_{94}(v).\eea

Based on the discussion above, we now analyze the various eigenvalue equations aiming to simplify them significantly beyond what we have done so far. 

\section{Eigenvalue equations and algebraic identities for elliptic modular graphs}

Let us briefly summarize an important feature of the eigenvalue equations that we have derived for the various modular graphs. The equations for the graphs $D^{(1,2,2)} (v)$ and $D^{(1,1,3)} (v)$ given by \C{D122} and \C{D113} respectively, are simple in the sense that they do not contain graphs whose links are given by derivatives of Green functions. No other eigenvalue equation for graphs with five links has this property. On the other hand, the equation for $D^{(2,2,1)}(v)$ in \C{D221} shows us that $\tau_2 \p_v G_3(v)\overline\p_v G_3(v)$ can be traded off for an expression involving no graphs with links given by derivatives of Green functions. Similar is the case for $\tau_2 \p_v G_2(v)\overline\p_v G_4(v)+c.c.$ using the equation for $D^{(1,3,1)} (v)$ in \C{D131}. Thus any simplification involving the other graphs which satisfy far more involved eigenvalue equations must involve several additional cancellations among each other. We do not have an algorithm to readily determine which linear combinations of graphs lead to such cancellations if any, so we shall proceed by guessing such combinations based on the eigenvalue equations.          

\subsection{Involving $D^{(1,1,1,2)}(v)$ and $D_5^{(2;2;1)}(v)$ among others}

Let us consider the eigenvalue equation for $D^{(1,1,1,2)} (v)$ given by \C{D1112}. We show in appendix C that the contribution involving $F_5(v)+c.c.$ can be expressed in terms of graphs without any derivatives on the links. Thus apart from the terms involving $F_7(v)$ and $P_1(v)P_2^*+c.c.$ on the right hand side of \C{D1112}, all the other terms can be expressed as graphs with no derivatives on the links.    

Next we consider the eigenvalue equation for $D_5^{(2;2;1)}$ in \C{D2;2;1}. Apart from the terms involving $\tau_2 \p_v G_3(v) \overline\p_v G_3(v)$ and $F_7(v)$, all the others can be expressed in terms of graphs without derivatives on the links. 

Thus it is natural to consider the combination
\be -\frac{1}{3} \Big(\Delta -6\Big) D^{(1,1,1,2)} (v)+\frac{1}{4} \Big(\Delta -2\Big)D_5^{(2;2;1)} (v)\ee
in which the contribution involving $F_7(v)$ cancels. Using the identities in appendices A and B, we observe that there is a striking simplification leading to
\bea &&\Big(\Delta -6\Big) \Big(-\frac{1}{3}D^{(1,1,1,2)}(v)+\frac{1}{4}D_5^{(2;2;1)}(v) + E_2 G_3(v)\Big) \non \\ &&= -\frac{4\tau_2}{\pi} \p_v G_3(v)\overline\p_v G_3(v) - 6 D^{(1,2,2)}(v) -3 D^{(2,2,1)} (v) \non \eea
\bea &&+ 8 E_2 G_3(v)  - 6 G_5(v) + 6 E_5 - 2G_2(v)G_3(v) + 2 E_3 G_2(v).\eea
Using the eigenvalue equations for $D^{(2,2,1)}(v)$ and $D^{(1,1,3)} (v)$, this yields the eigenvalue equation
\bea &&\Big(\Delta -6\Big) \Big(-\frac{1}{3}D^{(1,1,1,2)}(v)+\frac{1}{4}D_5^{(2;2;1)}(v)  -\frac{1}{2} D^{(2,2,1)}(v) + 2 D^{(1,1,3)}(v)+ E_2 G_3(v)\Big) \non \\ &&= 28 G_5(v)\eea
which does not have any graph with derivatives of the Green function as links. 
This immediately leads to
\bea \label{6}&&\Big(\Delta -6\Big)\Big(-\frac{1}{3}D^{(1,1,1,2)}(v)+\frac{1}{4}D_5^{(2;2;1)}(v)  -\frac{1}{2} D^{(2,2,1)}(v) \non \\ &&+ 2 D^{(1,1,3)}(v)+ E_2 G_3(v)-2 G_5(v)\Big) =0\eea
on using \C{Gs}. This equation is of the form 
\be \label{s}\Big(\Delta -s(s-1)\Big)\Phi_s(v)=0\ee
where $s=3$. We assume based on $SL(2,\mathbb{Z})$ invariance and the asymptotic properties for large $\tau_2$ that the solution is\footnote{In the remaining two subsections, we encounter \C{s} for $s=2$ and $s=0$. For $s=0$, we assume the answer is $\Phi_0(v)=$ constant. }
\be \label{sols}\Phi_s(v) = \a_s E_s +\b_s G_s(v)\ee
where $\a_s$ and $\b_s$ are constants. For modular graphs, the eigenfunction of $\Delta$ with eigenvalue $s(s-1)$ is $E_s$~\cite{Terras1}, and hence \C{sols} is a natural generalization to the elliptic case, as the choice of boundary conditions for large $\tau_2$ remains the same. It would be interesting to understand this issue in detail.

Thus proceeding \C{6} yields the algebraic identity
\bea \label{ID1} &&-\frac{1}{3}D^{(1,1,1,2)}(v)+\frac{1}{4}D_5^{(2;2;1)}(v)  -\frac{1}{2} D^{(2,2,1)}(v) + 2 D^{(1,1,3)}(v)+ E_2 G_3(v)-2 G_5(v) \non \\ &&= \a E_3 +\b G_3(v)\eea
between the various graphs. We now solve for $\a$ and $\b$.

On integrating over an unintegrated vertex, the left hand side vanishes, and we get $0=\a E_3$, hence $\a=0$.
On identifying $0$ and $v$, we get that
\bea \b E_3= -\frac{1}{3} \Big(D_{1,1,3}- 3 E_2E_3\Big) +\frac{D_{1,2,2}}{4} - \frac{C_{1,2,2}}{2} + 2 C_{1,1,3} - 2 E_5. \eea
Using the identities \C{a5}, the right hand side vanishes and hence $\b=0$. This algebraic identity is given in figure 99.

\begin{figure}[ht]
\begin{center}
\[
\mbox{\begin{picture}(440,100)(0,0)
\includegraphics[scale=.75]{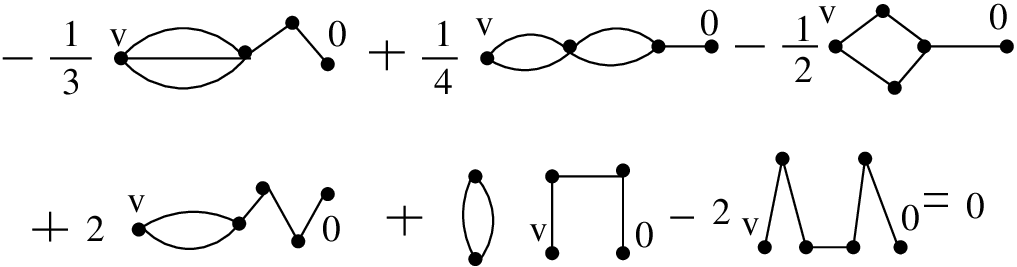}
\end{picture}}
\]
\caption{Algebraic identity \C{ID1}} 
\end{center}
\end{figure}

\subsection{Involving $D^{(1,1,2,1)}(v)$ and $D_5^{(1,2,2)}(v)$ among others}

We next start with the combination
\be \label{c1}\frac{1}{2} \Big(\Delta -2\Big)\Big(D^{(1,1,2,1)}(v)- D_5^{(1,2,2)}(v)\Big)\ee
where several terms involving graphs with derivatives as their links cancel. Among other terms, \C{c1} has the terms 
\be \frac{\tau_2}{\pi}E_2 \p_v G_2(v)\overline\p_v G_2(v)-\frac{\tau_2}{\pi} G_2 (v)\p_v G_2(v) \overline\p_v G_2(v)= \frac{\tau_2}{\pi} {\mathcal{F}}_2(v)\p_v G_2(v)\overline\p_v G_2(v)\ee
left in it, which can be cancelled by adding 
\be \label{c2} -\frac{1}{4} \Big(\Delta-2\Big) D_5^{(1;2;2)} (v)\ee 
to \C{c1}. The remaining terms include 
\be \frac{1}{\pi}\Big(F_{17}(v)+c.c.\Big)\ee
among others, which can be cancelled by further adding 
\be \label{c3}\frac{1}{2}\Big(\Delta -2\Big)D_5^{(1,2,1;1)}(v)\ee
to \C{c1} and \C{c2}. Among several other terms,
\be \frac{2}{\pi} F_7(v)\ee
remains, which can be cancelled by adding 
\be \label{c4}\frac{1}{3} \Big(\Delta -6\Big) D^{(1,1,1,2)}\ee
to \C{c1}, \C{c2} and \C{c3} as well. 

The remaining expression can be simplified using appendices A, B, C and D. We also use the algebraic identity \C{ID1} for $D_5^{(2;2;1)} (v)$ in the process. In fact, an intermediate step in the analysis yields
\bea &&\frac{1}{2} \Big(\Delta -2\Big) \Big(D^{(1,1,2,1)} (v) - D_5^{(1,2,2)} (v)  -\frac{1}{2}D_5^{(1;2;2)}(v) +D_5^{(1,2,1;1)} (v) +\frac{2}{3} D^{(1,1,1,2)}(v) \Big) \non \\&& -\frac{1}{\pi^2}\Big(9 P_1(v) P_2^* -6 P_1(v) P_3(v)^*- 3 P_2 P_4^*+6 P_3(v)P_4^* +c.c.\Big)\non \\ &&= 5 D^{(1,2,2)} (v) +12 D^{(1,1,3)} (v) -\frac{9}{2} D^{(2,2,1)}(v)+\frac{1}{2}C_{1,2,2} + 11 E_5 - 3 G_5(v) \non \\ &&-3 E_2 G_3(v) + G_2(v)G_3(v) - E_3 G_2(v) - 3 E_2 E_3 -\frac{2\tau_2}{\pi} \p_v G_3(v) \overline\p_v G_3(v).\eea

Finally, we use the eigenvalue equations for $D^{(2,2,1)}(v)$, $D^{(1,1,3)}(v)$ and $D^{(1,2,2)} (v)$. The resulting eigenvalue equation simplifies considerably and is given by
\bea  &&\Big(\Delta -2\Big) \Big(\frac{1}{2}D^{(1,1,2,1)} (v) - \frac{1}{2}D_5^{(1,2,2)} (v)  -\frac{1}{4}D_5^{(1;2;2)}(v) +\frac{1}{2}D_5^{(1,2,1;1)} (v) \non \\&& +\frac{1}{3} D^{(1,1,1,2)}(v) -\frac{1}{4} D^{(2,2,1)}(v)-3 D^{(1,1,3)}(v)- 4D^{(1,2,2)}(v) +\frac{1}{2} E_2E_3 \non \\ &&-\frac{3}{2} E_2G_3(v) -G_2(v)E_3 + G_2(v) G_3(v)\Big) = 16 E_5 -90 G_5(v) +\frac{1}{2}C_{1,2,2}.\eea

Thus easily leads to
\bea \label{2} &&\Big(\Delta -2\Big) \Big(\frac{1}{2}D^{(1,1,2,1)} (v) - \frac{1}{2}D_5^{(1,2,2)} (v)  -\frac{1}{4}D_5^{(1;2;2)}(v) +\frac{1}{2}D_5^{(1,2,1;1)} (v) \non \\&& +\frac{1}{3} D^{(1,1,1,2)}(v) -\frac{1}{4} D^{(2,2,1)}(v)-3 D^{(1,1,3)}(v)- 4D^{(1,2,2)}(v) +\frac{1}{2} E_2E_3 \non \\ &&-\frac{3}{2} E_2G_3(v) -G_2(v)E_3 + G_2(v) G_3(v) + 5 G_5(v) - \frac{9}{10} E_5 +\frac{\zeta(5)}{120}\Big) = 0,\eea
where we have used the expression for $C_{1,2,2}$ in \C{a5} as well as \C{Eisen} and \C{Gs}. Hence based on \C{s} and \C{sols}, we obtain the non--trivial algebraic identity
\bea  \label{ID2}&&\frac{1}{2}D^{(1,1,2,1)} (v) - \frac{1}{2}D_5^{(1,2,2)} (v)  -\frac{1}{4}D_5^{(1;2;2)}(v) +\frac{1}{2}D_5^{(1,2,1;1)} (v)  +\frac{1}{3} D^{(1,1,1,2)}(v) \non \\ &&-\frac{1}{4} D^{(2,2,1)}(v)-3 D^{(1,1,3)}(v)- 4D^{(1,2,2)}(v) +\frac{1}{2} E_2E_3 -\frac{3}{2} E_2G_3(v) \non \\ &&-G_2(v)E_3 + G_2(v) G_3(v) + 5 G_5(v) - \frac{9}{10} E_5 +\frac{\zeta(5)}{120}=\a E_2 +\b  G_2(v)\eea
where we now determine $\a$ and $\b$. 

To determine $\a$, on integrating over the unintegrated vertex, we get that
\be -4 \a E_2 = C_{1,2,2} -\frac{2}{5} E_5 -\frac{\zeta(5)}{30},\ee
and hence $\a=0$ using \C{a5}. 
To determine $\b$, we identify the vertices $0$ and $v$, leading to
\bea \b E_2 = \frac{1}{3}\Big(D_{1,1,3} - 3 E_2 E_3 \Big)+\frac{1}{4} D_{1,2,2} - 3 C_{1,1,3} -\frac{17}{4} C_{1,2,2} +\frac{41}{10} E_5  +\frac{\zeta(5)}{120}.\eea
Now the right hand side vanishes using the identities in \C{a5}, and hence $\b =0$. The identity \C{ID2} is given in figure 100.

\begin{figure}[ht]
\begin{center}
\[
\mbox{\begin{picture}(435,300)(0,0)
\includegraphics[scale=.75]{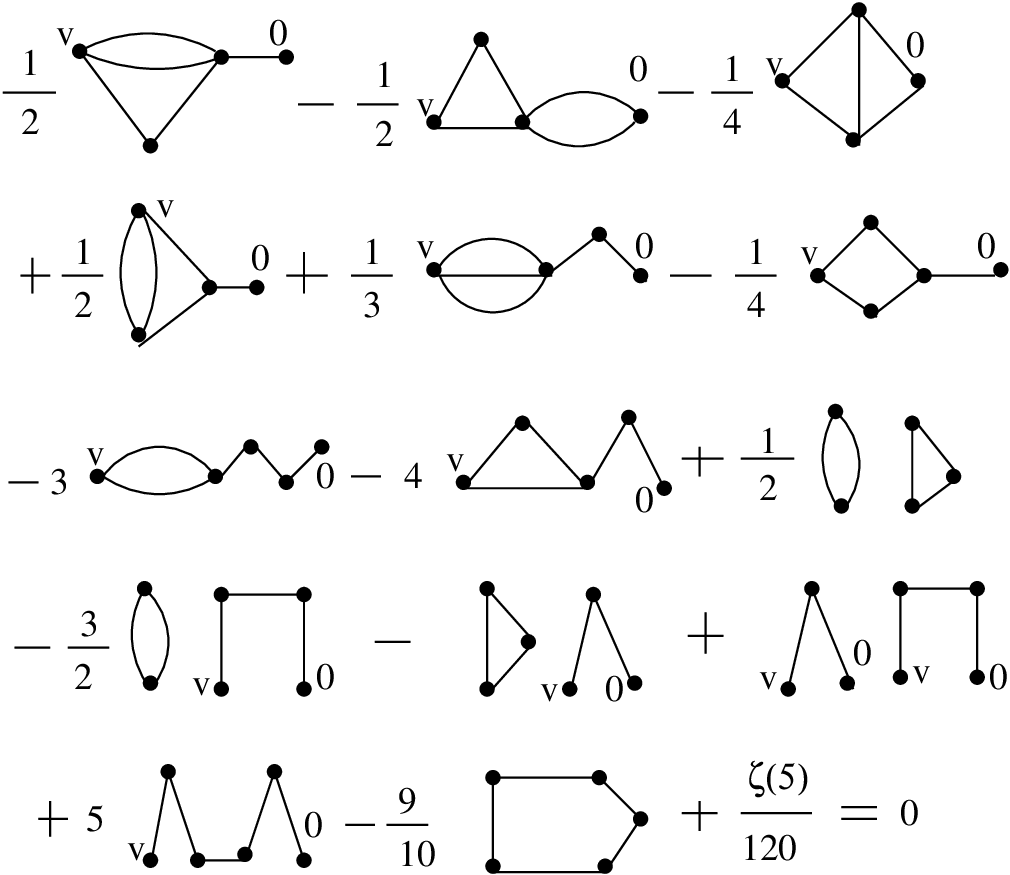}
\end{picture}}
\]
\caption{Algebraic identity \C{ID2}} 
\end{center}
\end{figure}

\subsection{Involving $D_5^{(1)}(v)$ and $D_5^{(2)}(v)$ among others}

Finally, we consider the eigenvalue equation
\be \label{x1}\Delta \Big(D_5^{(2)}(v)-\frac{3}{2} D_5^{(1)}(v)\Big).\ee
Among other contributions, it contains
\bea -\frac{48}{\pi} F_7(v) +\frac{48}{\pi}\Big(F_{17}(v)+c.c.\Big),\eea
which can be cancelled by adding
\be \label{x2} 24 \Delta D_5^{(1,2,1;1)}(v)\ee
to \C{x1}. 

Using the results in appendices A, B, C and D, as well as the identities \C{ID1} and \C{ID2}, we get that
\bea \label{big1}&&\Delta \Big(D_5^{(2)}(v)-\frac{3}{2} D_5^{(1)}(v)+24 D_5^{(1,2,1;1)}(v)\Big)= \frac{6}{\pi^2}\Big(P_2 P_6^*-6 P_3(v)P_6^*+c.c.\Big)\non \\&&-\frac{12}{\pi^2}\Big[\Big(P_2-P_3(v)\Big)\Big(2P_5(v)^*+2P_7(v)^*+P_8(v)^*\Big)+c.c.\Big]\non \\&&
-\frac{72\tau_2}{\pi}\Big(\p_v G_2(v) \overline\p_v G_4(v)+c.c.\Big) +\frac{12\tau_2}{\pi} \Big(E_2-G_2(v)\Big) \p_v G_2(v) \overline\p_v G_2(v) \non \\ &&+72 D^{(2,2,1)} (v) +144 D^{(1,3,1)} (v)+36 C_{1,1,3} -48 C_{1,2,2} -12 D_{1,1,3} + 9D_{1,2,2}\non \\ &&+\frac{384}{5} E_5 +432 G_5(v)-24 E_3 G_2(v)-24 E_2G_3(v) -12 D_3 G_2(v) \non \\ &&-72 G_2(v)G_3(v) -12 \Big(E_2 -G_2(v)\Big)D_3^{(1)}(v)+54 E_2E_3+2E_2D_3+\frac{2}{5}\zeta(5).\eea

On further using the relations in \C{deflap}, we see that \C{big1} gives us
\bea &&\Delta \Big(D_5^{(2)}(v)-\frac{3}{2} D_5^{(1)}(v)+24 D_5^{(1,2,1;1)}(v)-E_2D_3 + 6 D_3 G_2(v) -6 G_2(v)D_3^{(1)}(v)\non \\ &&+6 E_2 D_3^{(1)}(v)\Big) = 72  D^{(2,2,1)}(v)+144 D^{(1,3,1)}(v) +36C_{1,1,3} -48 C_{1,2,2} -12 D_{1,1,3}\non \\&&+9D_{1,2,2}+432G_5(v)+\frac{384}{5}E_5-96 G_2(v)G_3(v)+60 E_2E_3+\frac{2}{5}\zeta(5)\non \\ &&-\frac{72\tau_2}{\pi}\Big(\p_v G_2(v) \overline\p_v G_4(v)+c.c.\Big).\eea

Finally, on using the eigenvalue equations for the graphs $D^{(1,2,2)}(v)$ and $D^{(1,3,1)}(v)$ in \C{D122} and \C{D131} respectively, as well the relations \C{e5} and \C{a5}, we get the eigenvalue equation
\bea &&\Delta \Big(D_5^{(2)}(v)-\frac{3}{2} D_5^{(1)}(v)+24 D_5^{(1,2,1;1)}(v)-24 D^{(1,3,1)}(v)+24 D^{(1,2,2)}(v)\non \\ &&-\frac{3}{2}D_{1,2,2} -6 C_{1,1,3}+2D_{1,1,3}-E_2D_3 -6E_2E_3+ 6 D_3 G_2(v) \non \\ &&-6 G_2(v)D_3^{(1)}(v) +6 E_2 D_3^{(1)}(v)\Big) =480  G_5(v)-264 E_5\eea
which has simplified enormously. Now using \C{Eisen} and \C{Gs}, this easily yields   
\bea \label{Eg} &&\Delta \Big(D_5^{(2)}(v)-\frac{3}{2} D_5^{(1)}(v)+24 D_5^{(1,2,1;1)}(v)-24 D^{(1,3,1)}(v)+24 D^{(1,2,2)}(v)\non \\ &&-24 G_5(v)-\frac{3}{2}D_{1,2,2} -6 C_{1,1,3}+2D_{1,1,3}+\frac{66}{5}E_5 -E_2D_3 -6E_2E_3\non \\ &&+ 6 D_3 G_2(v)-6 G_2(v)D_3^{(1)}(v) +6 E_2 D_3^{(1)}(v)\Big)=0.\eea 
From \C{s} and \C{sols}, we see that \C{Eg} immediately gives us the algebraic identity
\bea \label{L}&&D_5^{(2)}(v)-\frac{3}{2} D_5^{(1)}(v)+24 D_5^{(1,2,1;1)}(v)-24 D^{(1,3,1)}(v)+24 D^{(1,2,2)}(v)\non \\ &&-24 G_5(v)-\frac{3}{2}D_{1,2,2} -6 C_{1,1,3}+2D_{1,1,3}+\frac{66}{5}E_5 -E_2D_3 -6E_2E_3\non \\ &&+ 6 D_3 G_2(v)-6 G_2(v)D_3^{(1)}(v) +6 E_2 D_3^{(1)}(v)=c,\eea
where $c$ is a constant. To determine $c$, we integrate over an unintegrated vertex to get that
\be 2D_{1,1,3} -\frac{3}{2}D_{1,2,2} -12 C_{1,1,3} +\frac{66}{5}E_5 -6 E_2E_3=c,\ee
which using the relations in \C{a5}, yields
\be \label{C}c=-\frac{\zeta(5)}{10}.\ee
As a non--trivial consistency check, on identifying the vertices $v$ and $0$, \C{L}
gives us
\bea -\frac{1}{2} \Big(D_5-10 E_2 D_3\Big) +2\Big(D_{1,1,3}-3 E_2E_3\Big) +\frac{45}{2}D_{1,2,2} -30 C_{1,1,3} +24 C_{1,2,2} -\frac{54}{5} E_5=c.\non \\\eea
Again on using \C{a5}, this precisely reproduces \C{C}. The identity \C{L} is given in figure 101. 

\begin{figure}[ht]
\begin{center}
\[
\mbox{\begin{picture}(450,300)(0,0)
\includegraphics[scale=.7]{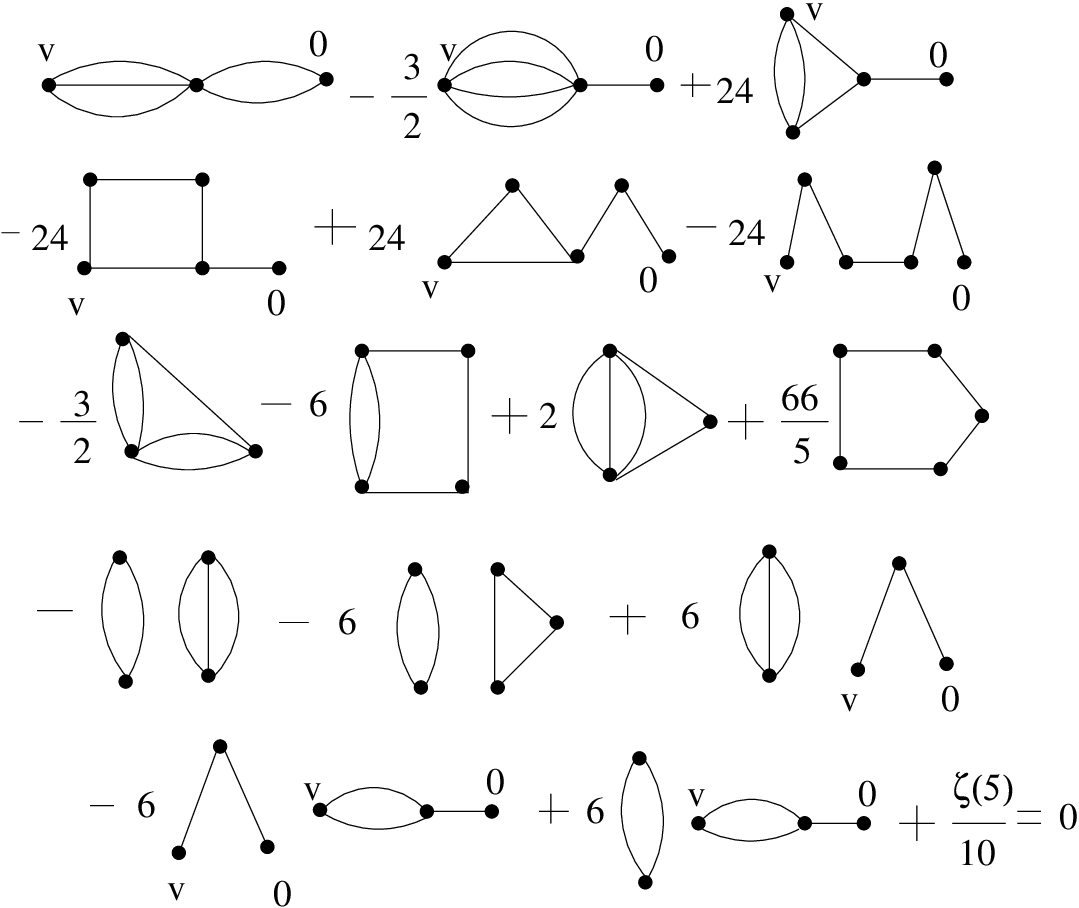}
\end{picture}}
\]
\caption{Algebraic identity \C{L}} 
\end{center}
\end{figure}

Thus we have obtained the eigenvalue equations \C{D122}, \C{D113}, \C{6}, \C{2} and \C{Eg} involving elliptic modular graphs with five links. These equations involve graphs whose links are given by the Green function. We have also obtained the algebraic identities \C{ID1}, \C{ID2} and \C{L} between the graphs. There is a notion of transcendentality which is preserved by these equations, where we assign transcendentality 1 to each link of the graph, and $s$ to $\zeta(s)$. This is not unexpected for elliptic modular graphs, since the plethora of known identities involving modular graphs all satisfy this property. In fact, one can define transcendentality of a graph by looking at the power behaved terms in the large $\tau_2$ expansion\footnote{This notion is not preserved for terms that are exponentially suppressed in this expansion.} and assigning transcendentality 1 to each factor of $\pi\tau_2$. Then from various examples (see~\cite{DHoker:2019blr} for modular graphs, and~\cite{DHoker:2018mys,Hidding:2022vjf} for some of their elliptic counterparts), one notes that transcendentality is indeed preserved.           
  
Given the non--trivial algebraic identities involving several elliptic modular graphs that we have systematically deduced, it is natural to expect this pattern to generalize to graphs with more links. Such identities should be obtainable by solving eigenvalue equations with eigenvalues of the form $s(s-1)$ for $s \geq 0$. A natural starting point is to consider the modular graphs with six or more links that arise in the low momentum expansion of the type II four graviton amplitude and cut them open in all possible ways as we have done. This structure should further generalize to include graphs where the links are also given by derivatives of the Green function. A natural starting point would be to consider higher point amplitudes in the type II theory and cut open the graphs. 

It would also be interesting to deduce these equations from equations involving genus two graphs on expanding them around the non--separating node in moduli space, and trying to understand the relation of these genus two graphs with string amplitudes. Finally, generalizing this analysis to $SL(2,\mathbb{Z})$ covariant elliptic modular graph forms is also expected to yield a rich structure, where the natural starting point would be to consider amplitudes in heterotic string theory.              

Finally, one can also analyze the action of the modular invariant expression 
\be 4\tau_2 \frac{\p^2}{\p v \p \overline{v}}\ee
on the graphs, which should lead to various interesting relations among them. Some such relations involving elliptic modular graphs with links also given by the derivatives of the Green function have been obtained in~\cite{Basu:2020iok}. 

\appendix

\section{Various identities}

In this appendix, we list various identities involving the graphs $F_i(v)$ that are used to simplify various expressions in the main text. We list only those identities that are relevant in obtaining the final eigenvalue equations having graphs with links given only by the Green functions, and not their derivatives. Thus several identities involving $F_i(v)$ which arise in the eigenvalue equation for each modular graph are irrelevant in the final sum, and so we do not list them. However, they can also be easily obtained like the ones listed below.   

\begin{figure}[ht]
\begin{center}
\[
\mbox{\begin{picture}(270,120)(0,0)
\includegraphics[scale=.65]{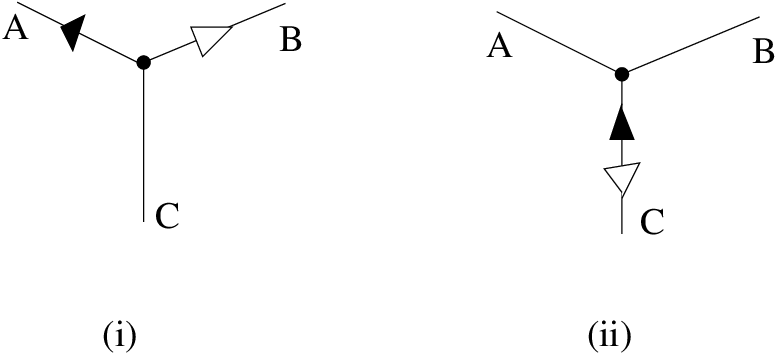}
\end{picture}}
\]
\caption{The simplification} 
\end{center}
\end{figure}

Now there are several graphs $F_i(v)$ that do not admit any simplification, while others can be simplified only with some effort, as discussed in the main text and the later appendices. However, many can be simplified easily, and are the ones given below. These are graphs which have one $\p G$ and one $\overline\p G$ along the links, which meet at a cubic vertex in the graph as given in figure 102(i)\footnote{To this we add the complex conjugate if the graph is not real.}. To express this in terms of graphs with all derivatives removed along the links, we start with the graph in figure 102(ii) where the two derivatives along the same link can be removed using \C{maineqn}. Alternatively this can be evaluated on integrating by parts, which reproduces the graph in figure 102(i) (plus complex conjugate if it is not real), as well as other two other graphs without any derivatives along the links, again on using \C{maineqn}. We now give the relations for the relevant graphs\footnote{The analysis for $F_{51}(v)+c.c.$ also proceeds as above, by moving the link on the right to the center, along the lines of what has been done for several graphs in the main text.}. 
\bea \frac{1}{\pi} \Big(F_1 (v) +c.c.\Big) &=& D^{(1,3,1)} (v) + D^{(2,2,1)} (v) + E_5 - G_2 (v) G_3 (v), \non \\ \frac{1}{\pi} \Big(F_2 (v) +c.c.\Big) &=& D^{(1,3,1)} (v) + D^{(1,2,2)} (v) + G_5 (v) - E_3 G_2 (v), \non \\  \frac{F_3(v)}{\pi}   &=& D^{(1,2,2)} (v) - \frac{1}{2} D^{(2,2,1)} (v), \non \\ \frac{1}{\pi} \Big(F_4 (v) +c.c.\Big) &=& D^{(1,1,3)} (v) - D^{(1,2,2)} (v) - G_5 (v) + E_2 G_3 (v), \non \\
\frac{1}{\pi}\Big(F_9 (v) + c.c.\Big) &=& D_5^{(1,2,1;1)} (v) + D^{(1,1,2,1)} (v) - E_2 G_3 (v) + C_{1,1,3} - G_2 (v) D_3^{(1)} (v),\non \\ \frac{1}{\pi} \Big(F_{10} (v)+c.c.\Big) &=& D_5^{(1,2,1;1)}(v) + D^{(1,1,1,2)} (v) + D^{(1,1,3)} (v)-E_2 G_3 (v) - D_3 G_2 (v),  \non \\  
\frac{1}{\pi} \Big(F_{41}(v)+c.c.\Big)&=& D_5^{(1;2;2)} (v) + D_5^{(1,2,2)} (v) + D^{(1,3,1)} (v) - G_2(v) G_3(v) - G_2(v) D_3^{(1)}(v), \non \\
\frac{1}{\pi}\Big(F_{51}(v)+c.c.\Big)&=& D_5^{(1,2,2)}(v)-D_5^{(2;2;1)} (v) - D^{(1,1,3)} (v) -E_2E_3+ E_2 D_3^{(1)}(v), \non \\  \frac{F_{57}(v)}{\pi}&=& D_5^{(1,2,1;1)} (v) - D^{(1,2,2)}(v)- \frac{1}{2} D_5^{(1;2;2)}(v)+\frac{1}{2} C_{1,2,2} , \non \\ \frac{1}{\pi} \Big(F_{63}(v)+c.c.\Big)&=& D_5^{(2;2;1)} (v)- D^{(2,2,1)} (v),\non \\
\frac{1}{\pi}\Big(F_{65}(v)+c.c.\Big)&=&D^{(1,3,1)} (v)+D^{(1,1,2,1)} (v) + D_5^{(1,2,1;1)} (v) - E_3 G_2(v) - E_2 D_3^{(1)} (v), \non \\ \frac{F_{79}(v)}{\pi} &=&\frac{1}{2} D_5^{(1;2;2)}(v).\eea

\section{More graphs and some identities}

The graphs $Q_1$, $Q_2$ and $Q_3$ in the main text, are given in figure 103.  
\begin{figure}[ht]
\begin{center}
\[
\mbox{\begin{picture}(280,100)(0,0)
\includegraphics[scale=.7]{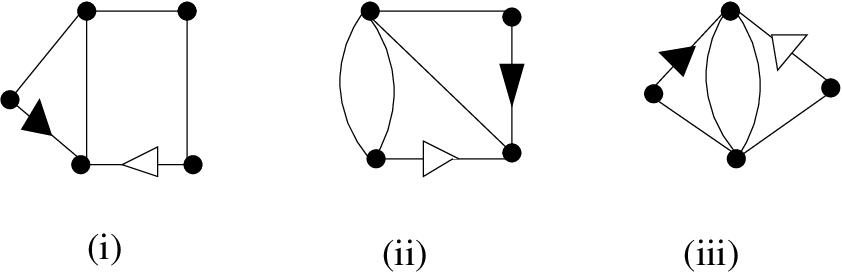}
\end{picture}}
\]
\caption{(i) $Q_1$, (ii) $Q_2$, (iii) $Q_{3}$} 
\end{center}
\end{figure}

They satisfy the equations~\cite{Basu:2016mmk}
\bea \frac{1}{\pi}\Big(Q_1 +c.c.\Big) &=& C_{1,1,3} + C_{1,2,2} + E_5 - E_2 E_3, \non \\ \frac{1}{\pi}\Big(Q_2 +c.c.\Big) &=& D_{1,1,3} + D_{1,2,2} + C_{1,1,3} - E_2 E_3 - E_2 D_3 , \non \\ \frac{Q_3}{\pi} &=& 4C_{1,2,2} +3C_{1,1,3} + D_{1,1,3} -2E_2 E_3 - E_2 D_3.\eea

Also the graphs $P_1(v), \ldots, P_8(v)$ in the main text are given in figure 104. 

\begin{figure}[ht]
\begin{center}
\[
\mbox{\begin{picture}(370,175)(0,0)
\includegraphics[scale=.75]{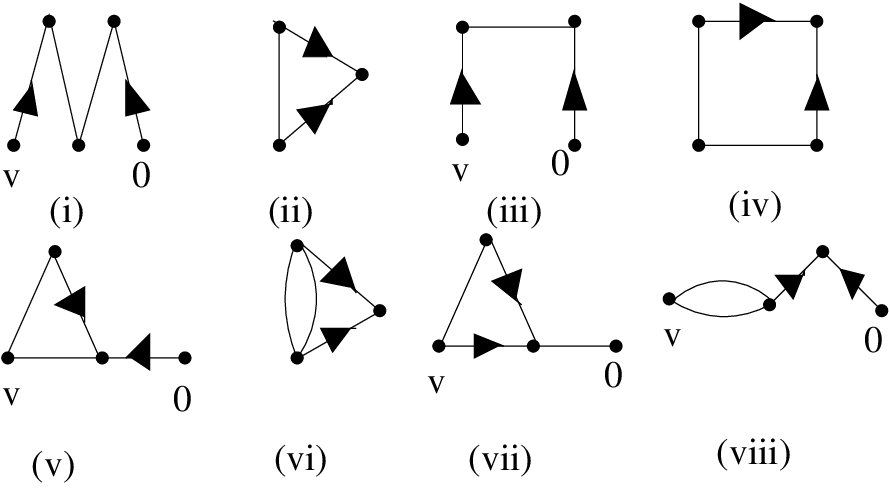}
\end{picture}}
\]
\caption{(i) $P_1 (v)$, (ii) $P_2$, (iii) $P_3 (v)$, (iv) $P_4$, (v) $P_5 (v)$, (vi) $P_6$, (vii) $P_7 (v)$, (viii) $P_8 (v)$} 
\end{center}
\end{figure}

We have that $P_4 =P_6$ starting from $D_3 = E_3 +\zeta(3)$ and acting with $\p_\m$.

\begin{figure}[ht]
\begin{center}
\[
\mbox{\begin{picture}(140,40)(0,0)
\includegraphics[scale=.7]{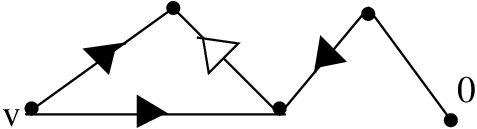}
\end{picture}}
\]
\caption{Auxiliary graph $P(v)$} 
\end{center}
\end{figure}

Also we find the identity
\be \label{imp} 2P_1(v) -2 P_7(v) -P_8(v)=0\ee
to be very useful, which follows starting from the auxiliary graph $P(v)$ in figure 105.

Finally, we list the eigenvalue equations satisfied by several products of two graphs.   
\bea \label{deflap}&&\Big(\Delta-8\Big) E_2 E_3 = \frac{6}{\pi^2}\Big(P_2 P_4^*+c.c.\Big), \non \\&& \Big(\Delta-8\Big) E_2 G_3(v) = \frac{6}{\pi^2}\Big(P_1(v)P_2^*+c.c.\Big),\non \\ &&\Big(\Delta -8\Big)G_2 (v)E_3 = \frac{6}{\pi^2}\Big(P_3(v)P_4^*+c.c.\Big), \non \\ &&\Big(\Delta -8\Big)G_2 (v)G_3(v) = \frac{6}{\pi^2}\Big(P_1(v)P_3(v)^*+c.c.\Big), \non \\&&\Big(\Delta -2\Big)E_2 D_3 = 6 E_2E_3 +\frac{6}{\pi^2}\Big(P_2P_6^*+c.c.\Big), \non \eea

\bea&&\Big(\Delta -2\Big)G_2 (v)D_3 = 6 G_2(v)E_3 +\frac{6}{\pi^2}\Big(P_3(v)P_6^*+c.c.\Big),\non \\ &&\Big(\Delta -2\Big) \Big(G_2(v)-E_2\Big)D_3^{(1)}(v)=2\Big(E_3+2 G_3(v)-\frac{\tau_2}{\pi}\p_v G_2(v) \overline\p_v G_2(v)\Big)\non \\ &&\times\Big(G_2(v)-E_2\Big)-\frac{2}{\pi^2}\Big[\Big(P_2-P_3(v)\Big)\Big(2 P_5(v)^*+2P_7(v)^*+P_8(v)^*\Big)+c.c.\Big],\eea
where we have used \C{Eisen}, \C{D3}, \C{Gs} and \C{D31}. Thus they are helpful in expressing a product of two graphs, one of which has two factors of $\p G$ and the other has two factors of $\overline\p G$ along their links, in terms of the action of the Laplacian on graphs which do not have derivatives of Green functions as their links.

\section{Simplifying $F_5(v) +c.c.$}

To simplify the expression involving $F_5(v)+c.c.$, we consider the auxiliary graph $A_1(v)$ given in figure 106, which leads to\footnote{This analysis is similar to the one in \cite{Basu:2019idd}.} 
\begin{figure}[ht]
\begin{center}
\[
\mbox{\begin{picture}(100,100)(0,0)
\includegraphics[scale=.65]{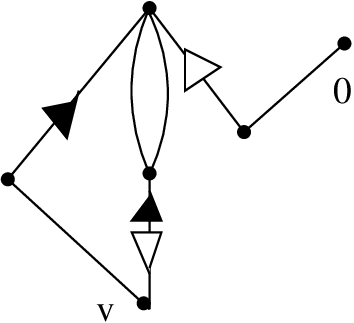}
\end{picture}}
\]
\caption{Auxiliary graph $A_{1}(v)$} 
\end{center}
\end{figure}
\bea F_5(v)= \pi E_2 G_3(v) - \pi D^{(2,2,1)} (v) - \frac{2}{\pi} A_2(v).\eea
Now the graph $A_2(v)$ is given in figure 107 which we simplify using the auxiliary graph $A_3(v)$ given in the same figure, to obtain 
\begin{figure}[ht]
\begin{center}
\[
\mbox{\begin{picture}(250,130)(0,0)
\includegraphics[scale=.7]{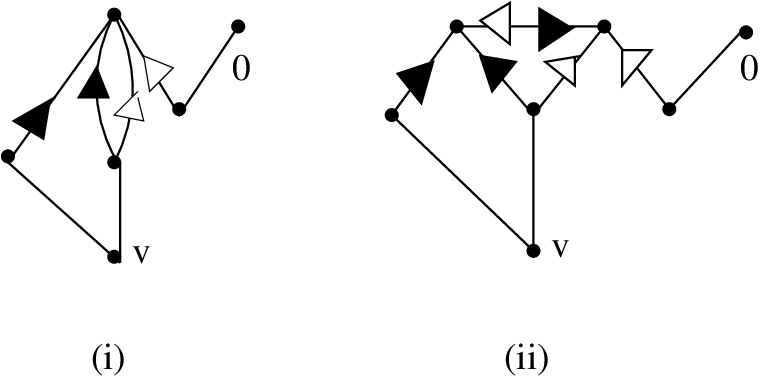}
\end{picture}}
\]
\caption{(i) $A_2(v)$ and (ii) auxiliary graph $A_{3}(v)$} 
\end{center}
\end{figure}
\bea \label{F5}F_5(v)+c.c.&=& -2 D^{(2,2,1)} (v) -8 G_5(v) + 5 E_2 G_3(v) +9 D^{(1,1,3)} (v) -2E_2E_3\non \\ &&+ D_5^{(1,2,1;1)}(v) - D_3 G_2(v) + D^{(1,1,1,2)} (v) + 2 D_5^{(1,2,2)} (v) \non \\&& +\frac{1}{\pi}\Big(2 F_1(v) -6 F_4(v) - F_{51}(v) - 2 F_{63}(v)+c.c.\Big).\eea

In an intermediate stage of the analysis, we need to calculate $A_4(v)+c.c.$, which we perform using
\be 2\Big(A_4(v)+c.c.\Big) = \Big(A_5(v)+c.c.\Big)- 2\pi D^{(1,1,1,2)}(v) +2\pi E_2G_3(v),\ee
where the graphs $A_4(v)$ and $A_5(v)$ are given in figure 108.

\begin{figure}[ht]
\begin{center}
\[
\mbox{\begin{picture}(230,115)(0,0)
\includegraphics[scale=.8]{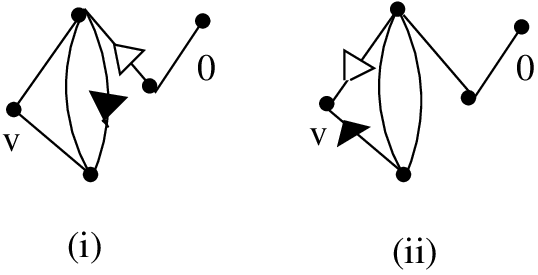}
\end{picture}}
\]
\caption{(i) $A_4(v)$, (ii)$A_5(v)$} 
\end{center}
\end{figure}

To analyze $A_5(v)+c.c.$, we consider the action of $\tau_2 \p_v\overline\p_v$ on the graph $A_6(v)$ in figure 109. While on interchanging the labels in $A_6(v)$ this leads to $-\pi D_5^{(1,2,1;1)}(v)$, we can evaluate it differently keeping the labels as they are and moving $\p_v$ and $\overline\p_v$ through the graph. Equating the two expressions gives us 
\bea A_5(v)+c.c.=-\pi D_5^{(1,2,1;1)} (v) +\pi D_3G_2(v) -\pi D^{(1,1,3)}(v) +\pi D^{(1,1,1,2)} (v) -\pi E_2 G_3(v).\eea

\begin{figure}[ht]
\begin{center}
\[
\mbox{\begin{picture}(110,75)(0,0)
\includegraphics[scale=.7]{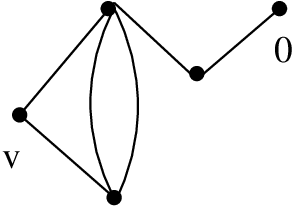}
\end{picture}}
\]
\caption{$A_6(v)$} 
\end{center}
\end{figure}

\begin{figure}[ht]
\begin{center}
\[
\mbox{\begin{picture}(110,100)(0,0)
\includegraphics[scale=.65]{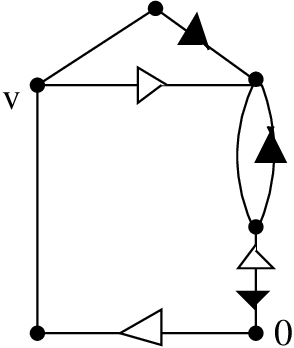}
\end{picture}}
\]
\caption{Auxiliary graph $B_1(v)$} 
\end{center}
\end{figure}

\section{Simplifying $F_{36}(v)$}

We have that
\bea F_{36}(v) = \pi E_2 E_3 - 2 F_3(v) - \frac{2}{\pi} F_{55}(v).\eea
Now using \C{F55} and appendix A, we get that
\bea \frac{F_{36}(v)}{\pi} &= & \frac{2}{\pi} \p_v G_3(v) \overline\p_v G_3 (v) + 2D^{(1,3,1)} (v) +2 D^{(1,2,2)} (v) + 3D^{(2,2,1)} (v)\non \\ &&- D_5^{(1,2,1;1)} (v) + D^{(1,1,2,1)}(v) + D_5^{(1;2;2)} (v)+ 2 G_5(v)-C_{1,2,2}\non \\&& + C_{1,1,3} -2 E_5 + E_2 E_3 - E_2 G_3(v) - 2 E_3G_2(v)-G_2(v)D_3^{(1)}(v).  \eea

\section{Simplifying $F_{111}(v) +c.c.$}

\begin{figure}[ht]
\begin{center}
\[
\mbox{\begin{picture}(220,140)(0,0)
\includegraphics[scale=.65]{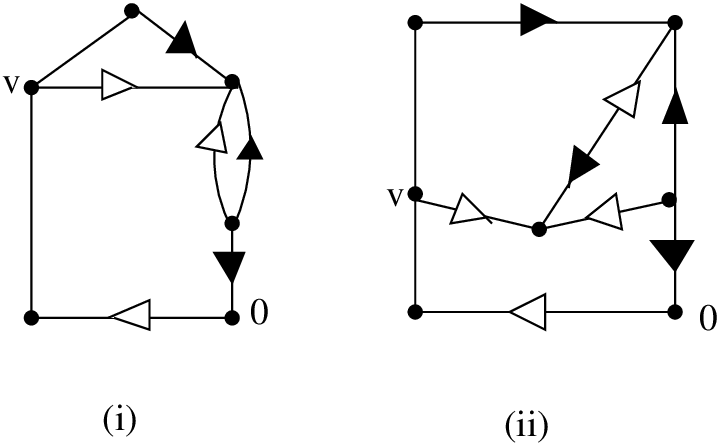}
\end{picture}}
\]
\caption{(i) $B_2(v)$ and (ii) auxiliary graph $B_3(v)$} 
\end{center}
\end{figure}

To simplify $F_{111}(v)$, we start with the auxiliary graph $B_1(v)$ in figure 110, to get
\bea F_{111}(v)+c.c. = \pi\Big(\frac{B_2(v)}{\pi^2}+F_{17}(v)-F_{40}(v)+c.c.\Big),\eea
where the graph $B_2(v)$ is in figure 111. To simplify it, we start with the auxiliary graph $B_3(v)$ in the same figure. This gives us 
\bea \label{111}&&F_{111}(v)+\tau_2P_3(v) \Big(\overline\p_v G_2(v)\Big)^2+c.c.= \pi \tau_2E_2 \p_v G_2(v)\overline\p_v G_2(v)\non \\ &&-\pi \tau_2 G(v)^2\p_v G_2 (v)\overline\p_v G_2(v) -\pi\tau_2 \Big(\p_v G_2(v) \overline\p_v G_4(v)+c.c.\Big) \non \\ &&-\pi\tau_2 G(v)\Big(\p_v G_2(v) \overline\p_v G_3(v)+c.c.\Big)-\pi\Big(F_{25}(v) -2 F_{35}(v)-\frac{F_{84}(v)}{2}+c.c.\Big).   \eea
\begin{figure}[ht]
\begin{center}
\[
\mbox{\begin{picture}(100,100)(0,0)
\includegraphics[scale=.65]{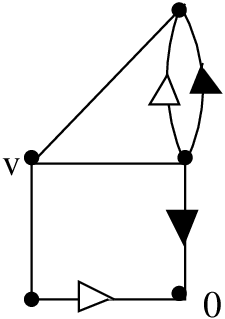}
\end{picture}}
\]
\caption{$B_4(v)$} 
\end{center}
\end{figure}
In an intermediate step, we have used the relation
\bea \frac{B_4(v)}{\pi} = \frac{\tau_2}{2} E_2 \p_v G_2(v) \overline\p_v G_2(v) -F_{35}(v)^*-\frac{F_{89}(v)^*}{2} ,\eea
where the graph $B_4(v)$ is given in figure 112. 

\begin{figure}[ht]
\begin{center}
\[
\mbox{\begin{picture}(100,70)(0,0)
\includegraphics[scale=.75]{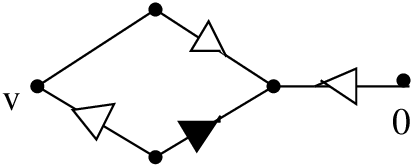}
\end{picture}}
\]
\caption{$B_5(v)$} 
\end{center}
\end{figure}
Now the second term on the left hand side of \C{111} is a graph which does not arise in the analysis elsewhere in the main text, and hence needs to be expressed in terms of the other graphs to see cancellations.        
To simplify this, we start with the graph $B_5(v)$ in figure 113 to get
\bea \tau_2\Big(\overline\p_v G_2(v)\Big)^2 = P_4(v)^* -2 P_5(v)^*,\eea 
leading to
\bea \tau_2P_3(v) \Big(\overline\p_v G_2(v)\Big)^2+c.c = P_3(v)P_4(v)^* -2 P_3(v)P_5(v)^* +c.c..\eea

\section{Simplifying graphs that arise in analyzing $D_5^{(2)}(v)$}

\subsection{Relating $F_{94}(v)$ and $F_{107}(v)$}

We see that
\be \label{94107}F_{94}(v)= F_{107}(v),\ee
which follows from manipulating the graph $C_1(v)$ in figure 114.
\begin{figure}[ht]
\begin{center}
\[
\mbox{\begin{picture}(90,95)(0,0)
\includegraphics[scale=.65]{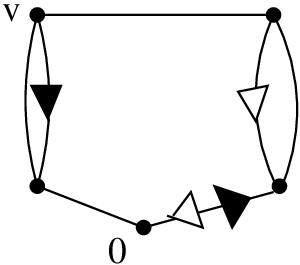}
\end{picture}}
\]
\caption{$C_1(v)$} 
\end{center}
\end{figure}

\subsection{Simplifying $F_{108}(v)$}
Now
\be F_{108}(v)= -\tau_2\overline\p_v G_2(v) C_2(v),\ee
where the graph $C_2(v)$ is given in figure 115.
\begin{figure}[ht]
\begin{center}
\[
\mbox{\begin{picture}(140,50)(0,0)
\includegraphics[scale=.7]{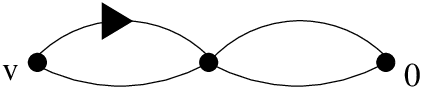}
\end{picture}}
\]
\caption{$C_2(v)$} 
\end{center}
\end{figure}

We have that
\be C_2(v)=\frac{2}{\pi}C_3(v),\ee
where the graph $C_3(v)$ is given in figure 116 which we simplify using the auxiliary graph $C_4(v)$ in the same figure.
\begin{figure}[ht]
\begin{center}
\[
\mbox{\begin{picture}(320,100)(0,0)
\includegraphics[scale=.7]{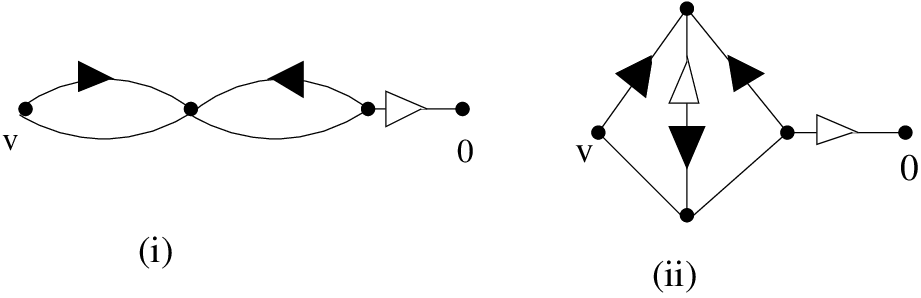}
\end{picture}}
\]
\caption{(i) $C_3(v)$ and auxiliary graph (ii) $C_4(v)$} 
\end{center}
\end{figure}

This leads to the relation
\bea F_{108}(v)&=& 2 \tau_2G_2(v) \p_v G_2(v)\overline\p_v G_2(v) + 4 \tau_2\p_v G_4(v) \overline\p_v G_2(v) \non \\ && -4 F_{40}(v)+\frac{2}{3} F_{89}(v)^* -2\tau_2E_2 \p_v G_2(v) \overline\p_v G_2(v).\eea

\subsection{Simplifying $F_{46}(v)$}
Starting from 
\be F_{46}(v)= C_5(v)\ee
where the graph $C_5(v)$ is given in figure 117, we get that
\begin{figure}[ht]
\begin{center}
\[
\mbox{\begin{picture}(200,50)(0,0)
\includegraphics[scale=.6]{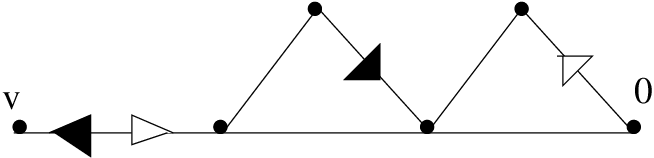}
\end{picture}}
\]
\caption{$C_5(v)$} 
\end{center}
\end{figure}
\bea F_{46}(v)= D_5^{(1,2,2)} (v) + D^{(1,1,3)} (v) - E_2 E_3 - E_2 D_3^{(1)} (v)+ F_{58}(v),\eea
which can be simplified using $F_{60}(v)$ to give us 
\bea F_{46}(v)&=& D_5^{(1,2,2)} (v) +D^{(1,1,3)} (v)- D^{(2,2,1)} (v)-2D^{(1,2,2)}(v)-E_2E_3  \non \\ &&-E_2 D_3^{(1)}(v) +E_5+ G_5(v) +\frac{1}{\pi}\Big(F_1(v)+F_2(v)+F_9(v)+c.c.\Big)\non \\ &&-\frac{F_7(v)}{\pi} +\frac{F_{61}(v)}{\pi}-\frac{\tau_2}{\pi} \p_v G_3(v)\overline\p_v G_3(v).\eea  

\section{Eigenvalue equation for $C_{a,b,c} (Z)$}

\begin{figure}[ht]
\begin{center}
\[
\mbox{\begin{picture}(90,105)(0,0)
\includegraphics[scale=.55]{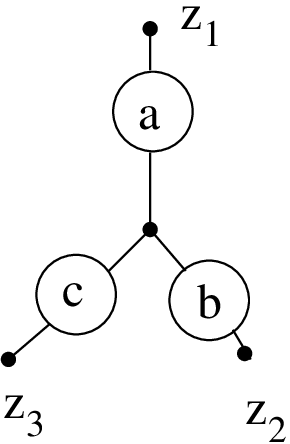}
\end{picture}}
\]
\caption{$C_{a,b,c}(Z)$} 
\end{center}
\end{figure}

Though we have only considered elliptic modular graphs with two unintegrated vertices in this paper, one can generalize the analysis to graphs with more unintegrated vertices. As a very simple example involving graphs with one cubic vertex where three vertices are unintegrated, consider the graph    
\bea 
C_{a,b,c}(Z) = {\mathcal{C}}^+\begin{bmatrix}
a&b&c\\
a&b&c\\
z_1&z_2&z_3
\end{bmatrix}
\eea
defined in section 3.5.1 of~\cite{DHoker:2020hlp}, which is given in figure 118, where a chain with $s$ links is defined in figure 119.    

\begin{figure}[ht]
\begin{center}
\[
\mbox{\begin{picture}(260,60)(0,0)
\includegraphics[scale=.6]{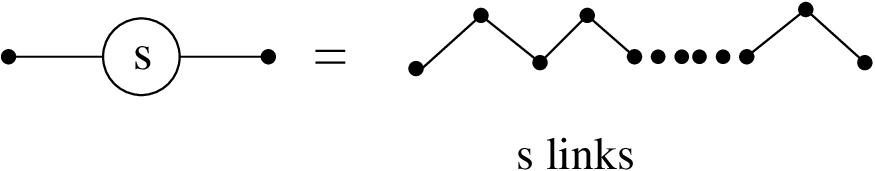}
\end{picture}}
\]
\caption{A chain with $s$ links} 
\end{center}
\end{figure}
Proceeding along the lines of the analysis in the main text, appendix A, and using the identity
\be \frac{1}{\pi}\Big(\widetilde{C}_{A,B,C} (Z)+c.c.\Big) = C_{A+1.B,C} (Z) +C_{A,B+1,C} (Z)- C_{A+1,B+1,C-1}(Z),\ee
where $\widetilde{C}_{A,B,C}(Z)$ is given in figure 120,  
we easily obtain the eigenvalue equation satisfied by $C_{a,b,c} (Z)$ 
\begin{figure}[ht]
\begin{center}
\[
\mbox{\begin{picture}(150,80)(0,0)
\includegraphics[scale=.6]{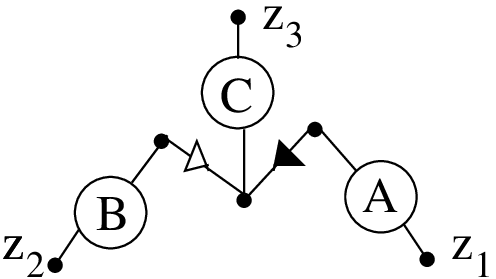}
\end{picture}}
\]
\caption{$\widetilde{C}_{A,B,C} (Z)$} 
\end{center}
\end{figure}
given by
\bea \label{Cabc}&&\Big(\Delta -a(a-1) -b(b-1)-c(c-1)\Big)C_{a,b,c}(Z) = \non \\ &&ab\Big[C_{a+1,b-1,c}+C_{a-1,b+1,c} + C_{a+1,b+1,c-2}-2 C_{a+1,b,c-1}-2 C_{a,b+1,c-1}\Big](Z)\non \\ &&+bc\Big[C_{a,b+1,c-1}+C_{a,b-1,c+1} + C_{a-2,b+1,c+1}-2 C_{a-1,b+1,c}-2 C_{a-1,b,c+1}\Big](Z)\non \\ &&+ca \Big[C_{a-1,b,c+1}+C_{a+1,b,c-1} + C_{a+1,b-2,c+1}-2 C_{a,b-1,c+1}-2 C_{a+1,b-1,c}\Big](Z),\eea
reproducing (3.46) of~\cite{DHoker:2020hlp}. Since $a,b,c \geq 1$, the right hand side of \C{Cabc} can contain terms where the label is 0 or $-1$. Such terms can be simplified using the identity in figure 121 for $n \geq 0$ where the right hand side stands for
\be \Big(\frac{\tau_2}{\pi}\Big)^{n+1} \p_z^{n+1}\overline\p_w^{n+1} G(z,w),\ee
which follows from \C{Green}~\cite{Basu:2019idd}.
\begin{figure}[ht]
\begin{center}
\[
\mbox{\begin{picture}(230,50)(0,0)
\includegraphics[scale=.6]{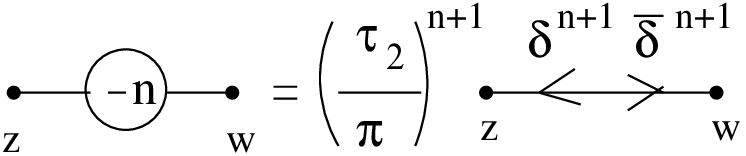}
\end{picture}}
\]
\caption{An identity for vanishing and negative labels} 
\end{center}
\end{figure}

This immediately leads to the identities
\bea &&C_{a,b,0}(Z) = G_a (z_1,z_3) G_b(z_2,z_3)-G_{a+b}(z_1,z_2),\non \\
&&C_{a,b,-1}(Z)=G_{a-1}(z_1,z_3) G_b(z_2,z_3) + G_a(z_1,z_3) G_{b-1}(z_2,z_3)\non \\&& -\frac{\tau_2}{\pi}\Big( \int_z \p_{z_3}G(z_3,z)G_{a-1}(z,z_1) \int_w\overline\p_{z_3} G(z_3,w)G_{b-1}(w,z_2)+c.c.\Big),\eea
reproducing (3.47) and (3.48) of~\cite{DHoker:2020hlp}.

\providecommand{\href}[2]{#2}\begingroup\raggedright\endgroup


\end{document}